\documentclass[%
 reprint,
 superscriptaddress,
 nofootinbib,
 amsmath,amssymb,
 aps,
 prb,
]{revtex4-2}

\usepackage{graphicx}
\usepackage{dcolumn}
\usepackage{bm}
\usepackage[dvipsnames]{xcolor}
\usepackage{physics}
\usepackage{soul}
\usepackage{subcaption}
\newcommand*{\rttensor}[1]{\overline{\overline{#1}}}

\usepackage{hyperref}
\hypersetup{
     colorlinks = true,
     citecolor  = blue,
     linkcolor  = blue,
     urlcolor   = blue
}

\begin{document}

\title{Classical and quantum theory of magnonic and magnetoelastic nonlinear dynamics in continuum geometries
}
\author{Marco Brühlmann}
 \email{marco.bruehlmann@tuwien.ac.at}
\affiliation{%
 Institute for Theoretical Physics and Vienna Center for Quantum Science and Technology, Vienna University of Technology, A-1040 Vienna, Austria
}%

\author{Yunyoung Hwang}

\affiliation{%
 CEMS, RIKEN, 2-1, Hirosawa, Wako 351-0198, Japan
}%
\affiliation{%
 Institute for Solid State Physics, University of Tokyo, Kashiwa 277-8581, Japan
}%

\author{Jorge Puebla}%
\affiliation{%
 CEMS, RIKEN, 2-1, Hirosawa, Wako 351-0198, Japan
}%
\affiliation{%
 Department of Electronic Science and Engineering, Kyoto University, Kyoto, Kyoto 615-8510, Japan
}%

\author{Carlos Gonzalez-Ballestero}%
 \email{carlos.gonzalez-ballestero@tuwien.ac.at}
\affiliation{%
 Institute for Theoretical Physics and Vienna Center for Quantum Science and Technology, Vienna University of Technology, A-1040 Vienna, Austria
}%

\date{\today}

\begin{abstract}
We provide a theory of spin and acoustic wave coupled nonlinear dynamics in continuum systems. Combining the Landau-Lifshitz-Gilbert equations with the magnetoelastic Hamiltonian, we derive classical equations of motion for the magnetization and acoustic wave amplitudes, which include magnonic nonlinearity -- both three- and four-magnon processes -- as well as linear and nonlinear magnetoelastic interactions. We focus on two-dimensional magnetic films sustaining surface acoustic waves, a geometry
where our model successfully reproduces our recent experimental observation of phonon-to-magnon down conversion under acoustic drive. We provide analytical expressions for all the rates in our equations, which make them particularly suitable for quantization. We then quantize our model, deriving Heisenberg-Langevin equations of motion for magnon and phonon operators, and show how to compute quantum expectation values in the mean-field approximation. Our work paves the way toward acoustic control of magnons in the quantum regime.
\end{abstract}

\maketitle

\section{Introduction}

Spin waves in ferromagnetic materials, or magnons, have recently been proposed as components in hybrid quantum platforms~\cite{Lachance-Quirion_2019}. This is due, among others, to their external frequency tunability, strong nonlinearity~\cite{Suhl1957,akhiezer1968spin,prabhakar2009spin} -- a key requisite to prepare  quantum states -- and their ability to couple, both linearly and nonlinearly, to multiple degrees of freedom such as qubits~\cite{WolskiPRL2020,Lachance2020,Bertelli_Magnetic_2020,FukamiPRXQ2021,gonzalez-ballestero_towards_2022,xu_quantum_2022,FukamiPNAS2024,DeyNanoletters2025,XueArxiv2025}, microwave and optical fields~\cite{RameshtiPhysRep2022,almpanis2021optomagnonic}, and acoustic phonons. Several works have explored the potential of magnon nonlinearity~\cite{WangPRB2016,WangPRL2018,ElyasiPRB2020,LeePRL2023,EtesamiradPRApp2023,ElyasiPRB2024,ElyasiPRB2024B,MookPRB2025,ArfiniArxiv2025} and magnon-phonon interaction~\cite{Xufeng_Cavity2016,gonzalez-ballestero_quantum_2020,gonzalez-ballestero_theory_2020,PottsPRX2021,KounalakisPRB2023,PottsPRB2023,HuangArxiv2025} for quantum magnonics, mostly in magnetic structures confined in three dimensions (e.g. spheres) and/or for single magnon modes (e.g., Kittel mode). 

Extending the scope of quantum nonlinear magnonic and magnon-phonon processes to systems where magnons form a continuum of modes (e.g., films or waveguides) would enable to import the complex structures developed in classical magnonics~\cite{ChumakNatComm2014,WangNatElect2020,roadmap} into the quantum regime, and facilitate their future hybridization with state-of-the-art acoustic quantum platforms, prominently based on surface acoustic waves~\cite{SchuetzPRX2015,ManentiNatComm2017,SatzingerNature2018}. Moreover, it would pave the way to preparing quantum magnonic states via nonlinear processes (e.g., parametric down conversion), without the need for qubits. This would significantly reduce the complexity of current quantum magnonics experiments. 
Recent theoretical advances \cite{RezendeIEEE1990,DreherPRB2012,KikkawaPRL2016,LIJAP2017,LisenkovPRB2019,ZhangPRL2020,BabuNanoletters2021,CerkaskiiArxiv2025}, as well as experimental achievements in the control of  linear~\cite{AnPRB2020,XuSciAdv2020,FungMQT2021,HatanakaPRApplied2022,SchlitzPRB2022,GaoAPL2022,HiokiCommPhys2022,HwangAMI2022,hwang_strongly_2024,MaezawaPRApplied2024,KunzAPL2024,JanderJAP2025} and nonlinear~\cite{ChangPRB2017,KraimiaPRB2020,GeilenAPR2025,HwangArxiv2025} magnetoelastic processes in continuum systems suggest that these goals are near reach. One of the roadblocks in this path is the lack of a full quantum theory of nonlinear magnon-phonon dynamics.

In this paper, we derive a nonlinear theory of magnon and phonon dynamics in continuum systems. As a case study we focus on a thin magnetic film lying on a substrate supporting surface acoustic waves. With quantization of the theory in mind, we first derive in Sec.~\ref{sec:system_and_linear} the linear dynamics of magnons and acoustic waves analytically for this geometry. Then, in Sec.~\ref{sec:nonlinear_EOM}, we include in our model linear and nonlinear magnetoelastic coupling as well as three- and four-magnon scattering. We use the derived equations in Sec.~\ref{sec:parametrc_instability} to explore the nonlinear magnetization response generated by an acoustic drive, and quantitatively predict threshold driving strengths for three-wave parametric instability (magnon-to-magnon or phonon-to-magnon parametric down-conversion).  We show how the competition between different processes leading to parametric instability critically depends on system parameters. Our theory  accurately reproduces our recent experimental observations~\cite{HwangArxiv2025}. 
In Sec.~\ref{sec:quantum}, we quantize the above theory and show how to compute expectation values of observables -- e.g., magnetization -- within mean-field theory. Finally, our conclusions are presented in Sec.~\ref{sec:conclusion}.

\section{System and independent linear dynamics}\label{sec:system_and_linear}
\begin{figure}
    \centering
    \includegraphics[width=1\linewidth]{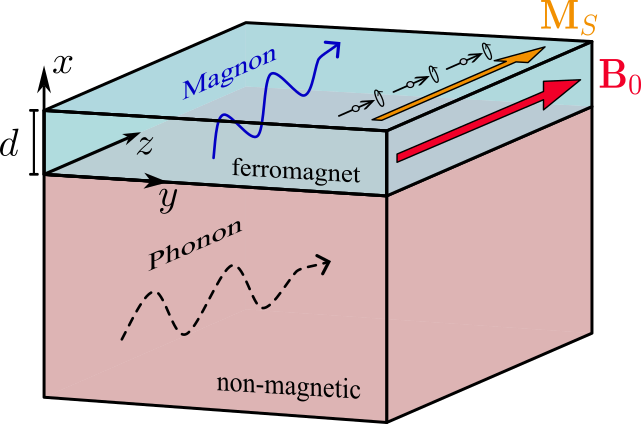}
    \caption{
    We consider an infinitely extended ferromagnetic thin film with thickness $d$ in the presence of a uniform magnetic field $\mathbf{B}_0$ which leads to saturation of the magnetization. Magnons in the thin film couple, both linearly and nonlinearly, to surface acoustic waves sustained in the whole film and susbtrate system. 
    }
    \label{fig:system_schematic}
\end{figure}
The system considered in this article is schematically depicted in Fig. \ref{fig:system_schematic}. It consists of an infinitely extended thin film of ferromagnetic material (thickness $d$) lying on top of a non-magnetic substrate. The latter is assumed to occupy all the region $x<0$. An in-plane magnetic field $\mathbf{B}_0=B_0\mathbf{e}_z$ is applied to saturate the magnetization of the ferromagnet along the $z$ axis. The magnetic film is able to sustain spin waves (magnons), and both the film and the substrate can sustain acoustic vibrations (phonons). In this section we briefly introduce the linear theory of magnons and phonons in this structure, which has been derived in detail in our work in preparation \cite{Bruehlmann_unpublished}. This theory, and in particular the magnonic and acoustic eigenmodes, are key ingredients to theoretically describe the nonlinear dynamics in the next section. 

\subsection{Magnon dynamics}\label{subsec:linear_magnon_dynamics}
\begin{figure*}[t]
    \centering
    \includegraphics[width=1\linewidth]{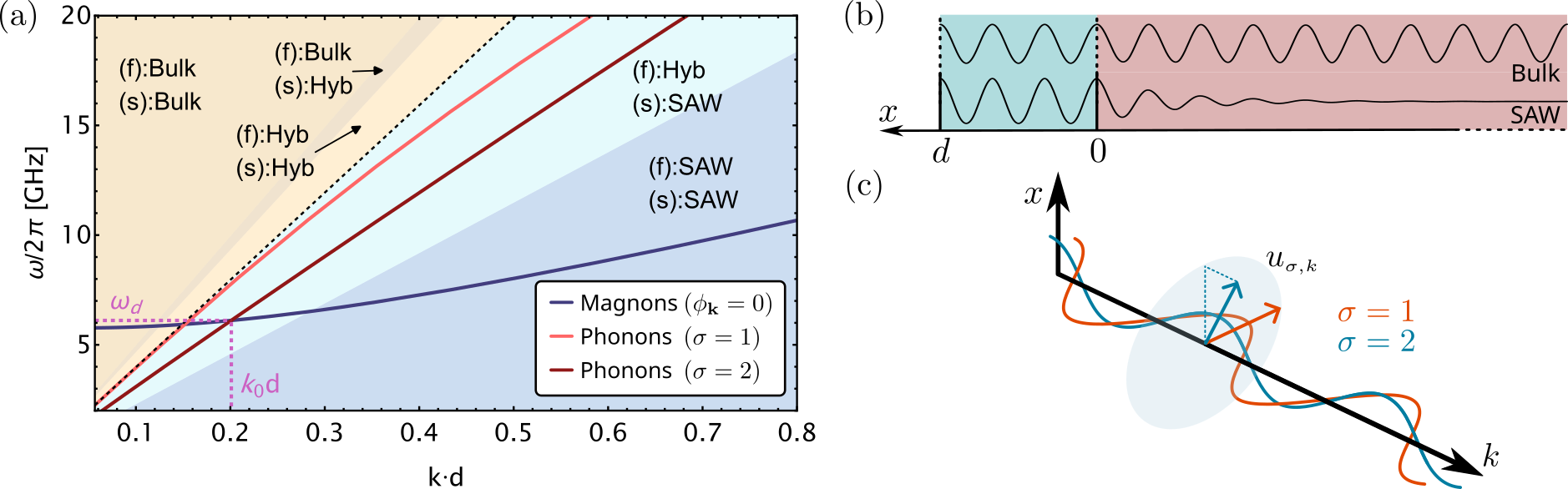}
    \caption{
    (a) Dispersion relation of lowest-band ($n=0$) phonon and magnon modes propagating parallel to the external magnetic field, for the parameters of Table~\ref{tab:parameters_1}. Background colors designate the different regimes of phonon mode types, i.e. bulk, hybrid, or SAW within the film (f) or substrate (s). Below the black dotted line modes are SAWs in the substrate, i.e., they decay exponentially inside the substrate. The pink dotted lines mark the driving frequency $\omega_\text{d}/2\pi=6.11\,\mathrm{GHz}$ and wave vector amplitude $k_0=2\pi/625\,\mathrm{nm}^{-1}$ of the acoustic mode we use for our results in Sec.~\ref{subsec:parametric_excitations_threshold_ordering}. (b) A schematic depiction of the difference between bulk (above) and SAW (below) phonon modes. (c) Schematic visualization of phonon mode families. Modes with polarization index $\sigma=1$ (orange) are purely transversal  $\mathbf{u}_{1,k}(\mathbf{r})\cdot\mathbf{k}=0$, while modes with $\sigma=2$ have a longitudinal component, $\mathbf{u}_{2,k}(\mathbf{r})\cdot\mathbf{k}\ne0$.}
    \label{fig:Dispersion_relation_and_SAW}
    \vspace{-0.3cm}
\end{figure*}
Let us start by describing the linear dynamics of the spin waves. The classical dynamics of the magnetization field $\textbf{M}(\textbf{r},t)$ within the magnetic film is described by the Landau-Lifshitz Gilbert (LLG) equation  \cite{prabhakar2009spin,landau1980statisticalII}
\begin{align}
\begin{split}\label{eq:LLG}
    \frac{d}{dt} &\textbf{M}(\textbf{r},t) = \gamma_g \mu_0 \textbf{M}(\textbf{r},t) \times \textbf{H}_{\mathrm{eff}}(\mathbf{M},\textbf{r},t) \\
    &+ \frac{\lambda}{M_S}(\gamma_g \mu_0) \mathbf{M}(\textbf{r},t)\times(\textbf{M}(\textbf{r},t) \times \textbf{H}_{\mathrm{eff}}(\mathbf{M},\textbf{r},t)).
\end{split}
\end{align}
The first term on the right-hand side is responsible for coherent precession of the magnetization, while the second term is responsible for damping. The strength of this Gilbert damping term is encoded in the dimensionless Gilbert damping coefficient $\lambda$. The quantities $\gamma_g$ and $\mu_0$ are the gyromagnetic ratio and the vacuum permeability, respectively. Finally, the effective magnetic field $\textbf{H}_{\mathrm{eff}}(\mathbf{r},t) = H_0 \textbf{e}_z + \textbf{H}_x(\mathbf{r},t) +  \textbf{H}_{\mathrm{dm}}(\mathbf{r},t) + \mathbf{H}_{\text{an}}(\mathbf{r},t)$ contains the uniform external field $H_0\equiv B_0/\mu_0$ and a set of effective fields capturing the effects of microscopic interactions. First, the exchange interaction is represented by $\textbf{H}_x(\textbf{M},\textbf{r},t) =  \alpha_x \nabla^2 \textbf{M}(\textbf{r},t)$, where the strength of exchange interactions is given by $\alpha_x = \frac{A_x}{\mu_0 M_S^2}$ with $A_x$ being the exchange stiffness and $M_S$ the saturation magnetization. Second, dipole-dipole interactions produce a demagnetizing field $\textbf{H}_{\text{dm}}(\textbf{M},\textbf{r},t) = \frac{1}{4\pi} \nabla \int d^3 \textbf{r}^\prime \, \frac{\nabla_{\textbf{r}^\prime} \textbf{M}(\textbf{r}^\prime,t) }{|\textbf{r}- \textbf{r}^\prime|}$. Third, magnetocrystalline anisotropy can be described by a field $\mathbf{H}_{\rm an}(\mathbf{r},t)$. Hereafter we ignore this anisotropy field as its effects are small in many  of the  magnetic materials used in magnetoelastic experiments (e.g., CoFeB~\cite{hwang_strongly_2024,HwangArxiv2025} or nickel~\cite{WeilerPRL2011,DreherPRB2012,HatanakaPRApplied2022}), although including it in our model is straightforward. Since the external magnetic field $\mathbf{B}_0$  saturates the magnetization in the ferromagnet, we can write the magnetization vector within the spin-wave approximation as 
\begin{align}\label{eq:magnetization}
    \mathbf{M}(\mathbf{r},t)=(M_S-\delta m(\mathbf{r},t)) \mathbf{e}_z+ \mathbf{m}(\mathbf{r},t),
\end{align} 
where $M_S$ is the saturation magnetization and $\mathbf{m}\ll M_S$ describes small oscillations of the magnetization in the $x,y$ plane. The eigenmodes of these oscillations are known as spin waves. The correction $\delta m \approx (\mathbf{m}\cdot\mathbf{m})/(2M_S)$ ensures that the magnetization amplitude $|\mathbf{M}|$ remains constant up to second order in $\mathbf{m}/M_S$. This correction will contribute significantly to nonlinear magnon processes.

To describe spin waves in the linear regime, we introduce Eq.~\eqref{eq:magnetization} into the LLG equation Eq.~\eqref{eq:LLG} and linearize it by neglecting any nonlinear terms in the small variable $\mathbf{m}(\mathbf{r},t)/M_S$. We obtain the equation 
\begin{align}\label{LLGlinearized}
    \frac{d}{dt} \mathbf{m}(\mathbf{r},t) =\mathcal{D}_0 [ \mathbf{m}(\mathbf{r},t) ] - \lambda\mathbf{e}_z \times \mathcal{D}_0 [ \mathbf{m}(\mathbf{r},t) ],
\end{align}
where we define the linear integro differential operator
\begin{multline}
    \mathcal{D}_0 [ \mathbf{m}(\mathbf{r},t) ] \equiv \mathbf{e}_z \times (\omega_H - \omega_M \alpha_x \nabla^2) \mathbf{m}(\mathbf{r},t) \\
    - \omega_M \mathbf{e}_z \times \frac{1}{4\pi} \nabla \int d^3 \textbf{r}^\prime \, \frac{\nabla_{\textbf{r}^\prime} \textbf{m}(\textbf{r}^\prime,t) }{|\textbf{r}- \textbf{r}^\prime|},
\end{multline}
with $\omega_M \equiv |\gamma_g|\mu_0 M_S$, and $\omega_H \equiv |\gamma_g|B_0$. 
Since Eq.~\eqref{LLGlinearized} is linear it allows to define magnetization eigenmodes, i.e., it allows for a decomposition of the magnetization field as
\begin{align}\label{eq:magnetization_decomposition}
    \mathbf{m}(\mathbf{r},t) = \int d^2\mathbf{k} s_\mathbf{k}(t) \mathbf{m}_\mathbf{k}(\mathbf{r}) + c.c. 
\end{align}
Here, $\mathbf{m}_\mathbf{k}(\mathbf{r})$ is the mode function of an eigenmode labeled by an in-plane wave vector $\mathbf{k} = k_y\mathbf{e}_y+k_z\mathbf{e}_z = k\sin(\phi_\mathbf{k})\mathbf{e}_y+k\cos(\phi_\mathbf{k})\mathbf{e}_z$, where $k$ is the wave vector modulus and $\phi_\mathbf{k}$ the angle between $\mathbf{k}$ and the applied field $\mathbf{B}_0$. Each eigenmode has an associated frequency $\omega^\text{m}_\mathbf{k}$. The coefficients $s_\mathbf{k}(t)$ in Eq.~\eqref{eq:magnetization_decomposition} describe the amplitudes associated to each mode and are the true dynamical variables of the spin waves in the eigenmode basis. 
Note that in general a film sustains an infinite number of discrete bands $n=0,1,2...$ due to the confinement along the $x$ direction \cite{kalinikos_theory_1986,gonzalez-ballestero_towards_2022}. For thin films, however, a single-band description is sufficient since all bands above the fundamental one $n=0$  are far detuned from the frequency range of interest. 

For convenience, we write the decomposition Eq.~\eqref{eq:magnetization_decomposition} in terms of eigenmodes of the \textit{lossless} LLG, i.e., we choose these eigenmodes to be the solutions to the eigenvalue equation
\begin{align}\label{eigenmodeEq}
    -i\omega^\text{m}_\mathbf{k} \mathbf{m}_\mathbf{k}(\mathbf{r}) =\mathcal{D}_0 [ \mathbf{m}_\mathbf{k}(\mathbf{r}) ].
\end{align}
The choice of these eigenmodes is just convenient in order to consistently quantize the theory later on~\cite{mills_quantum2006}. It has been shown that the resulting quantized theory (see Sec.~\ref{sec:quantum}) provides, in the classical regime, the same physical predictions obtained using the conventional (lossy) eigenmodes of the LLG, provided that an appropriate loss term is added to the quantum dynamical equations \cite{gonzalez-ballestero_theory_2020,gonzalez-ballestero_towards_2022}. We remark that despite our labeling of these modes as ``lossless'', Gilbert damping is fully included in our dynamical equations.  The eigenmode solutions of Eq.~\eqref{eigenmodeEq} can be found in the literature \cite{kalinikos_theory_1986,gonzalez-ballestero_towards_2022} and, for the lowest magnon band ($n=0$), they read
\begin{gather}\label{eq:magnon_modefunction}
    \mathbf{m}_\mathbf{k}(\mathbf{r}) = \mathcal{N}_\mathbf{k} e^{i\mathbf{kr}_{\scriptscriptstyle{\parallel}}} \begin{pmatrix}
        \sqrt{\nu_y(\mathbf{k})} \\ i \sqrt{\nu_x(\mathbf{k})} \\ 0 
    \end{pmatrix}, 
\end{gather}
with
\begin{align}
    \nu_x(\mathbf{k}) &\equiv \frac{\omega_H}{\omega_M} + \alpha_x k^2 - \frac{( e^{-k d } -1)}{k d}, \\
    \nu_y(\mathbf{k}) &\equiv \frac{\omega_H}{\omega_M} + \alpha_x k^2 + \frac{k_y^2}{k^2} \left( 1 +  \frac{( e^{-k d } -1)}{k d}\right),
\end{align}
The variable $\mathbf{r}_{\scriptscriptstyle{\parallel}} = (0,y,z)$ denotes an in-plane position vector and  $\mathcal{N}_\mathbf{k} = \sqrt{\omega_M/(2 d(2\pi)^2 \omega^\text{m}_\mathbf{k})}$ is a normalization factor. The eigenmode frequencies (magnon dispersion relation) are given by
\begin{align}\label{eq:magnon_eigenfreq}
    \omega^\text{m}_\mathbf{k} = \omega_M \sqrt{\nu_x(\mathbf{k})\nu_y(\mathbf{k})},
\end{align}
and are shown in Fig. \ref{fig:Dispersion_relation_and_SAW}\textcolor{blue}{(a)} for $\phi_\mathbf{k}=0$.

The eigenmode functions Eq.~\eqref{eq:magnon_modefunction} form an orthonormal basis,
\begin{equation}\label{magnonorthogonality}
    \bra{\mathbf{m}_\mathbf{k}}\ket{\mathbf{m}_{\mathbf{k}'}}= i \, \delta(\mathbf{k}-\mathbf{k}^\prime),
\end{equation}
with the inner product defined as
\begin{align}\label{innerproduct}
    \bra{\mathbf{g}}\ket{\mathbf{f}} &\equiv \int d\mathbf{r}\, (\mathbf{e}_z \times \mathbf{g}^*(\mathbf{r})) \cdot \mathbf{f}(\mathbf{r}).
\end{align}
Using this orthonormal basis the mode amplitudes $s_\mathbf{k}(t)$ are determined by taking the left overlap of Eq.~\eqref{eq:magnetization_decomposition} with a specific eigenmode, 
\begin{equation}
    s_\mathbf{k}(t) = -i \bra{\mathbf{m}_\mathbf{k}}\ket{\mathbf{m}(\mathbf{r},t)}.
\end{equation}
Introducing the decomposition Eq.~\eqref{eq:magnetization_decomposition} into the linearized LLG equation Eq.~\eqref{LLGlinearized} and using the above identities we obtain the following differential equation for the magnon amplitudes,
\begin{equation}\label{skeqlinear}
    \dot{s}_\mathbf{k} = (-i\omega^{\text{m}}_\mathbf{k} - \gamma^{\text{m}}_\mathbf{k}/2)s_\mathbf{k}.
\end{equation}
This equation is easily derived by noting that $\bra{\mathbf{m}_\mathbf{k}}\ket{\mathbf{m}_\mathbf{q}^*} = 0$ and $(\mathbf{e}_z\times\mathbf{m}_\mathbf{k}) \perp \mathbf{e}_z$, and contains the decay rate $\gamma^\text{m}_\mathbf{k}$ which originates from the Gilbert damping term, 
\begin{align}\label{eq:Gilbert}
\begin{split}
    \frac{\gamma_{\mathbf{k}}^{\text{m}}}{2} s_{\mathbf{k}} &\equiv -i\int d^2\mathbf{q} \bra{\mathbf{m}_\mathbf{k}} \ket{-\lambda \mathbf{e}_z \times (-i\omega_{\mathbf{q}}) s_{\mathbf{q}} \mathbf{m}_{\mathbf{q}} } \\
    &\approx \frac{\lambda}{2} \omega_M (\nu_x(\mathbf{k})+\nu_y(\mathbf{k})) s_{\mathbf{k}},
\end{split}
\end{align}
where in the last line we have introduced the damping rate expression obtained from phenomenological loss theory~\cite{gonzalez-ballestero_towards_2022,Bruehlmann_unpublished}.
The mode amplitude differential equation Eq.~\eqref{skeqlinear} can be  solved as
\begin{equation}\label{eq:sklinear}
    s_\mathbf{k}(t) = s_\mathbf{k}(0)e^{(-i\omega^{\text{m}}_\mathbf{k} - \gamma^{\text{m}}_\mathbf{k}/2) t},
\end{equation}
which fully determines the magnetization field at all times with given initial conditions.
This concludes the formulation of linear spin wave theory.

\begin{table}[t]
    \centering
    \begin{tabular}{l|l|l}
    \hline\hline
        Parameter & & Value\\ \hline
        Thickness & $d$ & $20\,\mathrm{nm}$\\
        Mass density  & $\{ \rho_{s}, \rho_{f} \}$ & $\{4650,8382 \}\,\mathrm{kg\,m^{-3}}$ \\
        Longitudinal sound velocity & $\{ c_{l,s}, c_{l,f}\}$ & $\{ 6247,5860 \}\,\mathrm{m\,s^{-1}}$ \\
        Transverse sound velocity & $\{ c_{t,s}, c_{t,f}\}$ & $\{ 5000,2885 \}\,\mathrm{m\,s^{-1}}$ \\
        Gyromagnetic ratio & $\gamma_g$ & $-1.76\times 10^{11}\, \mathrm{T^{-1}\,s^{-1}}$ \\
        External field & $\mu_0 H_0$ & $30\,\mathrm{mT}$ \\
        Saturation magnetization & $\mu_0 M_S$ & $1.4\, \mathrm{T}$ \\
        Gilbert damping coefficient & $\lambda$ & $7.319\times 10^{-3}$\\
        Phonon quality factor & $Q_0$ & $490$\\
        Exchange stiffness & $A_x$ & $7.132\times 10^{-11} \, \mathrm{J\,m^{-3}}$\\
        Magnetoelastic constant & $B_1$ & $-4\times 10^7 \, \mathrm{J\,m^{-3}}$\\
        Magnetoelastic constant & $B_2$ & $-2\times 10^7 \, \mathrm{J\,m^{-3}}$\\
    \hline\hline
    \end{tabular}
    \caption{
    Set of typical system parameters  consistent with a LiNbO3 substrate and a Co\textsubscript{20}Fe\textsubscript{60}B\textsubscript{20}
    thin film (indices $s$ and $f$) used for figures and numerical calculations throughout this work, unless stated otherwise. }
    \label{tab:parameters_1}
\end{table}

\subsection{Phonon dynamics}

Let us now focus on the dynamics of the acoustic waves, which we describe using linear elastodynamics theory. Specifically, we define elastic deformations  through the displacement field $\mathbf{u}(\mathbf{r},t)$~\cite{eringen_chapter_1975}. Contrary to the magnons, which exist only within the thin film, acoustic vibrations extend across both the film and the substrate. We thus need to compute the acoustic eigenmodes of the combined system of film and substrate. The displacement field in a homogeneous system is governed by Navier's equation,
\begin{align}\label{eq:Navier}
    \rho \ddot u_i &= \partial^j C_{ijkl} \partial^k u^l,
\end{align}
where the indices $i,j,k,l$ label Cartesian components ($\partial^k\equiv \partial/\partial r_k$), $\rho$ is the mass density, and $C_{ijkl}$ is the elasticity tensor, which for simplicity we assume can be locally approximated by that of an isotropic and homogeneous material,
\begin{align}
    C_{ijkl} = \rho c_t^2 \left( \delta_{ik} \delta_{jl} + \delta_{il} \delta_{jk} \right) + \rho(c_l^2 -2c_t^2) \delta_{ij} \delta_{kl},
\end{align}
where $c_l$ and $c_t$ are the longitudinal and transversal sound velocities, respectively. Since we are working with a piecewise homogeneous body, we need to introduce separate material parameters for the substrate ($\rho_{s}$, $c_{l,s}$, $c_{t,s}$) and the thin film ($\rho_{f}$, $c_{l,f}$, $c_{t,f}$), as well as appropriate boundary conditions at their interface (i.e. continuity of displacement and traction \cite{eringen_chapter_1975-2}). Similar to the magnon case we define acoustic eigenmodes $\mathbf{u}_{\sigma, \mathbf{k}}(\mathbf{r})$ as the solutions to the equation \cite{Bruehlmann_unpublished}
\begin{align}
    - \rho(\mathbf{r}) (\omega_{\sigma,\mathbf{k}}^{p} )^2 u_{\sigma, \mathbf{k}, i}(\mathbf{r}) =C_{ijkl} \partial^j \partial^k u^l_{\sigma, \mathbf{k}}(\mathbf{r})
\end{align}
in all space, with  $\rho(\mathbf{r}) = \rho_{s}$ for $\mathbf{r}$ in the substrate and $\rho(\mathbf{r}) = \rho_{f}$ for $\mathbf{r}$ in the film.
We then perform a mode decomposition of the displacement field of the form 
\begin{align}\label{eq:diplacement_field_decomposition}
    \textbf{u}(\textbf{r},t) = \sum_{\sigma=1}^2 \int d^2\mathbf{k}\, \frac{\mathbf{u}_{\sigma,\mathbf{k}}(\textbf{r})}{\sqrt{\rho(\textbf{r})}} a_{\sigma,\mathbf{k}}(t) + c.c.,
\end{align}
where $a_{\sigma,\mathbf{k}}(t)$ are the corresponding expansion coefficients.

Let us briefly summarize the properties of the acoustic eigenmodes for this geometry,  the detailed calculation of which will be available in \cite{Bruehlmann_unpublished}. Similarly to the spin wave eigenmodes, the eigenmode functions $\mathbf{u}_{\sigma,\mathbf{k}}(\textbf{r})$ in Eq.~\eqref{eq:diplacement_field_decomposition} are labeled by an in-plane wave vector $\mathbf{k}=(0,k_y,k_z)$. Also as for spin waves, we include in our model only the lowest-order acoustic band, since higher-order bands will be largely detuned from the frequency range of interest for sufficiently thin films. 
The acoustic modes are additionally labeled by a polarization index $\sigma=1,2$ which labels different families of orthogonal modes with the same in-plane wave vector 
[see Fig.~\ref{fig:Dispersion_relation_and_SAW}(c)]. The mode functions are formulated in a piece-wise expression and behave like plane waves in the $y,z$ plane.
\begin{align}\label{umodefunctions}
    &\mathbf{u}_{\sigma, \mathbf{k}}(\mathbf{r}) = \left\{ \begin{matrix}
            \mathbf{u}_{\sigma, \mathbf{k}}^{(i)}(x) e^{i\mathbf{k}\mathbf{r}_{\scriptscriptstyle{\parallel}}} ,\quad x<0 \\
            \mathbf{u}_{\sigma, \mathbf{k}}^{(ii)}(x) e^{i\mathbf{k}\mathbf{r}_{\scriptscriptstyle{\parallel}}} ,\quad0<x<d .
        \end{matrix} \right. 
\end{align} 
  These eigenmodes have a well-defined eigenfrequency $\omega^\text{p}_{\sigma,\mathbf{k}}$, whose derivation can be found in Appendix \ref{app:phonon_eigenmodes}, and form an orthonormal basis, 
\begin{align}
    \int d^3 \textbf{r} \; \mathbf{u}_{\sigma,\mathbf{k}}^*(\mathbf{r}) \cdot \mathbf{u}_{\sigma^\prime,\mathbf{k}^\prime}(\mathbf{r}) = \delta_{\sigma,\sigma^\prime} \, \delta(\mathbf{k-k}^\prime) .
\end{align}

An important property of acoustic eigenmodes is that their associated displacement field, $\mathbf{u}_{\sigma,\mathbf{k}}(\textbf{r})$, qualitatively changes depending on the value of their wave vector modulus $k$ and their frequency $\omega_{\sigma,\mathbf{k}}^p$, as we illustrate in Fig. \ref{fig:Dispersion_relation_and_SAW}. For each material $j=f,s$, if $\{k,\omega_{\sigma,\mathbf{k}}^p\}$ lies above both longitudinal and transverse sound cones, defined as the lines $\omega = c_{l,j} k$ and $\omega = c_{t,j} k$ respectively, the mode function extends across the whole $x<d$ region  and is called a bulk mode. Contrary to this, if $\{k,\omega_{\sigma,\mathbf{k}}^p\}$ lies below both sound cones, the mode function decays exponentially away from both the surface ($x=d$) and the interface ($x=0$), this is called a surface acoustic wave mode. A schematic comparison between bulk and substrate-SAW modes can be found in Fig. \ref{fig:Dispersion_relation_and_SAW}(b). Lastly, if $\{k,\omega_{\sigma,\mathbf{k}}^p\}$ lies in between the transverse and longitudinal soundcones, the mode becomes hybrid, i.e. it contains both extended and exponentially decaying components.
Throughout this work, we include in our mode decomposition Eq.~\eqref{eq:diplacement_field_decomposition} only modes lying below both soundcones of the substrate, i.e., only modes that are SAWs in the substrate [red curves in Fig.~\ref{fig:Dispersion_relation_and_SAW}(a)]. Any other modes extend across the whole film and substrate and thus have negligible overlap with the magnon mode functions, which are localized only within a small region $0\le x\le d$. As a consequence their coupling to magnons is negligible. 
The full expressions for the mode functions, Eq.~\eqref{umodefunctions}, of the SAW-like modes are given in Appendix \ref{app:phonon_eigenmodes}. Note that the first branch $\mathbf{u}_{\sigma=1,\mathbf{k}}(\mathbf{r})$ is characterized by being purely transverse modes ($\mathbf{k}\cdot\mathbf{u}_{\sigma=1,\mathbf{k}}(\mathbf{r})=0$) which are hybrid within the film, while the second branch $\mathbf{u}_{\sigma=2,\mathbf{k}}(\mathbf{r})$ is mixed longitudinal and transversal waves. 

As we will show in the next section, the eigenmode functions derived in this section are necessary for deriving analytical expressions for any magnetoelastic coupling rates in the system.

\section{Nonlinear equation of motion}\label{sec:nonlinear_EOM}

The Landau-Lifshitz-Gilbert equation (\ref{eq:LLG}) is inherently nonlinear, leading to interactions between different magnon modes. Also magnetoelastic interaction is a source of nonlinear effects. In this section we derive the classical magnon and phonon equations of motion and discuss the different types of interactions. First, in Sec.~\ref{subsec:nonlin_magnon_eom}, we discuss nonlinear magnon scattering processes. Second, in Sec.~\ref{subsec:magnetoelastic_interaction}, we cover linear and nonlinear magnetoelastic interactions.

\subsection{Nonlinear magnon scattering}\label{subsec:nonlin_magnon_eom}
For the derivation of the free magnon eigenmodes discussed above the linearized LLG equation was used. Nonlinear magnon scattering processes are included in the model by keeping higher orders in the  perturbative expansion in the small parameter $\mathbf{m}/M_S$. Specifically, the nonlinear magnon equations of motion are obtained in three steps, as follows.

\begin{figure}[]
    \centering
    \includegraphics[width=1\linewidth]{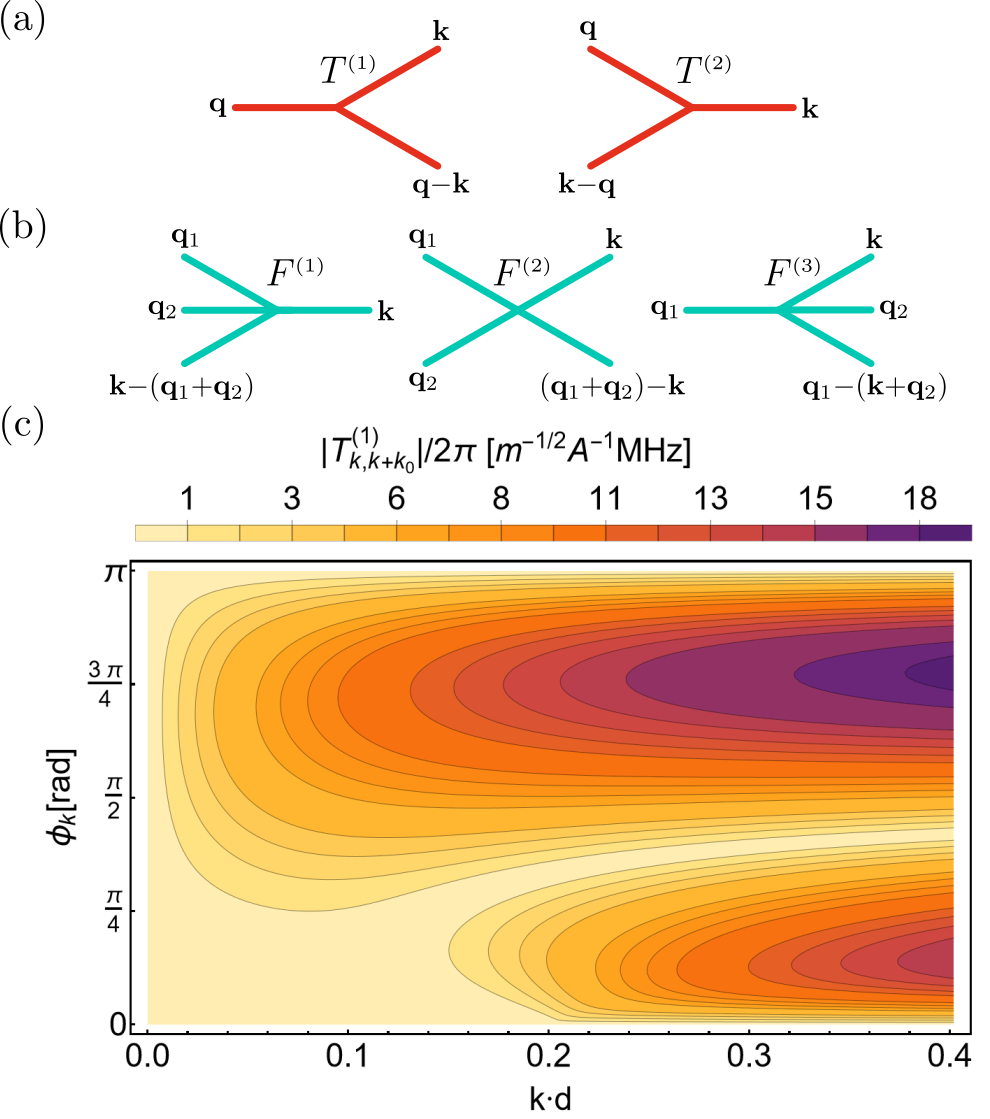}
    \caption{
    Schematic depiction of the different (a) three-magnon ($T^{(1,2)}$) and  (b) four-magnon ($F^{(1,2,3)}$) interactions appearing in the equations of motion Eq.~\eqref{eq:magnon_eom}. (c) Coupling rate density $T^{(1)}$ corresponding to a magnon of wavev ector $\mathbf{k}_0=(0,0,2\pi/625)\,\mathrm{nm^{-1}}$ scattering into two magnons ($\mathbf{k}=k(0,\sin\phi_k ,\cos\phi_k)$ and $\mathbf{k}_0-\mathbf{k}$), as a function of modulus and propagation angle of magnon $\mathbf{k}$.  System parameters are listed in Tab. \ref{tab:parameters_1}.
    }
    \label{fig:magnon_interaction_schematics}
\end{figure}

First, we introduce the second-order magnetization vector, Eq.~\eqref{eq:magnetization} with $\delta m = -(\mathbf{m}\cdot\mathbf{m})/(2M_S)$, into the Landau-Lifshitz Gilbert equation Eq.~\eqref{eq:LLG}. This includes introducing such second-order expression into the effective field $\textbf{H}_{\mathrm{eff}}(\mathbf{M},\textbf{r},t)$. For computational convenience, in this derivation we write the demagnetizing field as
\begin{equation}
    \textbf{H}_{\text{dm}}(\textbf{M},\textbf{r},t)  = \int d\mathbf{r}^\prime \mathcal{\bar{\bar G}}(\mathbf{r-r}^\prime) \mathbf{M}(\mathbf{r}^\prime,t),
\end{equation}
with the Green's function \cite{gonzalez-ballestero_towards_2022}
\begin{widetext}
\begin{align}\label{eq:Greens_function}
    2\rttensor{\mathcal{G}}(\textbf{r}-\textbf{r}^\prime) \equiv \delta(\textbf{r}_{\scriptscriptstyle{\parallel}}-\textbf{r}_{\scriptscriptstyle{\parallel}}^\prime)\mathbf{e}_x \mathbf{e}_x^T -\int d^2 \textbf{q} \, \frac{e^{i\textbf{q} (\textbf{r}_{\scriptscriptstyle{\parallel}}-\textbf{r}_{\scriptscriptstyle{\parallel}}^\prime)}}{(2\pi)^2q} e^{-q|x-x^\prime|} \mathbf{v}\mathbf{v}^T,
\end{align}
with in-plane wave vector $\mathbf{q}$ and the vectors $\mathbf{e}_x=(1,0,0)$ and $\mathbf{v}\equiv(-iq\,\text{sign}(x-x'),q_y,q_z)$. 
Second, we truncate the nonlinear LLG equation by keeping only terms up to third order in $\mathbf{m}/M_S$ in order to capture both three- and four-magnon processes.

Once we derive this nonlinear equation for the spin-wave field $\mathbf{m}(\mathbf{r},t)$, the third step is to write the magnetization field within this equation as a linear combination of eigenmodes of the linearized LLG equation $\mathbf{m}_\mathbf{k}(\mathbf{r})$, i.e., to write it in the form Eq.~\eqref{eq:magnetization_decomposition}. 
Then, we take the inner product of the resulting equation with an arbitrary eigenmode $\mathbf{m}_\mathbf{k}(\mathbf{r})$ in order to derive dynamical equations for the expansion coefficients $s_\mathbf{k}(t) \equiv \bra{\mathbf{m}_\mathbf{k}}\ket{\mathbf{m}(\mathbf{r},t)}$. As a result of nonlinearity, the resulting equations for the expansion coefficients $s_\mathbf{k}(t)$ contain interactions between different eigenmodes. Specifically, the expression we obtain reads:
\begin{multline}\label{eq:magnon_eom_intermediate}
    \dot s_\mathbf{k} = \left(-i \omega_\mathbf{k} -\frac{\gamma_\mathbf{k}^m}{2}\right) s_\mathbf{k} -i\iint d^2\mathbf{q}_1 d^2\mathbf{q}_2 \mspace{-4mu}  \sum_{\alpha,\beta=\pm} \Biggl[ \bra{\mathbf{m}_\mathbf{k}}\ket{D^{(3)}[\mathbf{m}_{\mathbf{q}_1}^\alpha,\mathbf{m}_{\mathbf{q}_2}^\beta]} s_{\mathbf{q}_1}^\alpha s_{\mathbf{q}_2}^\beta \\ + \mspace{-4mu} \int \mspace{-4mu} d^2\mathbf{q}_3 \mspace{-2mu}  \sum_{\gamma=\pm} \bra{\mathbf{m}_\mathbf{k}}\ket{D^{(4)}[\mathbf{m}_{\mathbf{q}_1}^\alpha,\mathbf{m}_{\mathbf{q}_2}^\beta,\mathbf{m}_{\mathbf{q}_3}^\gamma]}  s_{\mathbf{q}_1}^\alpha s_{\mathbf{q}_2}^\beta s_{\mathbf{q}_3}^\gamma \Biggr],
\end{multline}
where for compactness we define variables and their complex conjugates with a $\pm$ index, i.e., $\mathbf{m}_{k}^+(\mathbf{r})\equiv\mathbf{m}_{k}(\mathbf{r})$, $\mathbf{m}_{k}^-(\mathbf{r})\equiv\mathbf{m}_{k}^*(\mathbf{r})$, $s_{\mathbf{k}}^+\equiv s_{\mathbf{k}}$, and $s_{\mathbf{k}}^-\equiv s_{\mathbf{k}}^*$. 
The terms containing $D^{(3)}$ and $D^{(4)}$ correspond to three-magnon and four-magnon scattering processes respectively, and are given by\footnote{
\begin{minipage}[t]{0.9\columnwidth}
  Note also that in general structures $D^{(3)}$ has an extra term $+(\omega_M \alpha_{ex}/M_S) \mathbf{m}_{\mathbf{q}_1}\times \mathbf{m}_{\mathbf{q}_2}$, which here we omit as it is zero for thin film geometries.
\end{minipage}
}
\begin{align}
    D^{(3)}[\mathbf{m}_{\mathbf{q}_1},\mathbf{m}_{\mathbf{q}_2}] \equiv \frac{\omega_M}{2 M_S}  
    \biggl[ \mathbf{e}_z \times \Bigl(\int d\mathbf{r}^\prime \mathcal{\bar{\bar G}}(\mathbf{r\mspace{-4mu}-\mspace{-4mu}r}^\prime) \mathbf{e}_z \bigl( \mathbf{m}_{\mathbf{q}_1}(\mathbf{r}^\prime) \cdot \mathbf{m}_{\mathbf{q}_2}(\mathbf{r}^\prime)\bigr)\Bigr)
    - 2\mathbf{m}_{\mathbf{q}_1} \times \Bigl( \int d\mathbf{r}^\prime \mathcal{\bar{\bar G}}(\mathbf{r - r}^\prime) \mathbf{m}_{\mathbf{q}_2}(\mathbf{r}^\prime) \Bigr) \biggr] ,
\end{align}
\begin{multline}
    D^{(4)}[\mathbf{m}_{\mathbf{q}_1}, \mathbf{m}_{\mathbf{q}_2}, \mathbf{m}_{\mathbf{q}_3}] \equiv \frac{\omega_M }{2 M_S^2} \biggl[
    \alpha_{\text{ex}}\left( \mathbf{m}_{\mathbf{q}_1} \times \mathbf{e}_z \right) \cdot \bigl( \nabla^2 \left( \mathbf{m}_{\mathbf{q}_2} \cdot \mathbf{m}_{\mathbf{q}_3}\right)\bigr) 
    +\alpha_{\text{ex}}\left(\mathbf{m}_{\mathbf{q}_1} \cdot \mathbf{m}_{\mathbf{q}_2} \right) \mathbf{e}_z \times \left( \nabla^2  \mathbf{m}_{\mathbf{q}_3} \right) \\
    + \int d\mathbf{r}^\prime \Bigl( \mathbf{m}_{\mathbf{q}_1} \times \mathcal{\bar{\bar G}}(\mathbf{r-r}^\prime) \mathbf{e}_z \Bigr) \bigl( \mathbf{m}_{\mathbf{q}_2}(\mathbf{r}^\prime) \cdot \mathbf{m}_{\mathbf{q}_3}(\mathbf{r}^\prime)\bigr)
    +\left(\mathbf{m}_{\mathbf{q}_1} \cdot \mathbf{m}_{\mathbf{q}_2} \right)  \int d\mathbf{r}^\prime \mathbf{e}_z \times \Bigl(\mathcal{\bar{\bar G}}(\mathbf{r-r}^\prime) \mathbf{m}_{\mathbf{q}_3}(\mathbf{r}^\prime) \Bigr) \biggr].
\end{multline}
Note that in the above expressions we have omitted the position argument $\mathbf{r}$ for the magnetization mode functions except when it appears as the integration variable. Using these expressions the classical equation of motion Eq.~\eqref{eq:magnon_eom_intermediate} can be cast as
    \begin{multline}\label{eq:magnon_eom}
    \dot s_\mathbf{k} = \left( -i\omega^\text{m}_\mathbf{k}-\frac{\gamma^\text{m}_{\mathbf{k}}}{2} \right) s_\mathbf{k} 
    +\int d^2\mathbf{q}\; \Bigl( T^{(1)}_{\mathbf{q},\mathbf{k+q}}\, s_{\mathbf{q}}^* s_{\mathbf{k+q}} + T^{(2)}_{\mathbf{q},\mathbf{k-q}}\, s_{\mathbf{q}} s_{\mathbf{k-q}}\Bigr)\\
    +\int d^2 \mathbf{q}_1 \int d^2\mathbf{q}_2 \Bigl( F^{(1)}_{\mathbf{q}_1,\mathbf{q}_2,\tilde{\mathbf{q}}^{(1)} }\, s_{\mathbf{q}_1} s_{\mathbf{q}_2} s_{\tilde{\mathbf{q}}^{(1)} }
    + F^{(2)}_{\mathbf{q}_1,\mathbf{q}_2,\tilde{\mathbf{q}}^{(2)}}\, s_{\mathbf{q}_1} s_{\mathbf{q}_2} s_{\tilde{\mathbf{q}}^{(2)}}^*
    + F^{(3)}_{\mathbf{q}_1,\mathbf{q}_2,\tilde{\mathbf{q}}^{(3)}}\, s_{\mathbf{q}_1} s_{\mathbf{q}_2}^* s_{\tilde{\mathbf{q}}^{(3)}}^* \Bigr),
\end{multline}
\end{widetext}
with wave vectors $\tilde{\mathbf{q}}^{(1)} \equiv \mathbf{k}-(\mathbf{q}_1+\mathbf{q}_2)$, $\tilde{\mathbf{q}}^{(2)} \equiv (\mathbf{q}_1+\mathbf{q}_2)-\mathbf{k}$, and $\tilde{\mathbf{q}}^{(3)} \equiv \mathbf{q}_1-(\mathbf{q}_2+\mathbf{k})$. 
Note that total momentum is conserved in each interaction, a condition that arises naturally from the calculation of the coupling rate densities and that is due to the system's translational invariance. In the above equations we have applied a rotating wave approximation, thus neglecting the rapidly oscillating terms proportional to $s^* s^*$ and $s^* s^* s^*$.\footnote{Such terms correspond, in a quantum picture, to spontaneous creation (annihilation) of excitations from (to) vacuum.\vspace{-1cm}} Note also that the three- and four- magnon coupling strength coefficients $T_{\mathbf{k_1},\mathbf{k}_2}$ and $F_{\mathbf{k_1},\mathbf{k}_2,\mathbf{k}_3}$ (whose expressions are given in Appendix~\ref{app:magnon_couoplig_rates}) are coupling rate densities, i.e., they do not have frequency dimensions due to the continuum nature of our system. 
Figure \ref{fig:magnon_interaction_schematics}   shows the type of interactions each term in the above equation corresponds to, and, as an example, the behavior of nonlinear coupling rate density $T^{(1)}_{\mathbf{q},\mathbf{k}+\mathbf{q}}$ for a fixed wave vector of the incoming magnon. 
Note the nontrivial dependence of the coupling rate density on the direction of the outgoing magnons, which is caused by the suppressed three-mode overlap for magnons propagating along the two symmetry axis of the plane, $y$ and $z$. Note also that the coupling rate density tends to zero for $k_0d,k d\ll 1$ as the modes become approximately homogeneous and their associated dipolar field  vanishes inside the magnet.

\subsection{Magnetoelastic interaction}\label{subsec:magnetoelastic_interaction}

Deformations of the ferromagnetic body change the distances and orientations between spins leading to interactions between magnons and phonons. The lowest-order contribution to the magnetoelastic energy density is described by the phenomenological expression \cite{Xufeng_Cavity2016,gurevich_magnetization_1996,gonzalez-ballestero_theory_2020,yamamoto_interaction_2022}
\begin{multline}\label{Ume}
    U_{me}(\textbf{r},t) = \sum_{i,j=1}^3 \frac{B_2 + (B_1-B_2) \delta_{ij}}{M_S^2} \\
    \times M_i(\textbf{r},t) M_j(\textbf{r},t) \epsilon_{ij}(\textbf{r},t),
\end{multline}
which we use to derive the linear and nonlinear coupling rate densities. Here, $B_1$ and $B_2$ are cubic magnetoelastic coefficients of the material and $\epsilon_{ij} \equiv (\partial_i u_j + \partial_j u_i)/2$ the strain tensor. 
By introducing in Eq.~\eqref{Ume} the decomposition of the magnetization field $\mathbf{M}(\mathbf{r},t)$ as in Eq.~\eqref{eq:magnetization} with $\delta m \approx (\mathbf{m}\cdot\mathbf{m})/(2M_S)$ and neglecting terms of order $(\mathbf{m}/M_S)^3$ or higher, we obtain an expression for the energy density up to second order in $\textbf{m}/M_S$,
\begin{multline}
    U_{me}(\textbf{r},t) \approx B_2 \sum_{i=x,y} \frac{m_i}{M_S} \epsilon_{iz}     \\
    +\mspace{-6mu} \sum_{ij = x,y} \mspace{-4mu} ( B_2 \mspace{-2mu} + \mspace{-2mu} (B_1 \mspace{-2mu} - \mspace{-2mu} B_2) \delta_{ij} ) \frac{m_i}{M_S} \frac{m_j}{M_S} (\epsilon_{ij} - \epsilon_{zz} \delta_{ij} ) .
\end{multline}
By introducing now the decomposition of the magnetization and acoustic fields into their respective eigenmodes, Eqs. (\ref{eq:magnetization_decomposition}) and (\ref{eq:diplacement_field_decomposition}), and integrating over the whole volume we obtain the magnon-phonon interaction Hamiltonian 
\begin{multline}\label{eq:me_interaction_Hamiltonian}
    \mathcal{H}_{\text{mp}} = \sum_\sigma \int d^2\mathbf{q} \int d^2\mathbf{q}^\prime\, \biggl[ (I^{(1)}_{\sigma, \mathbf{q}, \mathbf{q}^\prime} \, a_{\sigma , \mathbf{q}} s_{\mathbf{q}^\prime}^* + cc. )\\
    +\int d^2\mathbf{q}^{\prime\prime} \, \Bigl[ (I^{(2)}_{\sigma, \mathbf{q}, \mathbf{q}^\prime,\mathbf{q}^{\prime\prime}} \, a_{\sigma , \mathbf{q}} s_{\mathbf{q}^\prime}^*  s_{\mathbf{q}^{\prime\prime}}+ cc. )\\
    \phantom{\int d^2\mathbf{q}^{\prime\prime} \, \Bigl[} + (I^{(2)}_{\sigma, \mathbf{q}, \mathbf{q}^{\prime\prime}, \mathbf{q}^\prime} \, a_{\sigma , \mathbf{q}} s_{\mathbf{q}^{\prime\prime}} s_{\mathbf{q}^\prime}^* + cc. )\\
    \phantom{\int d^2\mathbf{q}^{\prime\prime} \, \Bigl[} + (I^{(3)}_{\sigma, \mathbf{q}, \mathbf{q}^\prime, \mathbf{q}^{\prime\prime}} \, a_{\sigma , \mathbf{q}} s_{\mathbf{q}^\prime}^* s_{\mathbf{q}^{\prime\prime}}^* + cc. )\Bigr]\biggr].
\end{multline}
Expressions for the mode overlap integrals $I^{(i)}$ ($i\in\{1,2,3\}$) are calculated in Ref~\cite{Bruehlmann_unpublished} and are given in Appendix \ref{app:magnetoelastic_coupling}. Total momentum is also conserved in all magnetoelastic interactions. Here we have undertaken a rotating wave approximation neglecting rapidly oscillating terms proportional to $a^*s^*$, $a^*s^*s^*$, and their complex conjugates. 

Once the magnetoelastic Hamiltonian Eq.~\eqref{eq:me_interaction_Hamiltonian} has been derived, we can use it to compute its contribution to the dynamical equations for magnon and phonon amplitudes, $s_\mathbf{k}(t)$ and $a_{\sigma,\mathbf{q}}(t)$, using the principles of Hamiltonian mechanics. 
Let us illustrate the procedure in the simple example where no damping, nonlinearity, nor magnon-phonon interactions are present. In this case the total Hamiltonian of the magnon and phonon systems is $\mathcal{H}^0=\mathcal{H}_p+\mathcal{H}^0_{\rm m}$ with bare magnon and phonon Hamiltonians given by
\begin{align}
     \mathcal{H}_\text{m}^0 &= \int d^2\mathbf{k}\, |\mathcal{M}|^{-2} \omega^{\text{m}}_\mathbf{k} s_\mathbf{k}^* s_\mathbf{k},\label{eq:hamiltonian_magnon} 
\end{align}
\begin{align}
    \mathcal{H}_\text{p} &= \sum_\sigma \int d^2\mathbf{k}\, |\mathcal{U}_{\sigma,\mathbf{k}}|^{-2} \omega^{\text{p}}_{\sigma,\mathbf{k}} a_{\sigma,\mathbf{k}}^* a_{\sigma,\mathbf{k}}, \label{eq:hamiltonian_phonon} 
\end{align}
with $\mathcal{M}\equiv\sqrt{|\gamma_g| M_S}$ and $\mathcal{U}_{\sigma,\mathbf{k}}\equiv1/\sqrt{2 \omega^{\text{p}}_{\sigma,\mathbf{k}}}$.
From this Hamiltonian we obtain the equations of motion by applying the following Hamilton equations,
\begin{align}
    \dot{s}_\mathbf{k} &= -i|\mathcal{M}|^2 \frac{\partial \mathcal{H}^0}{\partial s_{\mathbf{k}}^*} = -i\omega^{\text{m}}_{\mathbf{k}} s_\mathbf{k} ,\\
    \nonumber\\
    \dot{a}_{\sigma,\mathbf{k}} &= -i|\mathcal{U}_{\sigma,\mathbf{k}}|^2 \frac{\partial \mathcal{H}^0}{\partial a_{\sigma,\mathbf{k}}^*} = -i\omega^{\text{p}}_{\sigma,\mathbf{k}} a_{\sigma,\mathbf{k}} .
\end{align}
The full equations of motion are obtained by applying the same principle to the full Hamiltonian for the magnon-phonon system,
\begin{align}\label{eq:full_Hamiltonian}
    \mathcal{H}= \mathcal{H}_{\text{m}} + \mathcal{H}_{\text{p}} + \mathcal{H}_{\text{mp}},
\end{align}
with $\mathcal{H}_{\text{m}}$ including all nonlinear interactions and with $\mathcal{H}_{\text{mp}}$ given in Eq.~\eqref{eq:me_interaction_Hamiltonian}.
Note that an expression for $\mathcal{H}_m$ does not need to be explicitly derived as the dynamics it induces are already known from Eq.~\eqref{eq:magnon_eom}. The full magnetoelastic coupled dynamics obtained from Hamilton's equations is given by the following system of coupled nonlinear equations,
\begin{widetext}
\begin{align}
\begin{split}\label{eq:EOM_sk}
    \dot{s}_{\mathbf{k}} = &\left( -i\omega^\text{m}_\mathbf{k}-\frac{\gamma^\text{m}_{\mathbf{k}}}{2} \right) s_\mathbf{k} + \sum_\sigma g_{\sigma,\mathbf{k}}\, a_{\sigma,\mathbf{k}} + \int d^2\mathbf{q}\; \Bigl( T^{(1)}_{\mathbf{q},\mathbf{k+q}}\, s_{\mathbf{q}}^* s_{\mathbf{k+q}} + T^{(2)}_{\mathbf{q},\mathbf{k-q}}\, s_{\mathbf{q}} s_{\mathbf{k-q}}\Bigr)\\
    &+ \sum_\sigma \int d^2\mathbf{q} \Bigl( G^{(1)}_{\sigma,\mathbf{q},\mathbf{k}} \, a_{\sigma,\mathbf{q}} s_{\mathbf{k}-\mathbf{q}} +G^{(2)}_{\sigma,\mathbf{q},\mathbf{k}} \, a_{\sigma,\mathbf{q}}^* s_{\mathbf{k}+\mathbf{q}} + G^{(3)}_{\sigma,\mathbf{q},\mathbf{k}} \, a_{\sigma,\mathbf{q}} s_{\mathbf{q}-\mathbf{k}}^* \Bigr) \\
    &+\int d^2 \mathbf{q}_1 \int d^2\mathbf{q}_2 \Bigl( F^{(1)}_{\mathbf{q}_1,\mathbf{q}_2,\mathbf{k}-(\mathbf{q}_1+\mathbf{q}_2)}\, s_{\mathbf{q}_1} s_{\mathbf{q}_2} s_{\mathbf{k}-(\mathbf{q}_1+\mathbf{q}_2)} + F^{(2)}_{\mathbf{q}_1,\mathbf{q}_2,(\mathbf{q}_1+\mathbf{q}_2)-\mathbf{k}}\, s_{\mathbf{q}_1} s_{\mathbf{q}_2} s_{(\mathbf{q}_1+\mathbf{q}_2)-\mathbf{k}}^*\\
    &\phantom{+\int d^2 \mathbf{q}_1 \int d^2\mathbf{q}_2} + F^{(3)}_{\mathbf{q}_1,\mathbf{q}_2,\mathbf{q}_1-(\mathbf{q}_2+\mathbf{k})}\, s_{\mathbf{q}_1} s_{\mathbf{q}_2}^* s_{\mathbf{q}_1-(\mathbf{q}_2+\mathbf{k})}^* \Bigr),
\end{split}
\end{align}
\begin{align}
\begin{split}\label{eq:EOM_ak}
    \dot a_{\sigma,\mathbf{k}} = &\left( -i\omega^\text{p}_{\sigma,\mathbf{k}} -\frac{\gamma^{\text{p}}_{\sigma,\mathbf{k}}}{2}\right) a_{\sigma,\mathbf{k}} + \tilde{g}_{\sigma,\mathbf{k}} \, s_{\mathbf{k}} + \int d^2\mathbf{q} \Bigl( \tilde{G}^{(1)}_{\sigma,\mathbf{k},\mathbf{q}} \, s_{\mathbf{q}}^* s_{\mathbf{k}+\mathbf{q}} + \tilde{G}^{(3)}_{\sigma,\mathbf{k},\mathbf{q}} \, s_{\mathbf{q}} s_{\mathbf{k}-\mathbf{q}} \Bigr) + i\alpha_\text{d} \delta(\mathbf{k} - \mathbf{k}_0) e^{-i\omega_\text{d} t}.
\end{split}
\end{align}
\end{widetext}
In the above expressions $g_{\sigma,\mathbf{k}}, \tilde{g}_{\sigma,\mathbf{k}}$ and $G^{(j)}_{\sigma,\mathbf{k},\mathbf{k}'},\tilde{G}^{(j)}_{\sigma,\mathbf{k},\mathbf{k}'}$ represent the linear and the nonlinear coupling rate densities, respectively, and their analytical expressions are given in Appendix \ref{app:magnetoelastic_coupling}. In the above equations of motion we have added a phonon damping rate $\gamma^{\text{p}}_{\sigma,\mathbf{k}}$
as well as a coherent driving term of the acoustic modes  $\sigma\in\{1,2\},\,\mathbf{k}=\mathbf{k}_0$ at a frequency $\omega_\text{d}$ with strength $\alpha_d$, where $\vert\alpha_d\vert^2\propto{P}$ with $P$ the drive power. This term models the SAW drive utilized in experiments~\cite{hwang_strongly_2024,HwangArxiv2025} to excite nonlinear responses. A potential magnon drive can be added to the magnon equation of motion in a similar way.
\begin{figure}[t]
    \centering
    \includegraphics[width=1\linewidth]{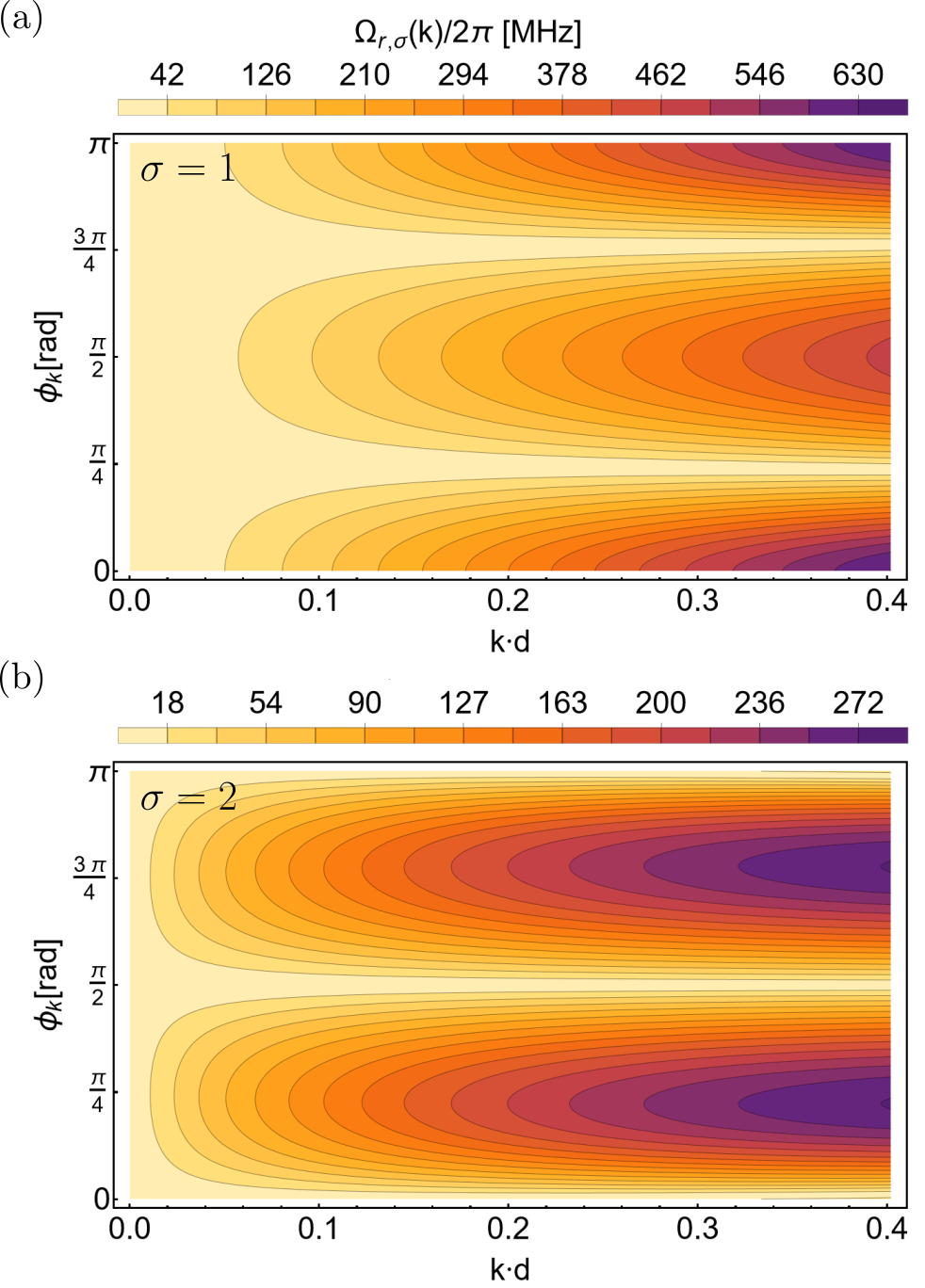}
    \caption{
     Resonant magnetoelastic linear coupling rates $\Omega_{\text{r},\sigma}$ [Eq.~\ref{EqRabi}]  vs mode wave vector and direction, for the $\sigma=1$ (a) and $\sigma=2$ (b) phonon families, respectively.  System parameters are listed in Tab. \ref{tab:parameters_1}. 
    }
    \label{fig:magnetoelastic_linear_interaction_rates}
\end{figure}

\begin{figure}[]
    \centering
    \includegraphics[width=1\linewidth]{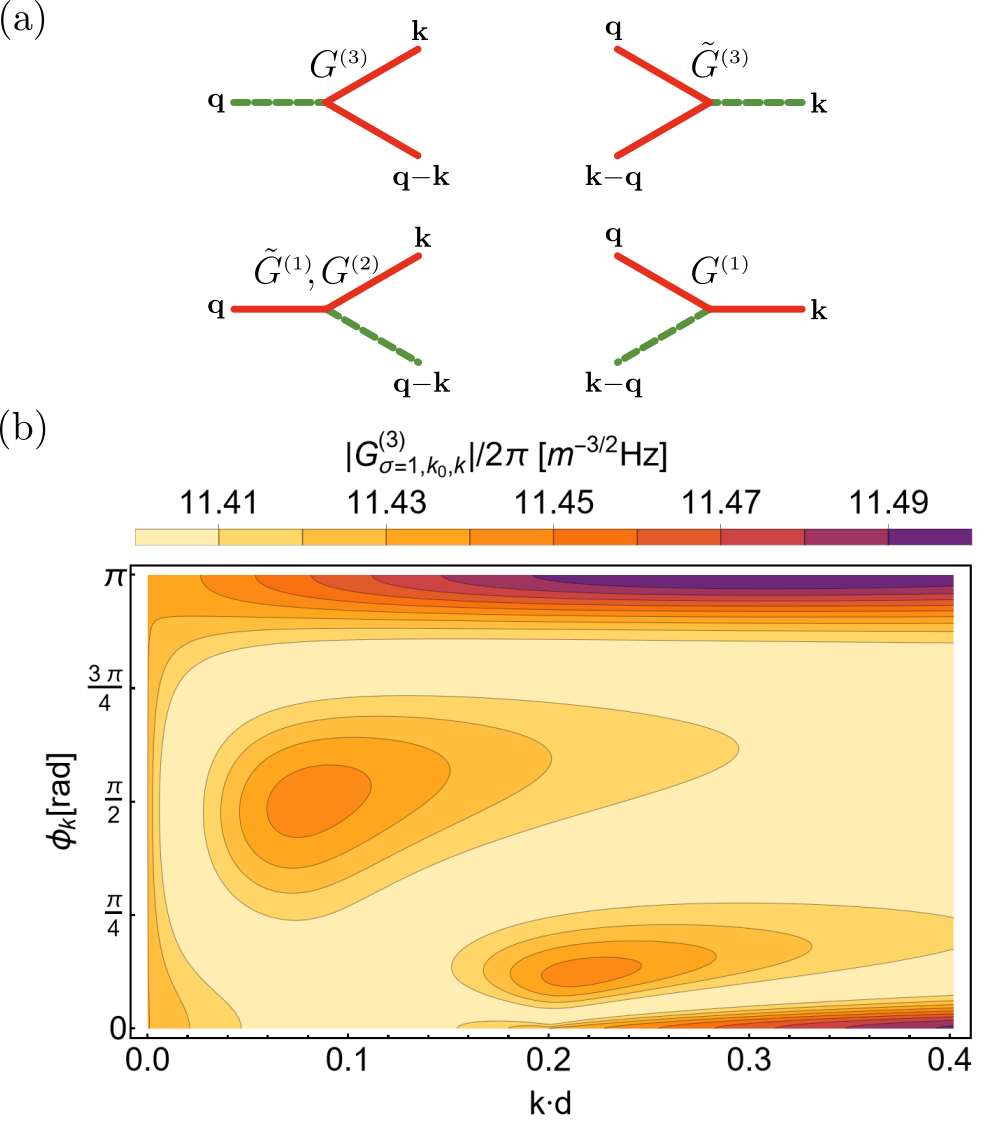}
    \caption{
     (a) Schematic depiction of the types of nonlinear magnetoelastic interactions appearing in the equations of motion (red/green lines indicate magnons and phonons, respectively). (b) Nonlinear coupling strength density $G^{(3)}$ corresponding to a phonon of mode $\{\sigma=1,\mathbf{k}_0=(0,0,2\pi/625)\,\mathrm{nm^{-1}}\}$ scattering into two magnons [$\mathbf{k}=k(0,\sin\phi_k ,\cos\phi_k)$ and $\mathbf{k}_0-\mathbf{k}$], as a function of modulus and propagation angle of magnon $\mathbf{k}$. System parameters are listed in Tab. \ref{tab:parameters_1}. 
    }
    \label{fig:magnetoelastic_interaction_schematics}
\end{figure}

As a final remark, note that the Rabi frequency splitting between magnon and phonon dispersion relations is, when magnon and phonon have resonant frequencies, given by
\begin{multline}\label{EqRabi}
    \Omega_{\text{r},\sigma}(\mathbf{k}) = \text{Im}\sqrt{\frac{(\gamma_\mathbf{k}^{\text{m}}-\gamma_{\sigma,\mathbf{k}}^{\text{p}})^2}{4}+4g_{\sigma,\mathbf{k}} \tilde{g}_{\sigma,\mathbf{k}}}
    \\
    \approx 2\sqrt{\vert g_{\sigma,\mathbf{k}} \tilde{g}_{\sigma,\mathbf{k}}\vert},
\end{multline}
and has units of frequency. When this Rabi splitting is larger than the damping rates of both components, $\Omega_{\text{r},\sigma}(\mathbf{k})>\gamma_{\sigma,\mathbf{k}}^\text{p},\gamma^\text{m}_\mathbf{k}$, the system is in the strong (linear) coupling regime~\cite{hwang_strongly_2024}. In this work we will focus on this regime, where the excitation of nonlinear dynamics is most efficient~\cite{HwangArxiv2025}. Figure~\ref{fig:magnetoelastic_linear_interaction_rates} shows the Rabi coupling rate Eq.~(\ref{EqRabi}) for both phonon bands. We see that at high-symmetry angles the coupling vanishes, a feature caused by the low magnon-strain overlap (see Appendix~\ref{app:magnetoelastic_coupling}) Specifically, for modes propagating parallel to the external magnetic field ($\phi_k=0$) magnons couple mainly to phonons of branch $\sigma=1$. The diagrams in Fig.~\ref{fig:magnetoelastic_interaction_schematics}\textcolor{blue}{(a)} depict the different types of nonlinear magnetoelastic interactions contained in the equations of motion Eqs.~\eqref{eq:EOM_sk} and \eqref{eq:EOM_ak}, and Fig.~\ref{fig:magnetoelastic_interaction_schematics}\textcolor{blue}{(b)}  shows as an example the coupling rate density corresponding to the process of a phonon of the resonantly driven mode $\{\sigma=1,\mathbf{k}_0=(0,0,2\pi/625)\,\mathrm{nm^{-1}}\}$ scattering into two magnons [$\mathbf{k}=k(0,\sin\phi_k ,\cos\phi_k)$ and $\mathbf{k}_0-\mathbf{k}$].

This section concludes the derivation of the classical coupled nonlinear dynamical equations for magnons and phonons. In the next section we will use this equation to characterize all the nonlinear processes arising from these equations.

\section{High-harmonic generation and parametric instability process}\label{sec:parametrc_instability}
In this section we detail our approach to extracting from the equations of motion derived in the previous section the information about the nonlinear response of the magnetization field, namely high-harmonic generation and, most importantly, sub-harmonic generation due to parametric instability, akin to the process of parametric down conversion in optics. All these processes result in well-defined peaks in the magnetization spectrum~\cite{HwangArxiv2025}. 
First, in Sec. \ref{subsec:parametric_excitations_threshold_ordering}, we show how to predict the instability thresholds, namely the driving strengths $\alpha_d$ at which each such peak appears, as well as the energy and wave vector of the modes excited by parametric instability. Then, in Sec. \ref{subsec:3mode_system}, we outline how to solve the full nonlinear problem to obtain the mode amplitudes at driving strengths above the threshold, focusing on three-magnon scattering as it allows for deeper analytical insight. Obtaining the mode amplitudes, in particular above threshold, is essential to derive a mean-field quantum theory, as we will see in Sec.~\ref{sec:quantum}.

\subsection{High-harmonic amplitudes and subharmonic driving thresholds}\label{subsec:parametric_excitations_threshold_ordering}
The simplest nonlinear feature of our equations of motion is high-harmonic generation. Here, the  drive of a phonon mode with wave vector $\mathbf{k}_0$ at frequency $\omega_\text{d}$ (which for the parameters of Table~\ref{tab:parameters_1} is resonant for the $\sigma=2$ phonon, i.e. $\omega_\text{d} = \omega^{\text{p}}_{2,\mathbf{k}_0}$) leads, through a combination of linear and nonlinear interactions, to the population of higher harmonic magnon and phonon modes $\xi \mathbf{k}_0$ ($\xi \in \{ 2,3,\dots\}$) driven respectively at frequencies $\xi\omega_d$. We can derive the amplitudes of these harmonics by introducing displaced mode amplitudes as
\begin{align}
    s_\mathbf{k} &\to s_\mathbf{k} + \sum_{\xi=1}^N \alpha_{\xi}^\text{m} \delta(\mathbf{k}-\xi\mathbf{k}_0) e^{-i(\xi\omega_\text{d}) t},\label{displacementS}\\
    a_{\sigma,\mathbf{k}} &\to a_{\sigma,\mathbf{k}} + \sum_{\xi=1}^N \alpha_{\xi,\sigma}^\text{p} \delta(\mathbf{k}-\xi\mathbf{k}_0) e^{-i(\xi\omega_\text{d}) t}.\label{displacementa}
\end{align}
Here, $N$ sets the number of harmonics one aims at determining and $\alpha_{\xi,\sigma}^\text{p}$ and $\alpha_{\xi}^\text{m}$ are the amplitudes of each harmonic, which need to be determined. 

We introduce the redefinition Eqs.~\eqref{displacementS} and \eqref{displacementa} into the equations of motion \eqref{eq:EOM_sk} and \eqref{eq:EOM_ak}, obtaining a new, lengthy system of equations. These equations retain the same structure but contain two families of additional terms: First, new linear coupling terms which represent nonlinear processes that involve the higher harmonics. For instance, one of the terms contains a linear coupling between magnon modes $s_\mathbf{k}$ and  $s_{2\mathbf{k}_0-\mathbf{k}}$, 
representing a three-magnon process where magnons of these two eigenmodes are generated by the splitting of the second-harmonic magnon $2\mathbf{k}_0$. Second, new driving terms of the form $C_\xi \delta(\mathbf{k}-\xi\mathbf{k}_0) e^{-i(\xi\omega_\text{d}) t}$. We determine the parameters $\alpha^\text{m}_{\xi}$ and $\alpha^\text{p}_{\xi,\sigma}$ by imposing that such terms $C_\xi$ vanish. This choice gives physical meaning to the redefinition Eqs.~\eqref{displacementS} and \eqref{displacementa} as a transformation into a frame where the new $s_\mathbf{k}$ represent additional oscillations on top of a magnetization field that is already populated by the drive and by the high harmonics. We remark that this procedure is exact and analogous to a displacement transformation in quantum optics \cite{scully1997quantum}. Note that because the magnon dispersion relation is curved and (in thin films) monotonically increasing [see Fig.~\ref{fig:Dispersion_relation_and_SAW}(a)], the detuning between the natural frequency of the $\xi$ th harmonic mode, $\omega^m_{\xi\mathbf{k}_0}$, and the frequency at which it is driven, $\xi\omega_d$, increases as a function of $\xi$, making the excitation of higher harmonics less and less efficient. For this reason in our calculations below we truncate the sum in Eqs.~\eqref{displacementS} and \eqref{displacementa} at the third harmonic, i.e., we set $N=3$. Explicit expressions for the computed $\alpha_{\xi,\sigma}^\text{p}$ and $\alpha_{\xi}^\text{m}$, as well as for the new equations of motion, are given in Appendix \ref{app:displacement}. Note that the high-harmonic generation is thresholdless, i.e., it already arises at any finite driving strength. The high harmonic peaks will thus appear in the spectrum as soon as their amplitude can rise above the noise floor.

Once the high harmonics have been taken into account, we analyze in the resulting equations the second family of nonlinear processes. These are parametric instability processes, where one mode spontaneously splits into two or more modes. The dynamics of these processes is more complex, e.g., they appear only after a defined driving threshold. Our first goal is to determine such threshold. We can do so by employing the following \emph{linear} equations of motion,
\begin{widetext}
\vspace{-0.5cm}
\begin{align}
\begin{split}\label{eq:parametric_amplification_EOM_s}
    \dot s_\mathbf{k} = &\left( -i\omega^\text{m}_\mathbf{k} - \frac{\gamma^\text{m}_{\mathbf{k}}}{2} \right) s_\mathbf{k} + \sum_{\xi,\zeta=1}^3 \left( F^{(2)}_{\zeta\mathbf{k}_0,\xi\mathbf{k}_0,(\xi+\zeta)\mathbf{k}_0-\mathbf{k}} \right) \alpha^{\text{m}}_\xi \alpha^{\text{m}}_\zeta e^{-i(\xi+\zeta) \omega_\text{d} t} \, s_{(\xi+\zeta)\mathbf{k}_0-\mathbf{k}}^*\\
   &+\sum_{\xi,\zeta=1}^3 \left( F^{(3)}_{\xi\mathbf{k}_0,\zeta\mathbf{k}_0,(\xi-\zeta)\mathbf{k}_0-\mathbf{k}} + F^{(3)}_{\xi\mathbf{k}_0,(\xi-\zeta)\mathbf{k}_0-\mathbf{k},\zeta\mathbf{k}_0} \right) \alpha^{\text{m}}_\xi \alpha^{\text{m}*}_\zeta e^{-i(\xi-\zeta)\omega_\text{d} t} \, s_{(\xi-\zeta)\mathbf{k}_0-\mathbf{k}}^*\\
    &-\sum_{\xi=1}^3 e^{-i\xi\omega_\text{d} t} \left( \Bigl( \sum_{\sigma =1,2} ( G^{(3)}_{\sigma, \xi \mathbf{k}_0, \mathbf{k}} \, \alpha^{\text{p}}_{\xi,\sigma}) + T^{(1)}_{ \xi\mathbf{k}_0-\mathbf{k}, \xi\mathbf{k}_0} \, \alpha^{\text{m}}_{\xi} \Bigr)  s_{\xi\mathbf{k}_0-\mathbf{k}}^*
    + \sum_{\sigma =1,2} G^{(2)}_{\sigma, \xi \mathbf{k}_0-\mathbf{k}, \mathbf{k}} \, \alpha^{\text{m}}_{\xi} a_{\sigma, \xi\mathbf{k}_0-\mathbf{k}}^* \right),
\end{split}
\end{align}
\begin{align}
\begin{split}\label{eq:parametric_amplification_EOM_a}
    \dot a_{\sigma,\mathbf{k}} = &\left( -i\omega^\text{p}_\mathbf{k} - \frac{\gamma^\text{p}}{2} \right) a_\mathbf{k}-\sum_{\xi=1}^3 \tilde{G}^{(1)}_{\sigma, \mathbf{k}, \xi \mathbf{k}_0-\mathbf{k}} \, \alpha^{\text{m}}_{\xi} e^{-i\xi\omega_\text{d} t} s_{\xi\mathbf{k}_0-\mathbf{k}}^*.
\end{split}
\end{align}
\end{widetext}
To derive these linearized expressions, we start with the equations of motion derived above [namely, the set of equations of motion \eqref{eq:EOM_sk} and \eqref{eq:EOM_ak} under the redefinition Eqs.~\eqref{displacementS} and \eqref{displacementa}] and neglect (i) linear terms that are known to be stable and thus lead to no parametric instability [e.g., terms proportional to $s_\mathbf{q}$ in the equation of motion for $\mathbf{s}_\mathbf{k}$], and (ii) any nonlinear terms. Note that (ii) does not remove the nonlinearity altogether since, as discussed above, the main nonlinear effects are encoded as linear terms in these equations. More details about this derivation are given in Appendix~\ref{app:displacement}.

\begin{figure}
    \centering
    \includegraphics[width=1\linewidth]{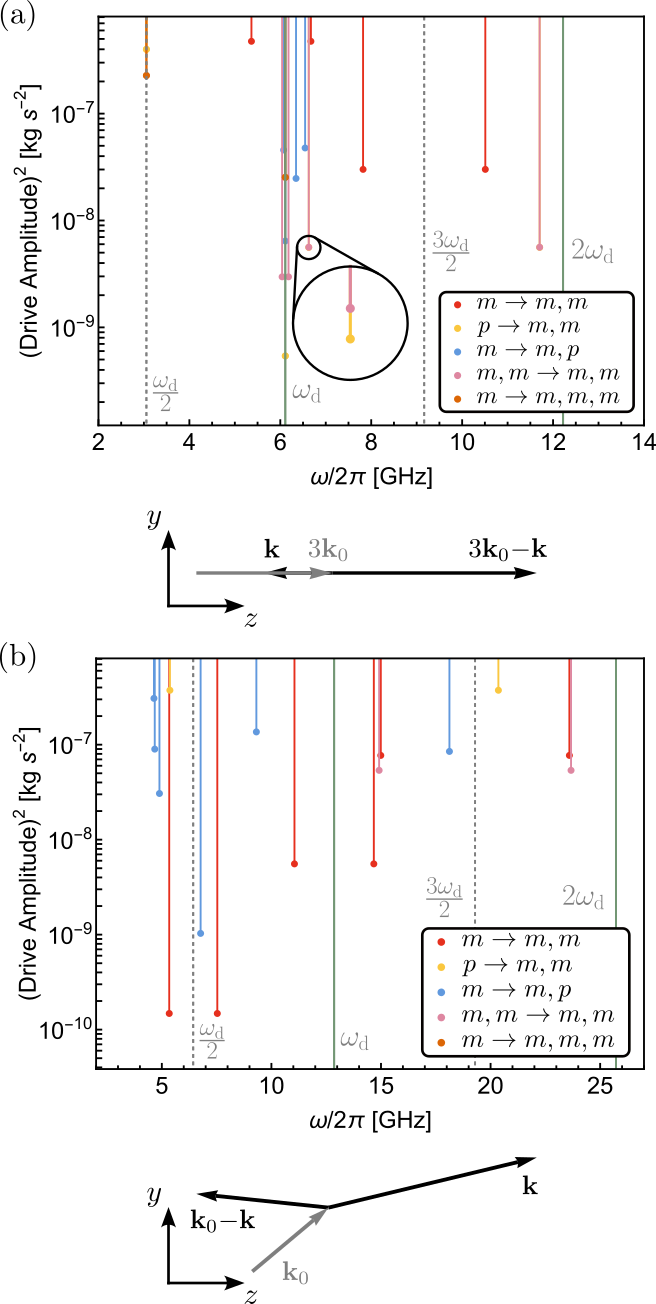}
    \caption{
    Peaks in the magnetization spectrum as a function of driving amplitude squared, $\vert \alpha_\text{d}\vert^2$ (proportional to driving power). Dots mark the parametric instability threshold for each process and the frequency of the newly populated modes. Green lines mark the frequency of the drive and its higher harmonics. For these calculations we have set the phonon decay rate proportional to its frequency, $\gamma^{\text{p}}_{\sigma,\mathbf{k}}=  \omega^\text{p}_{\sigma,\mathbf{k}}/Q_0$, with $Q_0$ the quality factor. The schematics below each panel depict the wave vector modulus and direction of the three modes involved in the lowest threshold parametric instability processes. (a) Subharmonic parametric instability with the lowest threshold originates from splitting of phonon into 2 magnons (yellow dot in zoomed view). System parameters: Tab.~\ref{tab:parameters_1}, driven at $\phi_{\mathbf{k}_0}=0$. (b) Subharmonic parametric instability with the lowest threshold originates from three-magnon scattering (red dots). System parameters: Tab.~\ref{tab:parameters 2}, driven at $\phi_{\mathbf{k}_0}=40^{\circ}$.} 
    \label{fig:Magentization_spectrum}
\end{figure}

Equation~\eqref{eq:parametric_amplification_EOM_s} captures three kinds of parametric instabilities, namely the ones induced by all the possible four-magnon processes (first and second lines), three-magnon processes (terms proportional to $T^{(1)}_{ \xi\mathbf{k}_0-\mathbf{k}, \xi\mathbf{k}_0}$), and magnon-phonon nonlinear interaction (remaining terms in the last line). In addition, for each of these mechanisms many different processes can reach parametric instability, namely processes where the amplified modes stem from the first, second, or third harmonic, or a combination of them.
Our first goal is thus to identify the most relevant of these processes, namely the one which reaches threshold at the lowest driving power. To this end it is sufficient to consider each process in Eqs.~\eqref{eq:parametric_amplification_EOM_s} and \eqref{eq:parametric_amplification_EOM_a} separately, which allows to split the problem into independent systems of two ordinary differential equations. Since the equations have an identical form for all the processes, let us illustrate their structure and solution with the specific example of the three-magnon process where magnons from the third harmonic mode $3 \mathbf{k}_0$ split into two new modes. As we will show below, this is the process with lowest threshold
for the parameters of Table~\ref{tab:parameters 2}.
This case is described by the closed system of equations
\begin{multline}\label{linearized1}
    \dot s_\mathbf{k} = \left( -i\omega^\text{m}_\mathbf{k} - \frac{\gamma^\text{m}_{\mathbf{k}}}{2} \right) s_\mathbf{k} \\
    - e^{-i3\omega_\text{d} t} T^{(1)}_{ 3\mathbf{k}_0-\mathbf{k}, 3\mathbf{k}_0} \, \alpha^{\text{m}}_{3} \,  s_{3\mathbf{k}_0-\mathbf{k}}^*,
\end{multline}
\begin{multline}\label{linearized2}
    \dot s_{3\mathbf{k}_0-\mathbf{k}}^* = \left( i\omega^\text{m}_{3\mathbf{k}_0-\mathbf{k}} - \frac{\gamma^\text{m}_{3\mathbf{k}_0-\mathbf{k}}}{2} \right) s_{3\mathbf{k}_0-\mathbf{k}}^* \\
    - e^{i3\omega_\text{d} t} T^{(1)*}_{ \mathbf{k}, 3\mathbf{k}_0} \, \alpha^{\text{m}*}_{3} \,  s_{\mathbf{k}}.
\end{multline}
This equation is solved by an exponential function,
\begin{align}
    s_\mathbf{k}(t)= s_\mathbf{k}(0) e^{(-i\bar\Omega_\mathbf{k}-\Gamma_\mathbf{k}/2)t},
\end{align}
with the real coefficients $\bar\Omega_\mathbf{k}$ and $\Gamma_\mathbf{k}$ being defined through the identity
\begin{multline}\label{eq:sk_solution_exponent}
    -i\bar\Omega_\mathbf{k}-\Gamma_\mathbf{k}/2 =
    -i\frac{3\omega_d-(\omega_{3\mathbf{k}_0-\mathbf{k}}^\text{m}-\omega_\mathbf{k}^\text{m})}{2} 
    -\frac{\gamma_\mathbf{k}^\text{m}+\gamma_{3\mathbf{k}_0-\mathbf{k}}^\text{m}}{4} \\
    \pm i\frac{\sqrt{\left(\delta +i\frac{\gamma_\mathbf{k}^\text{m}-\gamma_{3\mathbf{k}_0-\mathbf{k}}^\text{m}}{2}\right)^2-4|\alpha^{\text{m}}_3|^2  T^{(1)}_{3\mathbf{k}_0-\mathbf{k},3\mathbf{k}_0}\, T^{(1)*}_{k,3\mathbf{k}_0}}}{2},
\end{multline}
where $\delta \equiv 3\omega_d -(\omega_\mathbf{k}^\text{m}+\omega_{3\mathbf{k}_0-\mathbf{k}}^\text{m})$ is the frequency difference between the original mode and the two modes populated by instability. Considering that $\alpha^{\text{m}}_3$ is an increasing function of the drive $\alpha_\text{d}$, it is clear that there exists a threshold driving power at which $\Gamma_\mathbf{k}$ becomes negative and the solution becomes amplified instead of damped. This defines the threshold for parametric instability~\cite{braginsky_parametric_2001,Danilishin_time_2014}.
In addition, since in Eq.~\eqref{eq:sk_solution_exponent} the drive $|\alpha^{\text{m}}_3|^2$ is accompanied by the $\mathbf{k}-$dependent factor $T^{(1)}_{3\mathbf{k}_0-\mathbf{k},3\mathbf{k}_0}\, T^{(1)*}_{k,3\mathbf{k}_0}$, the instability threshold is different for every pair of modes $\{ \mathbf{k},3\mathbf{k}_0-\mathbf{k} \}$. The pair $\{ \mathbf{k},3\mathbf{k}_0-\mathbf{k} \}$ with the lowest threshold is obtained by performing a numerical maximization of the function $\text{Re}[\Gamma_\mathbf{k}]$ in the two-dimensional  variable $\mathbf{k}$ for increasing values of the driving strength $\alpha_d$, until a positive value is obtained. This method also provides a value for the threshold drive $\alpha_d$.

\begin{table}[ht!]
    \centering
    \begin{tabular}{l|l|l}
    \hline\hline
        Parameter & & Value\\ \hline
        Thickness & $d$ & $40\,\mathrm{nm}$\\
        Mass density  & $\{ \rho_{s}, \rho_{f} \}$ & $\{4650,7800 \}\,\mathrm{kg\,m^{-3}}$ \\
        Longitudinal sound velocity & $\{ c_{l,s}, c_{l,f}\}$ & $\{ 6247,6419 \}\,\mathrm{m\,s^{-1}}$ \\
        Transverse sound velocity & $\{ c_{t,s}, c_{t,f}\}$ & $\{ 5000,2621 \}\,\mathrm{m\,s^{-1}}$ \\
        Gyromagnetic ratio & $\gamma_g$ & $-1.76\times 10^{11}\, \mathrm{T^{-1}\,s^{-1}}$ \\
        External field & $\mu_0 H_0$ & $20\,\mathrm{mT}$ \\
        Saturation magnetization & $\mu_0 M_S$ & $1.4\, \mathrm{T}$ \\
        Gilbert damping coefficient & $\lambda$ & $1.73\times 10^{-2}$\\
        Phonon quality factor & $Q_0$ & $490$\\
        Exchange stiffness & $A_x$ & $2.3\times 10^{-11} \, \mathrm{J\,m^{-3}}$\\
        Magnetoelastic constant & $B_1$ & $-2.03\times 10^7 \, \mathrm{J\,m^{-3}}$\\
        Magnetoelastic constant & $B_2$ & $-1.35\times 10^7 \, \mathrm{J\,m^{-3}}$\\
    \hline\hline
    \end{tabular}
    \caption{
    Set of alternative system parameters employed for Figs.~\ref{fig:Magentization_spectrum}(b) and \ref{fig:dispersion_relation_2}. They lie within the range of typical parameters for a LiNbO3 substrate and a CoFeB thin film (indices $s$ and $f$ respectively). 
    }
    \label{tab:parameters 2}
\end{table}

\begin{figure}
    \centering
    \includegraphics[width=1\linewidth]{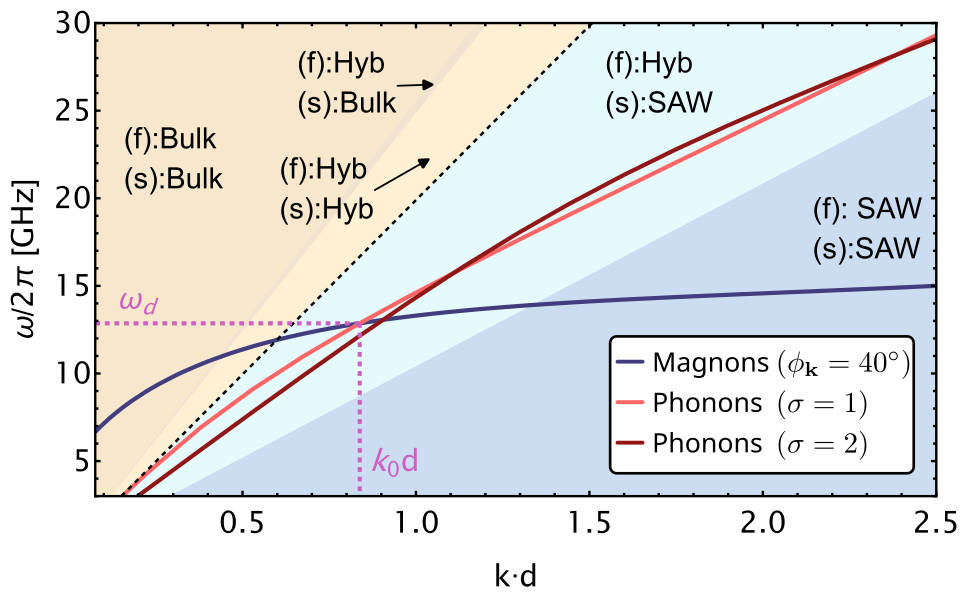}
    \caption{
     Dispersion relation of lowest band ($n=0$) phonon and magnon modes propagating at an angle of $40\deg$ to the external magnetic field, for the parameters in Table~\ref{tab:parameters 2}. Background colors designate the different regimes of phonon mode types, i.e. bulk, hybrid, or surface acoustic. The black dotted line marks the beginning of SAW behavior of phonon modes in the substrate. The pink dotted lines show the driving frequency $\omega_\text{d}/2\pi=12.86\,\mathrm{GHz}$ (on resonance with the $\sigma=1$ phonon mode) and wave vector $k_0=2\pi/300\,\mathrm{nm}^{-1}$ of the driven mode.}
    \label{fig:dispersion_relation_2}
\end{figure}

We undertake the same procedure for all the possible instability processes in Eqs.~\eqref{eq:parametric_amplification_EOM_s} and \eqref{eq:parametric_amplification_EOM_a}, displaying our results in Fig. \ref{fig:Magentization_spectrum}a. This figure displays the peaks in the magnetization spectrum (i.e. the Fourier transform of $\mathbf{m}(\mathbf{r},t)$) as a function of acoustic driving power, for the parameters of Table~\ref{tab:parameters_1}. The circles indicate the threshold drive for each parametric instability process. Several processes feed back into the driven mode $\mathbf{k}_0$ as evidenced by the points lying at the drive frequency $\omega_\text{d}$. This makes any of these potential processes difficult to observe above the high-intensity first harmonic peak and thus experimentally not relevant.
The first observable subharmonic excitation is therefore predicted to originate from the scattering of a third harmonic phonon into two magnons, indicating that the nonlinear response is dominated by magnetoelastic interaction. This is however not always the case, as the position and nature of the peaks in the magnetization spectrum is strongly dependent on several system parameters. To illustrate this we consider the slightly different parameters of Table \ref{tab:parameters 2}. As opposed to the case displayed in Fig.~\ref{fig:Magentization_spectrum}\textcolor{blue}{(a)}, where we considered resonant driving of the acoustic mode $\sigma=2$ along the direction of $\mathbf{B}_0$, here we consider drive of the acoustic mode $\sigma=1$ at an angle
$\phi_{\mathbf{k}_0}=40^\circ$ and resonant frequency $\omega_\text{d}/2\pi=12.86\,\mathrm{GHz}$ (see Fig. \ref{fig:dispersion_relation_2}). As shown in Fig.~\ref{fig:Magentization_spectrum}\textcolor{blue}{(b)} the magnetization spectrum is sharply different, with three-magnon nonlinearities becoming dominant. Specifically, the first observable subharmonic peaks correspond to splitting of the first harmonic $\mathbf{k}_0$ into two magnon modes.

Note that the current method based on Eqs.~\eqref{eq:parametric_amplification_EOM_s} and \eqref{eq:parametric_amplification_EOM_a} allows to predict the parametric instability threshold but not the mode dynamics at higher powers, where it predicts an exponential increase of the mode amplitudes. This is because this simplified model does not take into account the depletion of the driven mode that occurs when it starts pumping energy into the new modes. In the next section we detail how to refine the model and compute the mode amplitudes above threshold, a necessary requisite to build a mean-field quantum theory.

\subsection{Beyond-threshold dynamics for a three-mode system}\label{subsec:3mode_system}

As discussed above, in deriving the linearized equations Eqs.~\eqref{eq:parametric_amplification_EOM_s} and \eqref{eq:parametric_amplification_EOM_a} we neglected  small nonlinear terms. 
In this section we describe the dynamics beyond the threshold by explicitly including these terms. As in the previous section, we consider each parametric instability process independently. 
For the sake of illustration, for this section we focus on three-magnon processes for which most of the results are analytical. We comment on other processes at the end of the section.

Specifically, we consider the three magnon processes where a magnon of mode $\mathbf{k}_0$ excites two modes of wave vectors $\mathbf{k}$ and $\mathbf{k}_0-\mathbf{k}$. Since only these three modes are significantly populated we write $s_\mathbf{q} = \sum_{j=1}^3 \bar s_{\mathbf{q}_j} \delta(\mathbf{q}-\mathbf{q}_j)$ in
the full magnetoelastic equations of motion Eqs.~\eqref{eq:EOM_sk} and \eqref{eq:EOM_ak}, where $\mathbf{q}_j\in\{\mathbf{k}_0,\mathbf{k},\mathbf{k}_0-\mathbf{k}\}$. We obtain
a system of three coupled nonlinear equations for the renormalized amplitudes $\bar s_{\mathbf{q}_j}$
\cite{braginsky_parametric_2001,Danilishin_time_2014},

\begin{multline}\label{eq:3-mode_eq1}
        \dot{s}_{\mathbf{k}_0} = \left( -i\omega^{\text{m}}_{\mathbf{k}_0} - \frac{\gamma^{\text{m}}_{\mathbf{k}_0}}{2} \right) s_{\mathbf{k}_0} \\
        + (T^{(2)}_{\mathbf{k},\mathbf{k}_0-\mathbf{k}} + T^{(2)}_{\mathbf{k}_0-\mathbf{k},\mathbf{k}}) s_{\mathbf{k}} s_{\mathbf{k}_0-\mathbf{k}},
\end{multline}
\begin{multline}\label{eq:3-mode_eq2}
        \dot{s}_{\mathbf{k}} = \left( -i\omega^{\text{m}}_{\mathbf{k}} - \frac{\gamma^{\text{m}}_{\mathbf{k}}}{2} \right) s_{\mathbf{k}} \\
        + (s_{\mathbf{k}_0} - \alpha^\text{m}_0e^{-i\omega_\text{d}t}) T^{(1)}_{\mathbf{k}_0-\mathbf{k},\mathbf{k}_0} s_{\mathbf{k}_0-\mathbf{k}}^* ,
\end{multline}
\begin{multline}\label{eq:3-mode_eq3}
        \dot{s}_{\mathbf{k}_0-\mathbf{k}} = \left( -i\omega^{\text{m}}_{\mathbf{k}_0-\mathbf{k}} - \frac{\gamma^{\text{m}}_{\mathbf{k}_0-\mathbf{k}}}{2} \right) s_{\mathbf{k}_0-\mathbf{k}} \\
        + (s_{\mathbf{k}_0} - \alpha^\text{m}_0 e^{-i\omega_\text{d}t}) T^{(1)}_{\mathbf{k},\mathbf{k}_0} s_{\mathbf{k}}^* .
\end{multline}
where we have omitted the bar for simplicity. The three coupling coefficients in this case are real-valued and fulfill $T_{\mathbf{k},\mathbf{k'}}^{(2)}=T^{(2)}_{\mathbf{k'},\mathbf{k}}>0>T_{\mathbf{k},\mathbf{k'}}^{(1)}$. 
To simplify the above equations we first introduce another shift in the magnon mode amplitude $s_{\mathbf{k}_0}\rightarrow s_{\mathbf{k}_0} - \alpha^\text{m}_0 e^{-i\omega_\text{d}t}$. Then, we define oscillating amplitudes $s_j=\tilde{s}_j e^{-i\Omega_j t}$ with $j=\mathbf{k}_0,\mathbf{k},\mathbf{k}_0-\mathbf{k}$, where the oscillation frequencies satisfy
$\Omega_{\mathbf{k}_0}=\omega_{\text{d}}$ and $\Omega_{\mathbf{k}} + \Omega_{\mathbf{k}_0-\mathbf{k}}= \Omega_{\mathbf{k}_0}$.  The equations for these new variables become time-independent, 
\begin{align}
    \label{eq:3-mode_eq4}
    \dot{s}_{\mathbf{k}_0} =\mspace{-2mu} \left( \mspace{-4mu} -i\Delta_{\mathbf{k}_0} \mspace{-2mu}- \frac{\gamma^{\text{m}}_{\mathbf{k}_0}}{2} \right) s_{\mathbf{k}_0} + 2 T^{(2)}_{\mathbf{k},\mathbf{k}_0-\mathbf{k}} s_{\mathbf{k}} s_{\mathbf{k}_0-\mathbf{k}} + \alpha, 
\end{align}
\begin{align}
    \label{eq:3-mode_eq5}
    \dot{s}_{\mathbf{k}} = &\left( \mspace{-4mu} -i\Delta_{\mathbf{k}} - \frac{\gamma^{\text{m}}_{\mathbf{k}}}{2} \right) s_{\mathbf{k}} + T^{(1)}_{\mathbf{k}_0-\mathbf{k},\mathbf{k}_0} s_{\mathbf{k}_0-\mathbf{k}}^* s_{\mathbf{k}_0}, 
\end{align}
\begin{align}
\label{eq:3-mode_eq6}
    \dot{s}_{\mathbf{k}_0-\mathbf{k}} = &\mspace{-1mu} \left( \mspace{-4mu} -i\Delta_{\mathbf{k}_0-\mathbf{k}} \mspace{-2mu} - \frac{\gamma^{\text{m}}_{\mathbf{k}_0-\mathbf{k}}}{2} \right) s_{\mathbf{k}_0-\mathbf{k}} + T^{(1)}_{\mathbf{k},\mathbf{k}_0} s_{\mathbf{k}}^* s_{\mathbf{k}_0} ,
\end{align}
where we have dropped the tilde for notational simplicity, and where we have introduced $\Delta_j\equiv \omega^{\text{m}}_j - \Omega_j$ and the complex driving parameter $\alpha\equiv(-i\Delta_{\mathbf{k}_0}-\gamma^{\text{m}}_{\mathbf{k}_0}/2) \alpha^{\text{m}}_0$. Because the above system of equations is time-independent, the frequencies $\Omega_j$ correspond to the frequency at which mode $j$ oscillates in the steady-state. In analogy to a forced harmonic oscillator, these oscillation frequencies can be different than the natural frequencies $\omega_j^{\rm m}$.

We are interested in solutions to the above equations for which the amplitudes reach a stationary finite value. We thus set the left-hand-side to zero to obtain equations for the steady-state amplitudes. One of the steady-state solutions can be readily identified from the equations, and we refer to it as the ``trivial'' solution,
\begin{align}\label{eq:3mode_trivial_solution}
    s_{\mathbf{k}_0}= \frac{\alpha}{i\Delta_{\mathbf{k}_0} + \gamma^{\text{m}}_{\mathbf{k}_0}/2}\hspace{0.3cm} ; \hspace{0.3cm} \quad s_{\mathbf{k}}=s_{\mathbf{k}_0-\mathbf{k}}=0.
\end{align}
This solution, which exists for all values of the driving $\alpha$, corresponds to the driven mode being populated by the drive while the remaining two modes remain unpopulated. It can be shown by a linear stability analysis~\cite{Danilishin_time_2014} that the trivial solution is only stable for driving strengths $\alpha$   below the parametric instability threshold. 

A more involved calculation shows that there is a second solution for which $s_{\mathbf{k}},\;s_{\mathbf{k}_0-\mathbf{k}}\neq 0$ (see Appendix \ref{app:3mode_stability} for details). This solution exists only if the additional condition $\text{Im}[(-i\Delta_{\mathbf{k}}-\gamma^\text{m}_{\mathbf{k}}/2)(i\Delta_{\mathbf{k}_0-\mathbf{k}}-\gamma^\text{m}_{\mathbf{k}_0-\mathbf{k}}/2)]=0$ is satisfied. This condition, together with $\Omega_{\mathbf{k}} + \Omega_{\mathbf{k}_0-\mathbf{k}}= \Omega_{\mathbf{k}_0}$, fully determines the values of the two frequencies $\Omega_{\mathbf{k}}$ and $\Omega_{\mathbf{k}_0-\mathbf{k}}$.
The mode amplitudes for this solution read
\begin{equation}\label{s0constant}
    |s_{\mathbf{k}_0}|^2 = \frac{\Delta_{\mathbf{k}} \Delta_{\mathbf{k}_0-\mathbf{k}} + \gamma^{\text{m}}_{\mathbf{k}}\gamma^{\text{m}}_{\mathbf{k}_0-\mathbf{k}}/4}{T^{(1)}_{\mathbf{k}_0-\mathbf{k},\mathbf{k}_0} T^{(1)}_{\mathbf{k},\mathbf{k}_0}},
\end{equation}
\begin{equation}\label{s1constant}
    |s_{\mathbf{k}}|^2 = \frac{  (-i\Delta_{\mathbf{k}_0-\mathbf{k}}- \gamma^{\text{m}}_{\mathbf{k}_0-\mathbf{k}}/2) A(\alpha) }{2 T^{(2)}_{\mathbf{k},\mathbf{k}_0-\mathbf{k}}  T^{(1)}_{\mathbf{k},\mathbf{k}_0}} ,
\end{equation}
\begin{equation}\label{s2constant}
    |s_{\mathbf{k}_0-\mathbf{k}}|^2 = \frac{(-i\Delta_{\mathbf{k}} - \gamma^{\text{m}}_{\mathbf{k}}/2) A(\alpha) }{2 T^{(2)}_{\mathbf{k},\mathbf{k}_0-\mathbf{k}}  T^{(1)}_{\mathbf{k}_0-\mathbf{k},\mathbf{k}_0}} .
\end{equation}
with $A(\alpha)\equiv ( \alpha/s_{\mathbf{k}_0} -i\Delta_{\mathbf{k}_0} - \gamma^{\text{m}}_{\mathbf{k}_0}/2 )$. The complex phase of the driven mode is given by
\begin{equation}
    e^{i\arg(s_{\mathbf{k}_0})}= e^{i(\arg(\alpha)+\theta_{\mathbf{k}_0-\mathbf{k}})} \left(i B - \sqrt{-B^2+1} \right),
\end{equation}
with a factor $B\equiv |s_{\mathbf{k}_0}(i\Delta_{\mathbf{k}_0} \mspace{-2mu} + \gamma^{\text{m}}_{\mathbf{k}_0}/2)| \sin(\theta_{\mathbf{k}_0}  \mspace{-4mu} + \theta_{\mathbf{k}_0-\mathbf{k}})/|\alpha|$ and arguments $\theta_{\mathbf{q}}\equiv \arg(-i\Delta_{\mathbf{q}} - \gamma^{\text{m}}_{\mathbf{q}}/2)$ for wave vectors $\mathbf{q}$. In the case of resonant drive considered throughout this work $\Delta_{\mathbf{k}_0}=0$. The complex phases of the remaining two modes are not fully determined, but their sum needs to fulfill
\begin{equation}\label{phasecondition}
    \arg(s_{\mathbf{k}_0}) = \arg(s_{\mathbf{k}})+\arg(s_{\mathbf{k}_0-\mathbf{k}})+\theta_{\mathbf{k}_0-\mathbf{k}}+\pi.
\end{equation}
This solution exists only for driving strengths $\alpha$ above the parametric instability threshold, which is given by
\begin{equation}
    \alpha_{\rm threshold}=\frac{\left( \Delta_{\mathbf{k}} \Delta_{\mathbf{k}_0-\mathbf{k}} + \gamma^{\text{m}}_{\mathbf{k}}\gamma^{\text{m}}_{\mathbf{k}_0-\mathbf{k}}/4 \right)\left(\Delta_{\mathbf{k}_0}^2+\gamma_{\mathbf{k}_0}^2/4\right)}{T^{(1)}_{\mathbf{k},\mathbf{k}_0}T^{(1)}_{\mathbf{k}_0-\mathbf{k},\mathbf{k}_0}}.
\end{equation}

\begin{figure}
    \centering
    \includegraphics[width=1\linewidth]{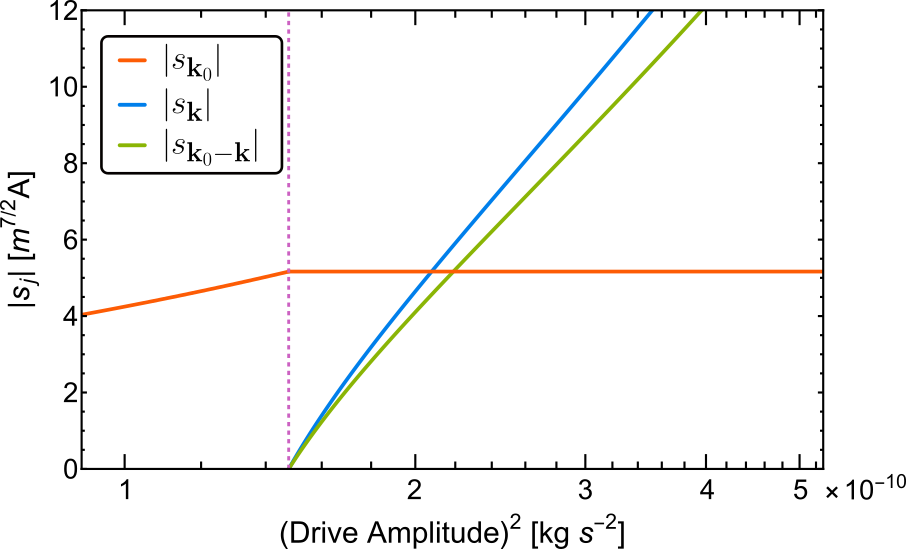}
    \caption{
    Modulus of mode amplitudes in three-mode system of coupled magnons. Below the threshold driving power (pink dashed line) only the trivial solution Eq.~(\ref{eq:3mode_trivial_solution}) exists. Above the threshold this solution becomes unstable and the parametric instability appears.
    }
    \label{fig:3mode}
\end{figure}

The mode amplitudes for both solutions are shown in Figure \ref{fig:3mode} for the parameters of Table~\ref{tab:parameters 2} with the dotted line marking the threshold power. As power increases, the occupation of mode $\mathbf{k}_0$ increases, until the power reaches the threshold. At this point the modes  $\mathbf{k}$ and $\mathbf{k}_0-\mathbf{k}$ become   parametrically amplified while the occupation of mode $\mathbf{k}_0$ remains constant, as also evidenced by Eq.~\eqref{s0constant}. This saturation of the driven mode is the reason why, in the general case where many processes can give rise to parametric instability (see Fig.~\ref{fig:Magentization_spectrum}), only the process with the lowest threshold appears in the spectrum~\cite{HwangArxiv2025}, at least for a range of powers above threshold.

We conclude this section by remarking that the same approach can be employed to determine the mode amplitudes above threshold for every nonlinear process considered in this work. While in the case of three-magnon interaction the solutions are analytical, this is likely not the case for four-magnon or magnon-phonon nonlinear processes, as in such cases the analog of Eqs.~\eqref{eq:3-mode_eq1} and \eqref{eq:3-mode_eq3} would be a system of more than three equations. The steady-state solutions in these cases can be computed numerically.

\section{Quantization of the theory}\label{sec:quantum}
In this section we quantize the magnetoelastic equations of motion Eqs.~\eqref{eq:EOM_sk} and \eqref{eq:EOM_ak} via canonical quantization. This method consists on promoting magnon and phonon amplitudes to ladder operators (see e.g., Refs.~\cite{mills_quantum2006,gonzalez-ballestero_quantum_2020,gonzalez-ballestero_theory_2020} and references therein) as
\begin{equation}
    s_\mathbf{k} \rightarrow \mathcal{M}_0 \hat s_\mathbf{k},
\end{equation}
\begin{equation}
    a_{\sigma,\mathbf{k}} \rightarrow \mathcal{U}_{0,\sigma,\mathbf{k}} \hat  a_{\sigma,\mathbf{k}},
\end{equation}
and imposing canonical bosonic commutation relations,
\begin{equation}\label{comm1}
    \left[\hat{s}_\mathbf{k},\hat{s}_\mathbf{q}\right]=\left[\hat{a}_\mathbf{k},\hat{s}_\mathbf{q}\right]=\left[\hat{a}_\mathbf{k},\hat{a}_\mathbf{q}\right]=\left[\hat{s}_\mathbf{k}^\dagger,\hat{a}_\mathbf{q}\right]=0,
\end{equation}
and
\begin{equation}\label{comm2}
    \left[\hat{s}_\mathbf{k},\hat{s}_\mathbf{q}^\dagger\right]=\left[\hat{a}_\mathbf{k},\hat{a}_\mathbf{q}^\dagger\right]=\delta(\mathbf{k}-\mathbf{q}).
\end{equation}
The factors $\mathcal{M}_0$ and $\mathcal{U}_{0,\sigma,\mathbf{k}}$ are the zero-point magnetization and displacement respectively, and are given by
\begin{equation}
    \mathcal{M}_0 \equiv  \sqrt{\hbar}\mathcal{M}=\sqrt{\hbar\vert \gamma_g\vert M_S},
\end{equation}
\begin{equation}
    \mathcal{U}_{0,\sigma,\mathbf{k}}\equiv\sqrt{\hbar}\mathcal{U_{\sigma,\mathbf{k}}}=\sqrt{\frac{\hbar}{2\omega^p_{\sigma,\mathbf{k}}}}.
\end{equation}
Note that according to this convention the zero-point magnetization is mode-independent.

The canonical quantization method is complete for linear and conservative Hamiltonians. In this work, however, we aim at directly quantizing 
the equations of motion Eqs.~\eqref{eq:EOM_sk} and \eqref{eq:EOM_ak}, which are nonlinear and dissipative. To do so we need to undertake two additional steps. First, we symmetrize any product of non-commuting operators -- that only appear due to the nonlinearity -- to ensure that the conservative dynamics of the Heisenberg operators $\hat{s}_\mathbf{k}(t)$ and $\hat{a}_{\sigma,\mathbf{k}}(t)$ is consistent with the dynamics generated by a Hermitian Hamiltonian~\cite{cohen2019quantum}. Second, we introduce Langevin force operators for each magnon and phonon mode, $\hat f^{\text{m}}_{\mathbf{k}}(t)$ and $\hat f^{\text{p}}_{\sigma, \mathbf{k}}(t)$, to ensure that the commutation relations Eqs.~\eqref{comm1} and \eqref{comm2} are preserved during time evolution~\cite{gardiner2004quantum}. The properties of these operators are described below.
The resulting Heisenberg-Langevin equations are given by
\begin{equation}\label{eq:quantum_EOM_s}
    \frac{d}{dt} \hat s_{\mathbf{k}} = \left( -i\omega^\text{m}_\mathbf{k}-\frac{\gamma^\text{m}_{\mathbf{k}}}{2} \right) \hat s_\mathbf{k} + \hat{\mathcal{D}}^{\text{m}}_{\rm int}(\mathbf{k}) + \hat f^{\text{m}}_{\mathbf{k}},
\end{equation}
\begin{multline}\label{eq:quantum_EOM_a}
    \frac{d}{dt} \hat a_{\sigma,\mathbf{k}} = \left( -i\omega^\text{p}_{\sigma,\mathbf{k}} -\frac{\gamma^{\text{p}}_{\sigma,\mathbf{k}}}{2}\right) \hat a_{\sigma,\mathbf{k}} + \hat{\mathcal{D}}^{\text{p}}_{\rm int}(\mathbf{k}) \\
    + \hat f^{\text{p}}_{\sigma,\mathbf{k}}+ i\frac{\alpha_\text{d}}{\mathcal{U}_{0,\sigma,\mathbf{k}}} \delta(\mathbf{k} - \mathbf{k}_0) e^{-i\omega_\text{d} t},
\end{multline}
where all operators are in the Heisenberg picture. The operators $\hat{\mathcal{D}}^{\text{m}}_{\rm int}(\mathbf{k})$ and $\hat{\mathcal{D}}^{\text{p}}_{\rm int}(\mathbf{k})$ capture all the linear and nonlinear interactions and take the following form:
\begin{widetext}
\begin{multline}
    \hat{\mathcal{D}}^{\text{m}}_{\rm int}(\mathbf{k}) = \sum_\sigma \frac{\mathcal{U}_{0,\sigma,\mathbf{k}}}{\mathcal{M}_0} g_{\sigma,\mathbf{k}}\, \hat a_{\sigma,\mathbf{k}} 
   + \sum_\sigma \int d^2\mathbf{q}\; \mathcal{U}_{0,\sigma,\mathbf{k}}\Bigl( G^{(1)}_{\sigma,\mathbf{q},\mathbf{k}} \, \hat a_{\sigma,\mathbf{q}} \hat s_{\mathbf{k}-\mathbf{q}} + G^{(2)}_{\sigma,\mathbf{q},\mathbf{k}} \, \hat a_{\sigma,\mathbf{q}}^\dagger \hat s_{\mathbf{k}+\mathbf{q}} + G^{(3)}_{\sigma,\mathbf{q},\mathbf{k}} \, \hat a_{\sigma,\mathbf{q}} \hat s_{\mathbf{q}-\mathbf{k}}^\dagger \Bigr) \\
    + \int d^2\mathbf{q}\; \mathcal{M}_0 \Bigl( T^{(1)}_{\mathbf{q},\mathbf{k+q}}\, \hat s_{\mathbf{q}}^\dagger \hat s_{\mathbf{k+q}} + T^{(2)}_{\mathbf{q},\mathbf{k-q}}\, \hat s_{\mathbf{q}} \hat s_{\mathbf{k-q}}\Bigr)
    +\int d^2 \mathbf{q}_1 \int d^2\mathbf{q}_2\; \mathcal{M}_0^2 \Bigl( F^{(1)}_{\mathbf{q}_1,\mathbf{q}_2,\mathbf{k}-(\mathbf{q}_1+\mathbf{q}_2)}\, \hat s_{\mathbf{q}_1} \hat s_{\mathbf{q}_2} \hat s_{\mathbf{k}-(\mathbf{q}_1+\mathbf{q}_2)} \\+ F^{(2)}_{\mathbf{q}_1,\mathbf{q}_2,(\mathbf{q}_1+\mathbf{q}_2)-\mathbf{k}}\, \frac{1}{3} ( 
    \hat s_{\mathbf{q}_1} \hat s_{\mathbf{q}_2} \hat s_{(\mathbf{q}_1+\mathbf{q}_2)-\mathbf{k}}^\dagger + \hat s_{\mathbf{q}_1} \hat s_{(\mathbf{q}_1+\mathbf{q}_2)-\mathbf{k}}^\dagger \hat s_{\mathbf{q}_2} + \hat s_{(\mathbf{q}_1+\mathbf{q}_2)-\mathbf{k}}^\dagger \hat s_{\mathbf{q}_1} \hat s_{\mathbf{q}_2})\\
    \phantom{+\int d^2 \mathbf{q}_1 \int d^2\mathbf{q}_2} + F^{(3)}_{\mathbf{q}_1,\mathbf{q}_2,\mathbf{q}_1-(\mathbf{q}_2+\mathbf{k})}\, \frac{1}{3} ( \hat s_{\mathbf{q}_1} \hat s_{\mathbf{q}_2}^\dagger \hat s_{\mathbf{q}_1-(\mathbf{q}_2+\mathbf{k})}^\dagger + \hat s_{\mathbf{q}_2}^\dagger \hat s_{\mathbf{q}_1} \hat s_{\mathbf{q}_1-(\mathbf{q}_2+\mathbf{k})}^\dagger + \hat s_{\mathbf{q}_2}^\dagger \hat s_{\mathbf{q}_1-(\mathbf{q}_2+\mathbf{k})}^\dagger \hat s_{\mathbf{q}_1} )\Bigr),
\end{multline}
\begin{equation}
    \hat{\mathcal{D}}^{\text{p}}_{\rm int}(\mathbf{k}) = \frac{\mathcal{M}_0}{\mathcal{U}_{0,\sigma,\mathbf{k}}} \tilde{g}_{\sigma,\mathbf{k}} \, \hat s_{\mathbf{k}} + \int d^2\mathbf{q} \; \frac{\mathcal{M}_0^2}{\mathcal{U}_{0,\sigma,\mathbf{k}}} \Bigl( \tilde{G}^{(1)}_{\sigma,\mathbf{k},\mathbf{q}} \, \hat s_{\mathbf{q}}^\dagger \hat s_{\mathbf{k}+\mathbf{q}} + \tilde{G}^{(3)}_{\sigma,\mathbf{k},\mathbf{q}} \, \hat s_{\mathbf{q}} \hat s_{\mathbf{k}-\mathbf{q}} \Bigr) .
\end{equation}
\end{widetext}
Note that the symmetrization of three-magnon interactions is omitted because it only affects the mode $\mathbf{k}=0$, which does not exist in realistic finite samples (since in such samples the lowest wave vector, $\vert \mathbf{k}\vert \approx \pi/L$ with $L$ the lateral dimension of the film, is always finite).  
The Langevin noise operators in Eqs.~\eqref{eq:quantum_EOM_s} and \eqref{eq:quantum_EOM_a} model fluctuations induced by an external bath, namely the same bath that is responsible for the damping in the equations of motion. We assume each mode is coupled to an independent, Gaussian, and thermal bath with white-noise correlations, i.e., 
\begin{equation}\label{eq:langevin_correlation_1}
    \langle \hat f^{\text{m}}_\mathbf{k} (t) \rangle = \langle \hat f^{\text{p}}_{\sigma,\mathbf{k}} (t) \rangle = 0,
\end{equation}
\begin{equation}\label{eq:langevin_correlation_2}
    \langle \hat f^{\text{m}}_\mathbf{k} (t) \hat f^{\text{m}}_{\mathbf{k}^\prime} (t^\prime) \rangle = \langle \hat f^{\text{p}}_{\sigma,\mathbf{k}} (t) \hat f^{\text{p}}_{\sigma,\mathbf{k}'} (t^\prime) \rangle=0,
\end{equation}
\begin{multline}\label{eq:langevin_correlation_3}
    \langle \hat f^{\text{m}}_\mathbf{k} (t) \hat f^{\text{m}\dagger}_{\mathbf{k}^\prime} (t^\prime) \rangle = 
    \delta(t-t^\prime) \delta(\mathbf{k}-\mathbf{k}^\prime) \\ \times \gamma^{\text{m}}_{\mathbf{k}} ( \bar{n}(\omega^\text{m}_\mathbf{k},T) + 1 ),
\end{multline}
\begin{align}\label{eq:langevin_correlation_4}
    \langle \hat f^{\text{m}\dagger}_\mathbf{k} (t) \hat f^{\text{m}}_{\mathbf{k}^\prime} (t^\prime) \rangle = \delta(t-t^\prime) \delta(\mathbf{k}-\mathbf{k}^\prime) \gamma^{\text{m}}_{\mathbf{k}} \bar{n}(\omega^\text{m}_\mathbf{k},T), 
\end{align}
\begin{multline}\label{eq:langevin_correlation_5}
    \langle \hat f^{\text{p}}_{\sigma,\mathbf{k}} (t) \hat f^{\text{p}\dagger}_{\sigma',\mathbf{k}'}(t^\prime) \rangle = \delta_{\sigma\sigma'}\delta(t-t^\prime) \delta(\mathbf{k}-\mathbf{k}^\prime) \\ \times \gamma^{\text{p}}_{\sigma,\mathbf{k}} \left( \bar{n}(\omega^\text{p}_{\sigma,\mathbf{k}},T) + 1 \right),
\end{multline}
\begin{multline}\label{eq:langevin_correlation_6}
    \langle \hat f^{\text{p}\dagger}_{\sigma,\mathbf{k}} (t) \hat f^{\text{p}}_{\sigma',\mathbf{k}'}(t^\prime) \rangle = \delta_{\sigma\sigma'}\delta(t-t^\prime) \delta(\mathbf{k}-\mathbf{k}^\prime) \\ \times \gamma^{\text{p}}_{\sigma,\mathbf{k}} \bar{n}(\omega^{\text{p}}_{\sigma,\mathbf{k}},T),
\end{multline}
where $\bar{n}(\omega,T)=(\exp(\hbar \omega/k_B T)-1)^{-1}$ is the Bose-Einstein distribution. The above equalities are equivalent to the fluctuation-dissipation theorem and fully determine the properties of the Langevin noise operators.
Note that the factors $\bar{n}(\omega,T)$ and $\bar{n}(\omega,T)+1$ account for both quantum statistics and vacuum fluctuations, which are key to correctly describe quantum effects arising at low temperatures.

 Equations~\eqref{eq:quantum_EOM_s} and \eqref{eq:quantum_EOM_a} are operator-valued and as such even more complex to solve than their classical counterparts Eqs.~\eqref{eq:EOM_sk} and \eqref{eq:EOM_ak}. A powerful tool to tackle these equations is the mean-field approximation~\cite{Kubo1962,Kadanoffmeanfield}. To apply it we first derive a system of equations of motion for first- and second-order moments $\langle \hat X_1 \rangle$ and $\langle \hat X_1 \hat X_2\rangle$, with $\hat{X}_j\in\{\hat{ s}_{\mathbf{k}},\hat a_{\sigma,\mathbf{k}},\hat{ s}_{\mathbf{k}}^\dagger,\hat a_{\sigma,\mathbf{k}}^\dagger\}$ arbitrary magnon or phonon ladder operators. The mean-field approximation consists in approximating all higher-order expectation values in these equations as
\begin{align}
\begin{split}\label{eq:mean_field_app}
    \langle \hat X_1 \hat X_2 \hat X_3 \rangle \approx &\langle \hat X_1 \hat X_2 \rangle\langle \hat X_3 \rangle +\langle \hat X_1 \hat X_3 \rangle\langle \hat X_2 \rangle\\ 
    &+ \langle \hat X_2 \hat X_3 \rangle\langle \hat X_1 \rangle - 2 \langle \hat X_1 \rangle\langle \hat X_2 \rangle\langle \hat X_3 \rangle,
\end{split}
\end{align}
so that the system of equations, which we refer to as the mean-field equations of motion, becomes closed.
This approximation is equivalent to neglecting the third-order cumulant and valid when correlations are small. Since magnons and phonons form a continuum the mean-field equations of motion still contain an infinite number of equations, making their solution challenging. We can further simplify them by noting that, as discussed in previous sections, any parametric instability process results in only a finite and small number of magnon and phonon modes becoming significantly occupied (for instance, in the three-magnon process of Sec.~\ref{subsec:3mode_system} only magnon modes $\mathbf{k}_0$, $\mathbf{k}$, and $\mathbf{k}_0-\mathbf{k}$ are significantly occupied, and no phonon modes are). We can thus write  
\begin{align}
    \langle \hat s_\mathbf{k} \rangle &= \sum_{\mathbf{q}\text{ occupied}} \delta(\mathbf{k}-\mathbf{q}) \bar{s}_{\mathbf{q}} ,\label{meanfield1} \\
    \langle \hat a_{\sigma,\mathbf{k}} \rangle &= \sum_{\sigma',\mathbf{q} \text{ occupied}}\delta_{\sigma\sigma'}\delta(\mathbf{k}-\mathbf{q}) \bar{a}_{\sigma',\mathbf{q}},\label{meanfield2}
\end{align}
for the first-order moments, and a similar expression for the second-order moments. For example, the expression for the magnon second-order moments is written as
\begin{multline}\label{meanfield3}
    \langle \hat s_{\mathbf{k}} \hat s_{\mathbf{k}^\prime} \rangle = \sum_{\mathbf{q}_1,\mathbf{q}_2\text{ occupied}} \delta(\mathbf{k}-\mathbf{q}_1) \delta(\mathbf{k}-\mathbf{q}_2) \\( \bar{s}_{\mathbf{q}_1} \bar{s}_{\mathbf{q}_2} + \langle \delta\hat{ s}_{\mathbf{q}_1} \delta\hat{ s}_{\mathbf{q}_2} \rangle ) ,
\end{multline}
where for convenience we have split the contribution from the first-order moments from the fluctuations, represented by the term $\langle\delta\hat{ s}_{\mathbf{q}_1} \delta\hat{ s}_{\mathbf{q}_2} \rangle$. 
We assume that unoccupied modes remain in a thermal state, and that the fluctuations for occupied modes are smaller than the expectation values, i.e., $\vert \langle\delta\hat{ s}_{\mathbf{q}_1} \delta\hat{ s}_{\mathbf{q}_2} \rangle\vert \ll \vert \bar{s}_{\mathbf{q}_1} \bar{s}_{\mathbf{q}_2}\vert$. Under these assumptions,
introducing Eqs.~\eqref{meanfield1}-\eqref{meanfield3} into the mean-field equations of motion leads to (i) a closed, nonlinear system of equations for the mean values $\langle \hat s_\mathbf{k} \rangle$, $\langle \hat a_{\sigma,\mathbf{k}} \rangle$ which, as expected for consistency, is equivalent to the classical equations Eqs.~\eqref{eq:EOM_sk} and \eqref{eq:EOM_ak}. (ii) An approximate \textit{linear} system of equations for the fluctuations, whose coefficients depend parametrically on the mean values $\langle \hat s_\mathbf{k} \rangle$, $\langle \hat a_{\sigma,\mathbf{k}} \rangle$.
 The mean-field approximation thus provides a simple route to evaluate quantum expectation values, provided that the full solution of the classical nonlinear equations (i) is known. In the next section we give a specific example for the three-magnon case treated in Sec. \ref{subsec:3mode_system}.

\subsection{Quantum fluctuations of magnetization across the three-magnon instability threshold}

As an example on how to compute quantum expectation values we consider again the system parameters in Table \ref{tab:parameters 2}, for which the origin of parametric instability is the three-magnon scattering process $\mathbf{k}_0 \to \{\mathbf{k},\mathbf{k}_0-\mathbf{k}\}$ (see Sec. \ref{subsec:3mode_system}). 
Following the procedure detailed in the previous section we obtain the following mean-field equations of motion for the first-order moments,
\begin{equation}\label{eq:mean_field_s_EOM}
    \frac{d}{dt} \langle \hat{\mathbf{v}} \rangle(t) = M (t) \langle \hat{\mathbf{v}} \rangle(t),
\end{equation}
where we have defined the vector of variables
\begin{equation}
    \hat{\mathbf{v}}(t) \equiv \left( \hat{\tilde{s}}_{\mathbf{k}_0} , \hat{\tilde{s}}_{\mathbf{k}_0}^\dagger , \hat{\tilde{s}}_{\mathbf{k}} ,  \hat{\tilde{s}}_{\mathbf{k}}^\dagger , \hat{\tilde{s}}_{\mathbf{k}_0-\mathbf{k}} , \hat{\tilde{s}}_{\mathbf{k}_0-\mathbf{k}}^\dagger \right)^T.
\end{equation}
Here we omit the time argument for simplicity and, in analogy to 
Sec.~\ref{subsec:3mode_system}, we have introduced rotating variables $\hat{s}_\mathbf{q}(t) \equiv \hat{\tilde{s}}_\mathbf{q}(t)e^{-i\Omega_\mathbf{q} t}$ and  $\bar{s}_\mathbf{q}(t) \equiv {\tilde{s}}_\mathbf{q}(t)e^{-i\Omega_\mathbf{q} t}$ 
for $\mathbf{q}\in \{\mathbf{k_0}, \mathbf{k}, \mathbf{k}_0 - \mathbf{k} \}$, with 
$\Omega_{\mathbf{k}_0}=\omega_{\text{d}}$ and whith the frequencies $\Omega_{\mathbf{k}}$ and  $\Omega_{\mathbf{k}_0-\mathbf{k}}$  given in Sec.~\ref{subsec:3mode_system} and Appendix~\ref{app:3mode_stability}.
The coefficient matrix $M(t)$ is given by
\begin{widetext}
    \begin{align}\label{Mdef}
    M(t)\equiv
    \begin{pmatrix}
        -i\Delta_1-\gamma^\text{m}_1 /2 & 0 & \mathcal{M}_0 2 T^{(2)}_{2,3} {\tilde{s}}_3(t) & 0 & \mathcal{M}_0 2 T^{(2)}_{2,3} {\tilde{s}}_2(t) & 0 \\
        0& i\Delta_1-\gamma^\text{m}_1 /2 & 0 & \mathcal{M}_0 2 T^{(2)*}_{2,3} {\tilde{s}}_3^*(t) & 0 & \mathcal{M}_0 2 T^{(2)*}_{2,3} {\tilde{s}}_2^*(t) \\
        \mathcal{M}_0 T^{(1)}_{3,1} {\tilde{s}}_3^*(t) & 0 & -i\Delta_2-\gamma^\text{m}_2 /2 & 0 & 0 & \mathcal{M}_0 T^{(1)}_{3,1} {\tilde{s}}_1(t) \\
        0 & \mathcal{M}_0 T^{(1)*}_{3,1} {\tilde{s}}_3(t) & 0 & i\Delta_2-\gamma^\text{m}_2 /2 & \mathcal{M}_0 T^{(1)*}_{3,1} {\tilde{s}}_1^*(t) & 0 \\
        \mathcal{M}_0 T^{(1)}_{2,1} {\tilde{s}}_2^*(t) & 0 & 0 & \mathcal{M}_0 T^{(1)}_{2,1} {\tilde{s}}_1(t) & -i\Delta_3-\gamma^\text{m}_3 /2 & 0 \\
        0 & \mathcal{M}_0 T^{(1)*}_{2,1} {\tilde{s}}_2(t) & \mathcal{M}_0 T^{(1)*}_{2,1} {\tilde{s}}_1^*(t) & 0 & 0 & i\Delta_3-\gamma^\text{m}_3 /2 \\
    \end{pmatrix},
\end{align}
\end{widetext}
where the labels $1,2,3$ indicate wave vectors $\mathbf{k}_0,\mathbf{k}$, and $\mathbf{k}_0-\mathbf{k}$ respectively.  The coefficients ${\tilde{s}}_\mathbf{q} (t)$ appearing in the above matrix can be calculated as $ {\tilde{s}}_\mathbf{q} (t) = s_{\mathbf{q}}(t)/\mathcal{M}_0$ with $s_\mathbf{q}$ the solutions of the classical equations Eqs.~\eqref{eq:3-mode_eq4}-\eqref{eq:3-mode_eq6}. 
The mean-field equations of motion for the second-order moments read
\begin{equation}
    \frac{d}{dt} \Sigma(t) = M(t) \Sigma(t) + \Sigma(t) M^T(t) + Q(t), \label{eq:mean_field_Covariance_EOM}
\end{equation}
where
\begin{equation}
    \Sigma(t) \equiv \Bigl\langle \bigl( \hat{\mathbf{v}}(t) - \langle \hat{\mathbf{v}}(t) \rangle \bigr) \bigl( \hat{\mathbf{v}}(t) - \langle \hat{\mathbf{v}}(t) \rangle \bigr)^T \Bigr\rangle , \label{eq:covariance_matrix}
\end{equation}
is the covariance matrix, and where we define the matrix
\begin{equation}\label{Qmatrixdef}
    Q(t) \equiv \Bigl\langle \hat{\mathbf{F}}^\dagger(t) \hat{\mathbf{v}}^T(t) + \hat{\mathbf{v}}^\dagger(t) \hat{\mathbf{F}}^T(t) \Bigr\rangle ,
\end{equation}
with the vector of Langevin forces 
\begin{equation}
    \hat{\mathbf{F}}(t) \equiv \left( \hat{\tilde{f}}^{\text{(m)}}_{\mathbf{k}_0} , \hat{\tilde{f}}_{\mathbf{k}_0}^{\text{(m)}\dagger} , \hat{\tilde{f}}^{\text{(m)}}_{\mathbf{k}} ,  \hat{\tilde{f}}_{\mathbf{k}}^{\text{(m)}\dagger} , \hat{\tilde{f}}^{\text{(m)}}_{\mathbf{k}_0-\mathbf{k}} , \hat{\tilde{f}}_{\mathbf{k}_0-\mathbf{k}}^{\text{(m)}\dagger} \right)^T,
\end{equation}
and $\hat{f}^{\text{m}}_{\mathbf{q}}(t) \equiv \hat{\tilde{f}}^{\text{m}}_{\mathbf{q}}(t)e^{-i\Omega_\mathbf{q} t}$ . In Eq.~\eqref{Qmatrixdef} the Hermitian conjugation applies only to the vector elements, but does not transpose the vector itself.

Although the above equations allow to compute any quantum expectation value as a function of time, hereafter we focus on the expectation values on the stable steady state, which are particularly relevant. That is, we take the long-time limit $\gamma_\mathbf{q}^{\text{m}} t\gg 1$ for $\mathbf{q}\in \{\mathbf{k_0}, \mathbf{k}, \mathbf{k}_0 - \mathbf{k} \}$, at which the classical solutions $s_{\mathbf{q}}(t)$ reach the stationary value given either in Eq.~\eqref{eq:3mode_trivial_solution} or in Eqs.~\eqref{s0constant}-\eqref{s2constant}, depending on whether the drive is below or above the parametric instability threshold. 
In this steady-state limit the coefficient matrices $M$ and $Q$ [Eqs.~\eqref{Mdef} and \eqref{Qmatrixdef}] become time-independent, and the covariance matrix can be calculated as
\begin{equation}
    \Sigma_{\rm ss} =\int_0^\infty d\tau e^{M\tau}Qe^{M^T \tau} = P N P^T,
\end{equation}
where the steady state matrix $Q$ is defined by $Q_{ij}\delta(t-t^\prime) = \langle \hat F_i(t) \hat F_j(t^\prime) \rangle$, $P$ is the $6\times6$ matrix that diagonalizes $M$, i.e. $M=P\Lambda P^{-1}$ with $\Lambda=\text{diag}[\lambda_1,...\lambda_6]$ the diagonal matrix containing the eigenvalues, and where
\begin{equation}
    N_{ij}\equiv\sum_{c,d=1}^6 \frac{ (P^{-1})_{ci} \, Q_{ij} \, (P^{-1})_{dj} }{-(\lambda_c + \lambda_d)}.
\end{equation}

The above expressions allow to compute experimentally accessible observables such as the magnetization operator, which in the Heisenberg picture reads
\begin{align}
    \hat{\mathbf{m}}(\mathbf{r},t) = \int d^2\mathbf{q}\; \mathcal{M}_0 \left(\mathbf{m}_{\mathbf{q}}(\mathbf{r}) \hat{\tilde{s}}_{\mathbf{q}(t)} e^{-i\Omega_\mathbf{q}t}+ \text{h.c.}\right).
\end{align}
As discussed above, the expectation value of such operator is fully described by the classical solution. Quantum fluctuations are however captured by the total variance of the magnetization,
\begin{equation}
    \Delta\hat{m}(\mathbf{r},t)\equiv \left\langle \big\vert\hat{\mathbf{m}}(\mathbf{r},t)-\langle\hat{\mathbf{m}}(\mathbf{r},t)\rangle\big\vert^2\right\rangle.
\end{equation}
This variance can be split into two contributions,
\begin{equation}\label{mvarsplit}
    \Delta\hat{m}(\mathbf{r},t)\equiv \Delta\hat{m}_{\rm thermal} +\Delta\hat{m}_{\rm d}(\mathbf{r},t),
\end{equation}
where the first term describes the variance at zero driving, given by thermal and vacuum fluctuations, and the second describes the modifications induced by the drive. The zero-drive contribution is given by
\begin{equation}\label{varMthermal}
    \Delta\hat{m}_{\rm thermal}\equiv \int d^2\mathbf{q}\; \mathcal{M}_0 \vert \mathbf{m}_\mathbf{q}(\mathbf{r})\vert^2(2\bar{n}(\omega_\mathbf{q}^{\text{m}},T)+1).
\end{equation}
This contribution, which has been studied in detail before (see Ref.~\cite{gonzalez-ballestero_towards_2022}), is time-independent and, within the single magnon band approximation, also position-independent. Note that Eq.~(\ref{varMthermal}) captures also the quantum fluctuations of the magnetization vacuum as it does not cancel at zero temperature. We are interested in modifications above this thermal background induced by the nonlinearity, so we consider only the first contribution in Eq.~(\ref{mvarsplit}). To eliminate the trivial time dependence we compute its time average, i.e.,
\begin{equation}\label{mvarfinal}
    \Delta\hat{m}_{\rm d, av}\equiv \left\langle
    \Delta\hat{m}_{\rm d}(\mathbf{r},t)
    \right\rangle_\tau,
\end{equation}
where the averaging time is chosen as $\tau \gg \vert \omega^{\rm m}_{\mathbf{k}_0}-\omega^{\rm m}_{\mathbf{k}}\vert^{-1}, \vert \omega^{\rm m}_{\mathbf{k}_0}-\omega^{\rm m}_{\mathbf{k}_0-\mathbf{k}}\vert^{-1},\vert \omega^{\rm m}_{\mathbf{k}}- \omega^{\rm m}_{\mathbf{k}_0-\mathbf{k}}\vert^{-1}$. After this averaging the position dependence also vanishes in the single-band approximation.

\begin{figure}
    \centering
    \includegraphics[width=1\linewidth]{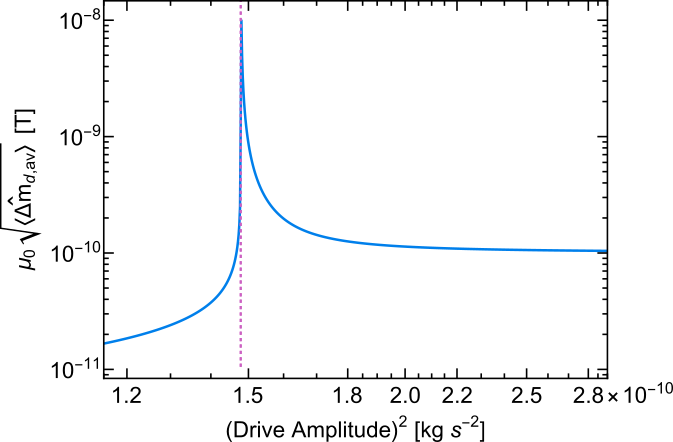}
    \caption{
    Time-average of the magnetization fluctuations above induced by the drive, Eq.~(\ref{mvarfinal}), as a function of driving power, for the parameters of Table~\ref{tab:parameters 2} and $T=1\,\mathrm{mK}$, corresponding to all magnon modes having negligible thermal occupation. The dashed pink line indicates the instability threshold, see Fig.~\ref{fig:3mode}.
    }
    \label{fig:magnetzation_variance}
\end{figure}

The variance Eq.~(\ref{mvarfinal}) is displayed in Fig.~(\ref{fig:magnetzation_variance}) as a function of driving power and at $T=1\,\mathrm{mK}$, a temperature where all the magnon modes can be considered in vacuum. The variance is peaked around the threshold power, indicating a sharp increase in the fluctuations. This rise has been previously explored in three-mode non-degenerate optical parametric amplifiers~\cite{PettiauxBook1989,DechoumPRA2004,Dechoum:16} as well as in lasers across the lasing transition~\cite{DeGiorgioPRA1970,GrahamZPhys1970}, where multiple similarities with second-order phase transitions have been drawn. Indeed, the increase in fluctuations is characteristic of such transitions, and can indicate the breakdown of the mean-field approximation~\cite{Kadanoffmeanfield}. We thus remark that -- like any other mean-field theory -- our model must be used cautiously at driving strengths near the parametric instability threshold.  Far from the threshold, the parametric instability regime is still characterized by magnetization fluctuations that are several times higher than below threshold. These results indicate a fundamental change in the quantum dynamics of the three-mode system across the parametric instability transition, and explicitly provide a method to experimentally probe such transition through the fluctuations of the magnetization (or equivalently, of the stray magnetic field above the film~\cite{gonzalez-ballestero_towards_2022}). As a final remark, 
note that below threshold the fluctuations of modes $\mathbf{k}$ and $\mathbf{k}_0-\mathbf{k}$ are not zero even though their expectation value is, $\langle \hat{s}_\mathbf{k}\rangle=\langle \hat{s}_{\mathbf{k}_0-\mathbf{k}}\rangle=0$. This is manifested in the increase of $\Delta\hat{m}_{\rm d, av}$ as a function of power before the threshold is reached, and indicates the onset of cross-correlations between these modes. Further exploration of such correlations, as well as of near-threshold phenomena in this system, are left for our future work.

\section{Conclusion}\label{sec:conclusion}
We have presented a full classical and quantum theory of coupled magnon-phonon dynamics in two-dimensional geometries. Our theory reproduces our recent experimental observation of parametric instability in the classical regime~\cite{HwangArxiv2025}.
We have shown how to rigorously quantize the classical theory and how to explicitly compute quantum expectation values within the mean-field approximation. As an illustrative example, we have predicted that the crossing of the parametric instability threshold in these systems results in an observable modification of the quantum fluctuations of the involved degrees of freedom.

Our theory paves the way to exploring nonlinear magnetoelasticity in the quantum regime. By providing analytical expressions for all the involved rates in terms of magnon and phonon eigenmode functions, our theory reduces the numerically demanding task of fully solving the classical dynamics to just computing the relevant eigenmodes (which in many cases, such as the film considered in this work, can be done fully analytically). Moreover, by working within the mean-field approximation we avoid the highly complex problem of simulating quantum nonlinear dynamics of a continuum of bosonic modes, each characterized by an infinite-dimensional Hilbert space. Our model thus provides an efficient and accessible toolbox to quantitatively model and tailor the coupled quantum nonlinear dynamics of magnons and phonons. Along this line, future extensions of our work will extend our model to other structures such as waveguides, or to include other effects such as surface spin pinning or magnetocrystalline anisotropy. A second future direction will be to exploit the nonlinearity to prepare purely quantum magnonic states without the need for qubits, significantly reducing the complexity of current quantum magnonics experiments.

\begin{acknowledgments}
C.G.B. and M.B. acknowledge the Austrian Science Fund FWF for the support with the Project No. PAT-1177623 ``Nanophotonics-inspired quantum magnonics''- J.P. acknowledges support of JSPS KAKENHI No. 24K00576 from MEXT, Japan.

\end{acknowledgments}

\section*{Data availability}
The data that support the findings of this article are not publicly available upon publication
because it is not technically feasible and/or the cost of preparing, depositing, and hosting the
data would be prohibitive within the terms of this research project. The data are available
from the authors upon reasonable request.

\appendix

\section{Phonon eigenmodes}\label{app:phonon_eigenmodes}
In this section of the appendix we provide the acoustic eigenmode functions and dispersion relations, which have been derived in detail in another work~\cite{Bruehlmann_unpublished}. To begin with we introduce the functions
\begin{gather}
    p_l(k,\omega) \equiv  i\sqrt{k^2-\frac{\omega^2}{c_{l,s}^2}}, \quad p_t(k,\omega) \equiv i\sqrt{k^2-\frac{\omega^2}{c_{t,s}^2}},\\
    p_l^{\prime}(k,\omega) \equiv \sqrt{\frac{\omega^2}{c_{l,f}^2} - k^2} \quad p_t^{\prime}(k,\omega) \equiv \sqrt{\frac{\omega^2}{c_{t,f}^2} - k^2},
\end{gather}
with $k=\vert \mathbf{k}\vert$ the modulus of an acoustic eigenmode. 
 We also define dimensionless factors
\begin{align}
    R_\rho \equiv \frac{\rho_{s}}{\rho_{f}}, \qquad R_t \equiv \frac{c_{t,s}}{c_{t,f}}, \qquad R_l \equiv \frac{c_{l,s}}{c_{l,f}}.
\end{align}
Solving Navier's equation [Eq.~\eqref{eq:Navier}] in 
the regime of surface acoustic modes in the substrate, i.e., $k>\omega/c_{l,s}$, reveals the existence of two branches of solutions. We focus on the lowest band of each branch. 

\subsection{First branch}
The first branch, which we denote by the index $\sigma=1$, is purely transverse. The dispersion relation of these modes, $\omega_{1,\mathbf{k}}^{p}$, is given by the solution of the equation
\begin{align}\label{eq:phonon_disp_rel_s1}
    p_t^{\prime}(k,\omega) \tan(p_t^{\prime}(k,\omega) d) = R_\rho R_t^2 |p_t(k,\omega)|.
\end{align}
The mode functions can be expressed piece-wise in each region, see Eq.~(\ref{umodefunctions}), with the exponentially decaying contribution
\begin{align}
    \mathbf{u}^{(i)}_{1,\mathbf{k}}(\textbf{r}) &= \tilde{C}_{\mathbf{k}} \sqrt{R_\rho} \cos(p_t^{\prime} d) e^{|p_t| x}  \begin{pmatrix} 0 \\ k_z/k \\ -k_y/k \end{pmatrix}
\end{align}
in the substrate, and the bulk-like mode function
\begin{align}\label{eq:phonon_modefunction_s1_film}
    \mathbf{u}^{(ii)}_{1,\mathbf{k}}(\textbf{r}) &= \tilde{C}_{\mathbf{k}}  \cos(p_t^{\prime} (x-d)) \begin{pmatrix} 0 \\ \cos(\phi_k) \\ -\sin(\phi_k) \end{pmatrix} 
\end{align}
in the thin film. The mode is normalized by the factor 
\begin{multline}
    \tilde C_{\mathbf{k}} = \sqrt{\frac{2}{(2\pi)^2 d }} \\
    \times \left( \frac{R_\rho \cos^2(p_{t}^\prime d)}{|p_{t} |d} + 1 + \frac{\sin(p_{t}^\prime  d) \cos(p_{t}^\prime  d)}{p_{t}^\prime  d} \right)^{-\frac{1}{2}}
\end{multline}
Note that in the above equations all the functions $p_l(k,\omega)$, $p_l'(k,\omega)$, $p_t(k,\omega)$, and $p_t'(k,\omega)$ are evaluated at the wave vector and mode eigenfrequency corresponding to the mode $\mathbf{k}$.

\begin{widetext}
\subsection{Second branch}
The second branch ($\sigma=2$) is mixed transverse and longitudinal. The 
dispersion relation $\omega_{2,\mathbf{k}}^{(p)}$ is given by the solution of
\begin{align}
    \det(\bar M)=0
\end{align}
with the $6\times 6$ matrix
\begin{align}\label{eq:phonon_branch2_Matrix}
   \bar{M}= \begin{pmatrix}
        |p_l| & -ik & -\sqrt{R_\rho} ip_l^{\prime} & \sqrt{R_\rho} ip_l^{\prime} & \sqrt{R_\rho} ik & \sqrt{R_\rho} ik \\
        ik & |p_t| & -\sqrt{R_\rho} ik & -\sqrt{R_\rho} ik & -\sqrt{R_\rho} ip_t^{\prime} & \sqrt{R_\rho} ip_t^{\prime} \\
        \sqrt{R_\rho} R_t^2 (2k^2 - \frac{\omega^2}{c_{t,f}^2}) & -\sqrt{R_\rho} R_t^2 2ik |p_t| & \frac{\omega^2}{c_{t,f}^2} - 2k^2 & \frac{\omega^2}{c_{t,f}^2} - 2k^2 & - 2k p_t^{\prime} & 2k p_t^{\prime} \\
        \sqrt{R_\rho} R_t^2 2ik |p_l| & \sqrt{R_\rho} R_t^2 (k^2 +|p_t|^2 ) & 2k p_l^{\prime} & -2k p_l^{\prime} & p_t^{\prime 2} - k^2 & p_t^{\prime 2} -k^2 \\
        0 & 0 & e^{i p_l^{\prime} d} (2k^2 - \frac{\omega^2}{c_{t,f}^2}) & e^{-i p_l^{\prime} d} (2k^2 - \frac{\omega^2}{c_{t,f}^2}) & e^{i p_t^{\prime} d} 2k p_t^{\prime} & -e^{-i p_t^{\prime} d} 2k p_t^{\prime} \\
        0 & 0 & e^{i p_l^{\prime} d} 2k p_l^{\prime} & -e^{-i p_l^{\prime} d} 2k p_l^{\prime} & e^{i p_t^{\prime} d} (p_t^{\prime 2} - k^2) & e^{-i p_t^{\prime} d} (p_t^{\prime 2} - k^2)
    \end{pmatrix}.
\end{align}
The mode functions take the form
\begin{align}\label{appu1}
    \mathbf{u}_{2, \mathbf{k}}^{(i)}(\mathbf{r})  = &\frac{B_{(i)} e^{|p_l|x}}{k} \begin{pmatrix} |p_l| \\ ik_y \\ ik_z \end{pmatrix} + \frac{ D_{(i)y} e^{|p_t|x}}{k} \begin{pmatrix} -ik \\ |p_t| \frac{k_y}{k} \\ |p_t| \frac{k_z}{k} \end{pmatrix} ,
\end{align}
and
\begin{equation}\label{appu2}
    \mathbf{u}_{2, \mathbf{k}}^{(ii)}(\mathbf{r})  = \frac{A_{(ii)} e^{ip_l^\prime x}}{k} \begin{pmatrix} ip_l^\prime \\ ik_y \\ ik_z \end{pmatrix} + \frac{B_{(ii)} e^{-ip_l^\prime x}}{k} \begin{pmatrix} -ip_l^\prime \\  ik_y \\ ik_z \end{pmatrix} 
        + \frac{C_{(ii)y} e^{ip_t^\prime x}}{k} \begin{pmatrix} -ik \\ ip_t^\prime \frac{k_y}{k}  \\ ip_t^\prime \frac{k_z}{k} \end{pmatrix} + \frac{D_{(ii)y} e^{-ip_t^\prime x}}{k} \begin{pmatrix} -ik \\ -ip_t^\prime \frac{k_y}{k}  \\ -ip_t^\prime \frac{k_z}{k}  \end{pmatrix}.
\end{equation}
The coefficients  $B_{(i)},$ $D_{(i)y} ,$ $A_{(ii)} ,$ $B_{(ii)} ,$ $C_{(ii)y} ,$and $D_{(ii)y}$ obey the system of equations 
\begin{align}
    \bar M \begin{pmatrix}
        B_{(i)} & D_{(i)y} & A_{(ii)} & B_{(ii)} & C_{(ii)y} & D_{(ii)y}
    \end{pmatrix}^T=0,
\end{align}
which determines all coefficients except for a normalization constant. Normalization is performed numerically.
We remark that in Eqs.~\eqref{appu1} and \eqref{appu2} all the functions $p_l(k,\omega)$, $p_l'(k,\omega)$, $p_t(k,\omega)$, and $p_t'(k,\omega)$ are evaluated at the wave vector and mode eigenfrequency corresponding to the mode $\mathbf{k}$.

\section{Magnon coupling rates} \label{app:magnon_couoplig_rates}
In this section we provide the explicit expressions of the nonlinear magnon coupling rates appearing in Eq.~(\ref{eq:magnon_eom}). We start with coupling rates corresponding to three-magnon scattering processes. By comparison of Eqs.~(\ref{eq:magnon_eom_intermediate}) and (\ref{eq:magnon_eom}) we identify
\begin{align}
    \int d^2\mathbf{q}\; T^{(1)}_{\mathbf{q},\mathbf{k+q}}\, s_{\mathbf{q}}^* s_{\mathbf{k+q}} = -i\int d^2\mathbf{q}_1 \int d^2 \mathbf{q}_2 
    \Bigl( \bra{\mathbf{m}_\mathbf{k}}\ket{D^{(3)}[\mathbf{m}_{\mathbf{q}_1},\mathbf{m}_{\mathbf{q}_2}^*]} s_{\mathbf{q}_1} s_{\mathbf{q}_2}^*  
    +\bra{\mathbf{m}_\mathbf{k}}\ket{D^{(3)}[\mathbf{m}_{\mathbf{q}_1}^*,\mathbf{m}_{\mathbf{q}_2}]} s_{\mathbf{q}_1}^* s_{\mathbf{q}_2} \Bigr),
\end{align}
and
\begin{align}
    \int d^2\mathbf{q}\; T^{(2)}_{\mathbf{q},\mathbf{k-q}}\, s_{\mathbf{q}} s_{\mathbf{k-q}} = 
    -i\int d^2\mathbf{q}_1 \int d^2 \mathbf{q}_2  \bra{\mathbf{m}_\mathbf{k}}\ket{D^{(3)}[\mathbf{m}_{\mathbf{q}_1},\mathbf{m}_{\mathbf{q}_2}]} \,s_{\mathbf{q}_1} s_{\mathbf{q}_2},
\end{align}
describing respectively the process of a magnon splitting into two ($\mathbf{k}+\mathbf{q}\to\{\mathbf{k},\mathbf{q}\}$), and two magnons combining into one ($\{\mathbf{q},\mathbf{k-q}\}\to\mathbf{k}$). Introducing the expressions for the magnon mode function Eq.~\eqref{eq:magnon_modefunction} and the Green's function Eq.~\eqref{eq:Greens_function} we obtain the expressions for the three-magnon coupling rates,
\begin{align}
    T^{(1)}_{\mathbf{q},\mathbf{k+q}} = \frac{d(2\pi)^2\omega_M}{M_S} \mathcal{N}_{\mathbf{k}} \mathcal{N}_{\mathbf{q}} \mathcal{N}_{\mathbf{k+q}}
    \times \left[J^+_{\mathbf{k}+\mathbf{q},\mathbf{q},\mathbf{k}}+J^+_{\mathbf{k},\mathbf{k}+\mathbf{q},\mathbf{q}}-J^-_{\mathbf{k},\mathbf{q},\mathbf{k}+\mathbf{q}}\right],
\end{align}
\begin{align}
    T^{(2)}_{\mathbf{q},\mathbf{k-q}} = \frac{d(2\pi)^2\omega_M}{M_S} \mathcal{N}_{\mathbf{k}} \mathcal{N}_{\mathbf{q}} \mathcal{N}_{\mathbf{k-q}}
    \times \left[J^-_{\mathbf{q},\mathbf{k}-\mathbf{q},\mathbf{k}}-J^+_{\mathbf{k},\mathbf{q},\mathbf{k}-\mathbf{q}}-J^+_{\mathbf{k},\mathbf{k}-\mathbf{q},\mathbf{q}}\right],
\end{align}
with the functions
\begin{align}
    J^\pm_{\mathbf{q}_1,\mathbf{q}_2,\mathbf{q}_3}\equiv\left( \pm\sqrt{\nu_x(\mathbf{q}_1) \nu_x(\mathbf{q}_2)} + \sqrt{\nu_y(\mathbf{q}_1) \nu_y(\mathbf{q}_2)} \right) 
    \times\sqrt{\nu_x(\mathbf{q}_3)}\biggl( -1 +\frac{1}{q_3d} (1-e^{-q_3d}) \biggr) \frac{q_{3y} q_{3z}}{q_3^2}.
\end{align}

The four-magnon scattering coupling rates are identified and calculated in an analogous way. The coupling rates correspond to physical processes as depicted in Figure \ref{fig:magnon_interaction_schematics}. 
We define the short-hand notations $\tilde{\mathbf{q}}_1 = \mathbf{k}-(\mathbf{q}_1+\mathbf{q}_2)$, $\tilde{\mathbf{q}}_2=(\mathbf{q}_1+\mathbf{q}_2)-\mathbf{k}$, and $\tilde{\mathbf{q}}_3=\mathbf{q}_1-(\mathbf{q}_2+\mathbf{k})$ and find the three defining equalities
\begin{equation}
    \int d^2 \mathbf{q}_1 \int d^2\mathbf{q}_2 \, F^{(1)}_{\mathbf{q}_1,\mathbf{q}_2,\tilde{\mathbf{q}}_1}\, s_{\mathbf{q}_1} s_{\mathbf{q}_2} s_{\tilde{\mathbf{q}}_1} = -i\int d^2\mathbf{q}_1 \int d^2 \mathbf{q}_2 \int d^2 \mathbf{q}_3 \bra{\mathbf{m}_\mathbf{k}} \ket{D^{(4)}[\mathbf{m}_{\mathbf{q}_1},\mathbf{m}_{\mathbf{q}_2},\mathbf{m}_{\mathbf{q}_3}]} s_{\mathbf{q}_1} s_{\mathbf{q}_2}  s_{\mathbf{q}_3},
\end{equation}
\begin{multline}
    \int d^2 \mathbf{q}_1 \int d^2\mathbf{q}_2 \, F^{(2)}_{\mathbf{q}_1,\mathbf{q}_2,\tilde{\mathbf{q}}_2}\, s_{\mathbf{q}_1} s_{\mathbf{q}_2} s_{\tilde{\mathbf{q}}_2}^* = -i\int d^2\mathbf{q}_1 \int d^2 \mathbf{q}_2 \int d^2 \mathbf{q}_3  \Bigl( \bra{\mathbf{m}_\mathbf{k}} \ket{D^{(4)}[\mathbf{m}_{\mathbf{q}_1}^*,\mathbf{m}_{\mathbf{q}_2},\mathbf{m}_{\mathbf{q}_3}]} s_{\mathbf{q}_1}^* s_{\mathbf{q}_2}  s_{\mathbf{q}_3} \\
        +\bra{\mathbf{m}_\mathbf{k}} \ket{D^{(4)}[\mathbf{m}_{\mathbf{q}_1},\mathbf{m}_{\mathbf{q}_2^*},\mathbf{m}_{\mathbf{q}_3}]} s_{\mathbf{q}_1} s_{\mathbf{q}_2}^*  s_{\mathbf{q}_3} 
        +\bra{\mathbf{m}_\mathbf{k}} \ket{D^{(4)}[\mathbf{m}_{\mathbf{q}_1},\mathbf{m}_{\mathbf{q}_2},\mathbf{m}_{\mathbf{q}_3}^*]} s_{\mathbf{q}_1} s_{\mathbf{q}_2}  s_{\mathbf{q}_3}^* \Bigr),
\end{multline}
and
\begin{multline}
    \int d^2 \mathbf{q}_1 \int d^2\mathbf{q}_2 \, F^{(3)}_{\mathbf{q}_1,\mathbf{q}_2,\tilde{\mathbf{q}}_3}\, s_{\mathbf{q}_1} s_{\mathbf{q}_2}^* s_{\tilde{\mathbf{q}}_3}^* = -i\int d^2\mathbf{q}_1 \int d^2 \mathbf{q}_2 \int d^2 \mathbf{q}_3 \Bigl( \bra{\mathbf{m}_\mathbf{k}} \ket{D^{(4)}[\mathbf{m}_{\mathbf{q}_1}^*,\mathbf{m}_{\mathbf{q}_2}^*,\mathbf{m}_{\mathbf{q}_3}]} s_{\mathbf{q}_1}^* s_{\mathbf{q}_2}^*  s_{\mathbf{q}_3} \\
        +\bra{\mathbf{m}_\mathbf{k}} \ket{D^{(4)}[\mathbf{m}_{\mathbf{q}_1}^*,\mathbf{m}_{\mathbf{q}_2},\mathbf{m}_{\mathbf{q}_3}^*]} s_{\mathbf{q}_1}^* s_{\mathbf{q}_2}  s_{\mathbf{q}_3}^* +\bra{\mathbf{m}_\mathbf{k}} \ket{D^{(4)}[\mathbf{m}_{\mathbf{q}_1},\mathbf{m}_{\mathbf{q}_2}^*,\mathbf{m}_{\mathbf{q}_3}^*]} s_{\mathbf{q}_1} s_{\mathbf{q}_2}^*  s_{\mathbf{q}_3}^*
    \Bigr).
\end{multline}
    From these expressions we derive the coupling coefficients as
\begin{multline}
    F^{(1)}_{\mathbf{q}_1,\mathbf{q}_2,\tilde{\mathbf{q}}_1} = i\frac{d(2\pi)^2 \omega_M}{2 M_S^2} \mathcal{N}_{\mathbf{k}} \mathcal{N}_{\mathbf{q}_1} \mathcal{N}_{\mathbf{q}_2} \mathcal{N}_{\tilde{\mathbf{q}}_1} Z^-_{\mathbf{q}_1\mathbf{q}_2}\biggl[ Z^+_{\mathbf{k}\tilde{\mathbf{q}}_1} \biggl( \alpha_\text{ex}(|\tilde{\mathbf{q}}_1|^2 + |\mathbf{k}-\tilde{\mathbf{q}}_1|^2) - Q_{\mathbf{k}-\tilde{\mathbf{q}}_1} \frac{(k_z-\tilde{q}_{1,z})^2}{|\mathbf{k}-\tilde{\mathbf{q}}_1|^2} \biggr) \\
    \phantom{=\biggl[}+ \Bigl( \sqrt{ \nu_y(\mathbf{k}) \nu_y(\tilde{\mathbf{q}}_1) } + \frac{ (\tilde{q}_{1,y})^2 }{ |\tilde{\mathbf{q}}_1|^2 } \sqrt{ \nu_x(\mathbf{k}) \nu_x(\tilde{\mathbf{q}}_1) } \Bigr) Q_{\tilde{\mathbf{q}}_1} - \sqrt{ \nu_y(\mathbf{k}) \nu_y(\tilde{\mathbf{q}}_1) } \Biggr],
\end{multline}
\begin{align}
\begin{split}
    F^{(2)}_{\mathbf{q}_1,\mathbf{q}_2,\tilde{\mathbf{q}}_2} &= i\frac{d(2\pi)^2 \omega_M}{2 M_S^2} \mathcal{N}_{\mathbf{k}} \mathcal{N}_{\mathbf{q}_1} \mathcal{N}_{\mathbf{q}_2} \mathcal{N}_{\tilde{\mathbf{q}}_2} \Biggl[ 
    Z^-_{\mathbf{q}_1\mathbf{q}_2} Z^-_{\mathbf{k}\tilde{\mathbf{q}}_2}
     \biggl( \alpha_\text{ex} |\mathbf{k}+\tilde{\mathbf{q}}_2|^2 - Q_{\mathbf{k}+\tilde{\mathbf{q}}_2} \frac{(k_z+\tilde{q}_{2,z})^2}{|\mathbf{k}+\tilde{\mathbf{q}}_2|^2} \biggr)\\
    &\phantom{=}+ 2 Z^+_{\mathbf{q}_2\tilde{\mathbf{q}}_2} Z^+_{\mathbf{q}_1\mathbf{k}} \biggl( \alpha_\text{ex} |\mathbf{q}_1-\mathbf{k}|^2 - Q_{\mathbf{q}_1-\mathbf{k}} \frac{(q_{1,z}-k_z)^2}{|\mathbf{q}_1-\mathbf{k}|^2} \biggr)\\
    &\phantom{=} +2 Z^+_{\mathbf{q}_1\tilde{\mathbf{q}}_2} \biggl[ Z^+_{\mathbf{k}\mathbf{q}_2} \alpha_\text{ex} |\mathbf{q}_2|^2 +\Bigl( \sqrt{ \nu_y(\mathbf{k}) \nu_y(\mathbf{q}_2) } + \frac{ (q_{2,y})^2 }{ |\mathbf{q}_2|^2 } \sqrt{ \nu_x(\mathbf{k}) \nu_x(\mathbf{q}_2) } \Bigr) Q_{\mathbf{q}_2} - \sqrt{ \nu_y(\mathbf{k}) \nu_y(\mathbf{q}_2) } \biggr]\\
    &\phantom{=} +Z^-_{\mathbf{q}_1\mathbf{q}_2} \biggl[ Z^-_{\mathbf{k}\tilde{\mathbf{q}}_2} \alpha_\text{ex} |\tilde{\mathbf{q}}_2|^2  +\Bigl( \sqrt{ \nu_y(\mathbf{k}) \nu_y(\tilde{\mathbf{q}}_2) } - \frac{ (\tilde{q}_{2,y})^2 }{ |\tilde{\mathbf{q}}_2|^2 } \sqrt{ \nu_x(\mathbf{k}) \nu_x(\tilde{\mathbf{q}}_2) } \Bigr) Q_{\tilde{q}_2} - \sqrt{ \nu_y(\mathbf{k}) \nu_y(\tilde{\mathbf{q}}_2) } \biggr] \Biggr],
\end{split}
\end{align}
and
\begin{align}
\begin{split}
    F^{(3)}_{\mathbf{q}_1,\mathbf{q}_2,\tilde{\mathbf{q}}_3} &= i\frac{d(2\pi)^2 \omega_M}{2 M_S^2} \mathcal{N}_{\mathbf{k}} \mathcal{N}_{\mathbf{q}_1} \mathcal{N}_{\mathbf{q}_2} \mathcal{N}_{\tilde{\mathbf{q}}_3} \Biggl[ Z^-_{\mathbf{q}_2\tilde{\mathbf{q}}_3}Z^+_{\mathbf{k}\mathbf{q}_1}  \biggl( \alpha_\text{ex} |\mathbf{q}_2+\tilde{\mathbf{q}}_3|^2 - Q_{\mathbf{q}_2+\tilde{\mathbf{q}}_3} \frac{(q_{2,z}+\tilde{q}_{3,z})^2}{|\mathbf{q}_2+\tilde{\mathbf{q}}_3|^2} \biggr)\\
    &\phantom{=}+ 2 Z^+_{\tilde{\mathbf{q}}_3\mathbf{q}_1} Z^-_{\mathbf{k}\mathbf{q}_2}  \biggl( \alpha_\text{ex} |\mathbf{q}_1-\tilde{\mathbf{q}}_3|^2 - Q_{\mathbf{q}_1-\tilde{\mathbf{q}}_3} \frac{(q_{1,z}-\tilde{q}_{3,z})^2}{|\mathbf{q}_1-\tilde{\mathbf{q}}_3|^2} \biggr)\\
    &\phantom{=} +2 Z^+_{\mathbf{q}_1\mathbf{q}_2} \biggl[ Z^-_{\mathbf{k}\tilde{\mathbf{q}}_3} \alpha_\text{ex} |\tilde{\mathbf{q}}_3|^2 +\Bigl( \sqrt{ \nu_y(\mathbf{k}) \nu_y(\tilde{\mathbf{q}}_3) } - \frac{ (\tilde{q}_{3,y})^2 }{ |\tilde{\mathbf{q}}_3|^2 } \sqrt{ \nu_x(\mathbf{k}) \nu_x(\tilde{\mathbf{q}}_3) } \Bigr) Q_{\tilde{\mathbf{q}}_3} - \sqrt{ \nu_x(\mathbf{k}) \nu_x(\tilde{\mathbf{q}}_3) } \biggr]\\
    &\phantom{=} +Z^-_{\mathbf{q}_2\tilde{\mathbf{q}}_3} \biggl[ Z^+_{\mathbf{k}\mathbf{q}_1} \alpha_\text{ex} |\mathbf{q}_1|^2  +\Bigl( \sqrt{ \nu_y(\mathbf{k}) \nu_y(\mathbf{q}_1) } + \frac{ (q_{1,y})^2 }{ |\mathbf{q}_1|^2 } \sqrt{ \nu_x(\mathbf{k}) \nu_x(\mathbf{q}_1) } \Bigr) Q_{\mathbf{q}_1} - \sqrt{ \nu_y(\mathbf{k}) \nu_y(\mathbf{q}_1) } \biggr] \Biggr],
\end{split}
\end{align}
\end{widetext}
where for compactness we have defined the functions
\begin{equation}
    Z^\pm_{\mathbf{q}\mathbf{q}'}\equiv \sqrt{ \nu_y(\mathbf{q}) \nu_y(\mathbf{q}') } \pm \sqrt{ \nu_x(\mathbf{q}) \nu_x(\mathbf{q}') },
\end{equation}
and
\begin{equation}
    Q_\mathbf{q}\equiv -1+\frac{1}{\vert\mathbf{q}\vert d}\left(1-e^{-\vert\mathbf{q}\vert d}\right).
\end{equation}

\section{Magnetoelastic coupling rates}\label{app:magnetoelastic_coupling}
In this appendix we provide the magnetoelastic coupling rates introduced in Sec.~\ref{subsec:magnetoelastic_interaction}. In that section we expressed the magnon-phonon interaction Hamiltonian Eq.~(\ref{eq:me_interaction_Hamiltonian}) with the help of the integral expressions $I^{(j)}$ ($j\in\{1,2,3\}$). These expressions read
\begin{multline}
    I^{(1)}_{\sigma, \mathbf{q}, \mathbf{q}^\prime} = \int d^3\mathbf{r} \frac{B_2}{2M_S \sqrt{\rho(\mathbf{r})}} \\ \times \bigl( L_{x}(\sigma,\mathbf{q},\mathbf{q}^\prime,\mathbf{r}) + L_{y}(\sigma,\mathbf{q},\mathbf{q}^\prime,\mathbf{r})\bigr),
\end{multline}
\begin{multline}
    I^{(2)}_{\sigma, \mathbf{q}, \mathbf{q}^\prime, \mathbf{q}^{\prime\prime}} = \int d^3\mathbf{r} \sum_{i,j\in\{x,y\}}\frac{B_2+(B_1-B_2)\delta_{i,j}}{2 M_S^2 \sqrt{\rho(\mathbf{r})}} \\
    \times  L_{i,j}^{(2)}(\sigma,\mathbf{q},\mathbf{q}^\prime,\mathbf{q}^{\prime\prime},\mathbf{r}),
\end{multline}
\begin{multline}
    I^{(3)}_{\sigma, \mathbf{q}, \mathbf{q}^\prime, \mathbf{q}^{\prime\prime}} = \int d^3\mathbf{r} \sum_{i,j\in\{x,y\}}\frac{B_2+(B_1-B_2)\delta_{i,j}}{2 M_S^2 \sqrt{\rho(\mathbf{r})}} \\
    \times L_{i,j}^{(3)}(\sigma,\mathbf{q},\mathbf{q}^\prime,\mathbf{q}^{\prime\prime},\mathbf{r}),
\end{multline}
where for compactness we defined
\begin{align}
    L_{i}^{(1)}(\sigma,\mathbf{q},\mathbf{q}^\prime,\mathbf{r}) \equiv m_{\mathbf{q}^\prime,i}^*(\mathbf{r})(\partial_i u_{\sigma,\mathbf{q},z}(\mathbf{r}) + \partial_z u_{\sigma,\mathbf{q},i}(\mathbf{r})),
\end{align}
\begin{multline}
    L_{i,j}^{(2)}(\sigma,\mathbf{q},\mathbf{q}^\prime,\mathbf{q}^{\prime\prime},\mathbf{r}) \equiv m_{\mathbf{q}^\prime,i}^* (\mathbf{r})m_{\mathbf{q}^{\prime\prime},j}(\mathbf{r}) \\(\partial_i u_{\sigma,\mathbf{q},j}(\mathbf{r}) + \partial_j u_{\sigma,\mathbf{q},i}(\mathbf{r}) - 2\delta_{i,j} \partial_z u_{\sigma,\mathbf{q},z}(\mathbf{r}) ),
\end{multline}
\begin{multline}
    L_{i,j}^{(3)}(\sigma,\mathbf{q},\mathbf{q}^\prime,\mathbf{q}^{\prime\prime},\mathbf{r}) \equiv m_{\mathbf{q}^\prime,i}^* (\mathbf{r})m_{\mathbf{q}^{\prime\prime},j}^*(\mathbf{r}) \\(\partial_i u_{\sigma,\mathbf{q},j}(\mathbf{r}) + \partial_j u_{\sigma,\mathbf{q},i}(\mathbf{r}) - 2\delta_{i,j} \partial_z u_{\sigma,\mathbf{q},z}(\mathbf{r}) ),
\end{multline}
with indices $i,j,z$ denoting Cartesian components. Plugging in the respective mode functions [Eqs.~\eqref{eq:magnon_modefunction}, \eqref{eq:phonon_modefunction_s1_film}, and \eqref{appu2}] and solving the integrals yields
\begin{widetext}
    \begin{multline}\label{Imp1}
        I^{(1)}_{1, \mathbf{q}, \mathbf{q}^{\prime}} = \delta(\mathbf{q}-\mathbf{q}^{\prime}) \tilde I^{(1)}_{1, \mathbf{q}, \mathbf{q}^{\prime}}\\\equiv \delta(\mathbf{q}-\mathbf{q}^{\prime})\frac{B_2(2\pi)^2 \mathcal{N}_{\mathbf{q}} \tilde{C}_{\mathbf{q}}}{2 M_S \sqrt{\rho_f}} \left( \sqrt{\nu_y(\mathbf{q})} \sin(\phi_q) (\cos(p_t^\prime d) -1 ) + \sqrt{\nu_x(\mathbf{q})} \frac{k}{p_t^\prime} \cos(2\phi_k) \sin(p_t^\prime d) \right) ,
    \end{multline}
    \begin{multline}\label{Imp2}
        I^{(1)}_{2, \mathbf{q}, \mathbf{q}^{\prime}} = \delta(\mathbf{q}-\mathbf{q}^{\prime}) \tilde I^{(1)}_{2, \mathbf{q}, \mathbf{q}^{\prime}}\equiv \delta(\mathbf{q}-\mathbf{q}^{\prime})\frac{B_2(2\pi)^2 \mathcal{N}_{\mathbf{q}} }{2 M_S \sqrt{\rho_f}} \biggl( \sqrt{\nu_y(\mathbf{q})} i \cos(\phi_q) \Bigl( 2(\tilde{A}_{(ii)} + \tilde{B}_{(ii)})  \\+ \left(\frac{p_t^\prime}{q} -\frac{q}{p_t^\prime} \right) ( \tilde{C}_{(ii)y} - \tilde{D}_{(ii)y} ) \Bigr) + \sqrt{\nu_x(\mathbf{q})} 2 \cos(\phi_q) \sin(\phi_q) \Bigl( \frac{q}{p_l^\prime} (\tilde{A}_{(ii)} - \tilde{B}_{(ii)}) +  \tilde{C}_{(ii)y} + \tilde{D}_{(ii)y}  \Bigr) \biggr) ,
    \end{multline}
    \begin{multline}\label{Imp3}
        I^{(j)}_{1, \mathbf{q}, \mathbf{q}^\prime, \mathbf{q}^{\prime\prime}} = \delta(\mathbf{q}^{\prime\prime} - \mathbf{q}_j) \tilde  I^{(j)}_{1, \mathbf{q}, \mathbf{q}^\prime, \mathbf{q}^{\prime\prime}} \equiv \delta(\mathbf{q}^{\prime\prime} - \mathbf{q}_j)\frac{(2\pi)^2 \mathcal{N}_{\mathbf{q}^\prime} \mathcal{N}_{\mathbf{q}^{\prime\prime}} \tilde{C}_{\mathbf{q}} }{2 M_S^2 \sqrt{\rho_f}} \biggl( i B_2 \Bigl(S_j\sqrt{ \nu_y(\mathbf{q}^\prime) \nu_x(\mathbf{q}^{\prime\prime}) } - \sqrt{ \nu_x(\mathbf{q}^\prime) \nu_y(\mathbf{q}^{\prime\prime}) }\Bigr)  \\ \times \cos(\phi_q) (1-\cos(p_t^\prime d))
        -i B_1 \Bigl(\sqrt{ \nu_y(\mathbf{q}^\prime) \nu_y(\mathbf{q}^{\prime\prime}) } + S_j 2 \sqrt{ \nu_x(\mathbf{q}^\prime) \nu_x(\mathbf{q}^{\prime\prime}) }\Bigr) 2\cos(\phi_q) \sin(\phi_q)\frac{q}{p_t^\prime} \sin(p_t^\prime d) \biggr) ,
    \end{multline}
    \begin{multline}\label{Imp4}
        I^{(j)}_{2, \mathbf{q}, \mathbf{q}^\prime, \mathbf{q}^{\prime\prime}} = \delta(\mathbf{q}^{\prime\prime} - \mathbf{q}_j) \tilde I^{(j)}_{2, \mathbf{q}, \mathbf{q}^\prime, \mathbf{q}^{\prime\prime}} \equiv \delta(\mathbf{q}^{\prime\prime} - \mathbf{q}_j) \frac{(2\pi)^2 \mathcal{N}_{\mathbf{q}^\prime} \mathcal{N}_{\mathbf{q}^{\prime\prime}} }{2 M_S^2 \sqrt{\rho_f}} \biggl( B_2 \Bigl(-\sqrt{ \nu_y(\mathbf{q}^\prime) \nu_x(\mathbf{q}^{\prime\prime}) } +S_j \sqrt{ \nu_x(\mathbf{q}^\prime) \nu_y(\mathbf{q}^{\prime\prime}) }\Bigr) \sin(\phi_q) \\ \times \Bigl( 2(\tilde{A}_{(ii)} + \tilde{B}_{(ii)}) + \left(\frac{p_t^\prime}{q} -\frac{q}{p_t^\prime} \right) ( \tilde{C}_{(ii)y} - \tilde{D}_{(ii)y} )\Bigr) \\
        + i B_1 2 \sqrt{ \nu_y(\mathbf{q}^\prime) \nu_y(\mathbf{q}^{\prime\prime}) } \Bigl( \left(\frac{p_t^\prime}{q} -\frac{q}{p_t^\prime} \cos^2(\phi_q) \right) (\tilde{A}_{(ii)} - \tilde{B}_{(ii)}) - (1+\cos^2(\phi_q)) (\tilde{C}_{(ii)y} + \tilde{D}_{(ii)y}) \Bigr)\\
        - S_j i B_1 2 \sqrt{ \nu_x(\mathbf{q}^\prime) \nu_x(\mathbf{q}^{\prime\prime}) } \cos(2\phi_q)\Bigl( \frac{q}{p_l^\prime} (\tilde{A}_{(ii)} - \tilde{B}_{(ii)}) + (\tilde{C}_{(ii)y} + \tilde{D}_{(ii)y}) \Bigr) \biggr) ,
    \end{multline}
\end{widetext}
for $j \in\{2,3\}$. We have defined wave vectors $\mathbf{q}_2\equiv(\mathbf{q}^\prime - \mathbf{q})$ and $\mathbf{q}_3\equiv(\mathbf{q} - \mathbf{q}^\prime)$, sign factors $S_2=1$, $S_3=-1$, and coefficients $\tilde{A}_{(ii)}= A_{(ii)}(e^{ip_l^\prime d}-1)$, $\tilde{B}_{(ii)}= B_{(ii)}(e^{-ip_l^\prime d}-1)$, $\tilde{C}_{(ii)y}= C_{(ii)y}(e^{ip_t^\prime d}-1)$, and $\tilde{D}_{(ii)y}= D_{(ii)y}(e^{-ip_t^\prime d}-1)$. We have also used the factors $p_l^\prime$, $p_l^\prime$ defined in Appendix~\ref{app:phonon_eigenmodes}, i.e., 
\begin{gather}
    p_l^{\prime} \equiv \sqrt{\frac{(\omega^{\text{m}}_\mathbf{q})^2}{c_{l,f}^2} - \mathbf{q}^2} \qquad p_t^{\prime} \equiv \sqrt{\frac{(\omega^{\text{m}}_\mathbf{q})^2}{c_{t,f}^2} - \mathbf{q}^2}.
\end{gather}

Introducing Eqs.~\eqref{Imp1}-\eqref{Imp4} into the magnetoelastic Hamiltonian~\eqref{eq:me_interaction_Hamiltonian} and using Hamilton's equations
\begin{align}
    \dot{s}_\mathbf{k} &= -i|\mathcal{M}|^2 \frac{\partial \mathcal{H}}{\partial s_{\mathbf{k}}^*} ,\\[5pt]
    \dot{a}_{\sigma,\mathbf{k}} &= -i|\mathcal{U}_{\sigma,\mathbf{k}}|^2 \frac{\partial \mathcal{H}}{\partial a_{\sigma,\mathbf{k}}^*} ,
\end{align}
with the total Hamiltonian~\eqref{eq:full_Hamiltonian}, we derive the magnetoelastic contribution to the equations of motion~\eqref{eq:EOM_sk} and \eqref{eq:EOM_ak}, with magnetoelastic coupling strengths given by
\begin{align}
    g_{\sigma,\mathbf{k}} &=-i|\mathcal{M}|^2 \tilde I^{(1)}_{\sigma,\mathbf{k},\mathbf{k}} , \\[5pt]
    G^{(1)}_{\sigma,\mathbf{q},\mathbf{k}} &=-i2|\mathcal{M}|^2 (\tilde I^{(2)}_{ \sigma, \mathbf{q}, \mathbf{k}, (\mathbf{k}-\mathbf{q}) } +\tilde I^{(2)}_{ \sigma, \mathbf{q}, (\mathbf{k}-\mathbf{q}), \mathbf{k} } ) , \\[5pt]
    G^{(2)}_{\sigma,\mathbf{q},\mathbf{k}} &=-i2|\mathcal{M}|^2 (\tilde I^{(2)}_{ \sigma, \mathbf{q}, \mathbf{k}, (\mathbf{k}+\mathbf{q}) } +\tilde I^{(2)}_{ \sigma, \mathbf{q}, (\mathbf{k}+\mathbf{q}), \mathbf{k} } ) , \\[5pt]
    G^{(3)}_{\sigma,\mathbf{q},\mathbf{k}} &=-i2|\mathcal{M}|^2 2\tilde I^{(3)}_{ \sigma, \mathbf{q}, \mathbf{k}, (\mathbf{q}-\mathbf{k}) } ,\label{G31}\\[5pt]
    \tilde{g}_{\sigma,\mathbf{k}} &=-i|\mathcal{U}_{\sigma,\mathbf{k}}|^2 \tilde I^{(1)*}_{\sigma,\mathbf{k},\mathbf{k}} , \\[5pt]
    \tilde{G}^{(1)}_{\sigma,\mathbf{k},\mathbf{q}} &=-i2|\mathcal{U}_{\sigma,\mathbf{k}}|^2 (\tilde I^{(2)*}_{ \sigma, \mathbf{k}, (\mathbf{q}+\mathbf{k}) , \mathbf{q} } +\tilde I^{(2)*}_{ \sigma, \mathbf{k}, \mathbf{q}, (\mathbf{q}+\mathbf{k}) } ) , \\[5pt]
    \tilde{G}^{(3)}_{\sigma,\mathbf{k},\mathbf{q}} &=-i2|\mathcal{U}_{\sigma,\mathbf{k}}|^2 \tilde I^{(3)*}_{ \sigma, \mathbf{k}, \mathbf{q}, (\mathbf{k}-\mathbf{q}) } \label{G32}.
\end{align}
In writing Eqs.~\eqref{G31}-\eqref{G32} we have made use of the symmetry $I^{(3)}_{\sigma,\mathbf{k},\mathbf{q}_1,\mathbf{q}_2} = I^{(3)}_{\sigma,\mathbf{k},\mathbf{q}_2,\mathbf{q}_1}$.

\section{Mode amplitude of harmonics and derivation of parametric instability equations}\label{app:displacement}

In this Appendix we provide expressions for the mode amplitude displacements 
$\alpha^\text{m}_{\xi}$ and $\alpha^\text{p}_{\xi,\sigma}$ defined in Eqs.~\eqref{displacementS} and \eqref{displacementa}, and outline how we simplify the equations of motion in order to simplify them to the expressions Eqs.~\eqref{eq:parametric_amplification_EOM_s} and \eqref{eq:parametric_amplification_EOM_a} given in the main text.

Regarding the mode amplitude displacements, by following the steps indicated in Sec.~\ref{subsec:parametric_excitations_threshold_ordering} we find explicit expressions of $\alpha^\text{m}_{\xi}$ and $\alpha^\text{p}_{\xi,\sigma}$ as polynomials of the original acoustic drive 
$\alpha_\text{d}$. 
In the coefficients of each of these polynomials we retain only leading-order contributions in any nonlinear coupling rate density (magnon-phonon, three-magnon, and four-magnon). Note that this ``weak nonlinear coupling'' approximation further justifies the truncation of the sum in Eqs.~\eqref{displacementS} and \eqref{displacementa} to $N\le 3$ discussed in the main text, as $\{\alpha^\text{m}_{j},\alpha^\text{p}_{j,\sigma}\}$ are,  for $j>3$, of order $3$ or higher in the nonlinear coupling rates.  We obtain the following expressions for the magnon displacement parameters,
 \begin{align}\label{alpham1}
     \alpha^{\text{m}}_1 = \frac{\sum_\sigma g_{\sigma,\mathbf{k}_0}\, \alpha_{\sigma,1}^\text{p}}{-i(\omega_d - \omega_{\mathbf{k}_0}^\text{m})-\gamma_{\mathbf{k}_0}^\text{m}/2 },  
 \end{align}
 
 \begin{multline}\label{alpham2}
     \alpha^{\text{m}}_2 = \frac{\sum_\sigma (-G^{(1)}_{\sigma,\mathbf{k}_0,2\mathbf{k}_0}\, \alpha_{\sigma,1}^\text{p}) \alpha_{1}^\text{m} - T^{(2)}_{\mathbf{k}_0,\mathbf{k}_0} \,\alpha_\text{d}^2 }{-i(2\omega_d - \omega_{2\mathbf{k}_0}^\text{m})-\gamma_{2\mathbf{k}_0}^\text{m}/2 }\\ 
     +\frac{\sum_\sigma g_{\sigma,2\mathbf{k}_0}\, \alpha_{\sigma,2}^\text{p}}{-i(2\omega_d - \omega_{2\mathbf{k}_0}^\text{m})-\gamma_{2\mathbf{k}_0}^\text{m}/2 }, 
 \end{multline}
 \begin{multline}
     \alpha^{\text{m}}_3 = \frac{-\sum_\sigma (G^{(1)}_{\sigma,\mathbf{k}_0,3\mathbf{k}_0}\, \alpha_{\sigma,1}^\text{p} \alpha_{2}^\text{m} + G^{(1)}_{\sigma,2\mathbf{k}_0,3\mathbf{k}_0}\, \alpha_{\sigma,2}^\text{p} \alpha_{1}^\text{m} )} { -i(3\omega_d - \omega^{\text{m}}_{3\mathbf{k}_0})-\gamma^{\text{m}}_{3\mathbf{k}_0}/2 }\\     
     +  \frac{-2 T^{(2)}_{\mathbf{k}_0,2\mathbf{k}_0}\, \alpha_{1}^\text{m} \alpha_{2}^\text{m} + F^{(1)}_{\mathbf{k}_0,\mathbf{k}_0,\mathbf{k}_0} (\alpha_{0}^\text{m})^3 + \sum_\sigma g_{\sigma,3\mathbf{k}_0}\, \alpha_{\sigma,3}^\text{p}} { -i(3\omega_d - \omega^{\text{m}}_{3\mathbf{k}_0})-\gamma^{\text{m}}_{3\mathbf{k}_0}/2 },
 \end{multline}
 and for the phonon displacement parameters,
\begin{align}\label{alphap1}
     \alpha^{\text{p}}_{\sigma,1} &= \frac{-i \alpha_\text{d} + \tilde{g}_{\sigma,\mathbf{k}_0}\,\alpha^{\text{m}}_1}{-i(\omega_d - \omega_{\sigma,\mathbf{k}_0}^\text{p})-\gamma_{\sigma,\mathbf{k}_0}^\text{p}/2 } ,\\
     \alpha^{\text{p}}_{\sigma,2} &= \frac{-\tilde{G}^{(3)}_{\sigma,2\mathbf{k}_0,\mathbf{k}_0}\, (\alpha^{\text{m}}_1)^2 + \tilde{g}_{\sigma,2\mathbf{k}_0}\, \alpha^{\text{m}}_2}{-i(2\omega_d - \omega_{\sigma,2\mathbf{k}_0}^\text{p})-\gamma_{\sigma,2\mathbf{k}_0}^\text{p}/2 } ,\label{alphap2}\\
     \alpha^{\text{p}}_{\sigma,3} &= \frac{-2 \tilde{G}^{(3)}_{\sigma,3\mathbf{k}_0,\mathbf{k}_0}\, \alpha^{\text{m}}_1 \alpha^{\text{m}}_2 +\tilde{g}_{\sigma,3\mathbf{k}_0}\, \alpha^{\text{m}}_3 }{-i(3\omega_d - \omega_{\sigma,3\mathbf{k}_0}^\text{p})-\gamma_{\sigma,3\mathbf{k}_0}^\text{p}/2 } .
\end{align}
These equations are coupled pairwise and can be solved in a recursive manner. First, we solve the $2\times 2$ system of equations~(\ref{alpham1}) and (\ref{alphap1}) to obtain the coefficients $\alpha_1^{\text{m}}$ and $\alpha_1^{\text{p}}$. Introducing these expressions into the next-order equations.~(\ref{alpham2}) and (\ref{alphap2}) we obtain a new $2\times 2$ system for $\alpha_2^{\text{m}}$ and $\alpha_2^{\text{p}}$. By repeating this procedure we determine $\alpha_3^{\text{m}}$ and $\alpha_3^{\text{p}}$.

Once the mode amplitude displacements are calculated, we can write the new equations of motion for the magnon and phonon amplitudes, namely Eqs.~\eqref{eq:EOM_sk} and \eqref{eq:EOM_ak} under the redefinition Eqs.~\eqref{displacementS} and \eqref{displacementa}:
\begin{widetext}
\begin{multline}\label{eq:eom0}
    \dot a_{\sigma,\mathbf{k}} = \left( -i\omega^\text{p}_\mathbf{k} - \frac{\gamma^\text{p}}{2} \right) a_{\sigma,\mathbf{k}} + \tilde{g}_{\sigma,\mathbf{k}} s_{\mathbf{k}} + \int d^2\mathbf{q} \Bigl( \tilde{G}^{(1)}_{\sigma,\mathbf{k},\mathbf{q}} \, s_{\mathbf{q}}^* s_{\mathbf{k}+\mathbf{q}} + \tilde{G}^{(3)}_{\sigma,\mathbf{k},\mathbf{q}} \, s_{\mathbf{q}} s_{\mathbf{k}-\mathbf{q}} \Bigr)
    \\
    -\sum_{\xi=1}^3 \bigg\lbrace 
    \alpha^{\text{m}}_{\xi} e^{-i\xi\omega_\text{d} t} \bigg((\tilde{G}^{(3)}_{\sigma,\mathbf{k}, \xi \mathbf{k}_0}+\tilde{G}^{(3)}_{\sigma,\mathbf{k}, \mathbf{k} - \xi \mathbf{k}_0}) \, s_{\mathbf{k}-\xi\mathbf{k}_0}+\tilde{G}^{(1)}_{\sigma, \mathbf{k}, \xi \mathbf{k}_0}s_{\mathbf{k}+\xi\mathbf{k}_0}\bigg)+\tilde{G}^{(1)}_{\sigma, \mathbf{k}, \xi \mathbf{k}_0-\mathbf{k}} \, \alpha^{\text{m}}_{\xi} e^{-i\xi\omega_\text{d} t} s_{\xi\mathbf{k}_0-\mathbf{k}}^*\bigg\rbrace,
\end{multline}
and
\begin{align}
\begin{split}
    \dot s_\mathbf{k} = &\left( -i\omega^\text{m}_\mathbf{k} - \frac{\gamma^\text{m}_{\mathbf{k}}}{2} \right) s_\mathbf{k} +\sum_{\sigma=1,2} g_{\sigma,\mathbf{k}} a_{\sigma,\mathbf{k}} + \int d^2\mathbf{q}\; \Bigl( T^{(1)}_{\mathbf{q},\mathbf{k+q}}\, s_{\mathbf{q}}^* s_{\mathbf{k+q}} + T^{(2)}_{\mathbf{q},\mathbf{k-q}}\, s_{\mathbf{q}} s_{\mathbf{k-q}}\Bigr)
    \\
    &+\sum_{\sigma=1,2} \int d^2\mathbf{q}\; \Bigl( 
    G^{(1)}_{\sigma,\mathbf{q}, \mathbf{k}} \, a_{\sigma,\mathbf{q}} s_{\mathbf{k}-\mathbf{q}} + 
    G^{(2)}_{\sigma,\mathbf{q}, \mathbf{k}} \, a^{*}_{\sigma,\mathbf{q}} s_{\mathbf{k}+\mathbf{q}} + 
    G^{(3)}_{\sigma,\mathbf{q}, \mathbf{k},} \, a_{\sigma,\mathbf{q}} s_{\mathbf{q}-\mathbf{k}}^{*} \Bigr) 
    \\
    &+\int d^2 \mathbf{q}_1 \int d^2\mathbf{q}_2 \Bigl( 
    F^{(1)}_{\mathbf{q}_1,\mathbf{q}_2,\tilde{\mathbf{q}}_1}\, s_{\mathbf{q}_1} s_{\mathbf{q}_2} s_{\tilde{\mathbf{q}}_1} + 
F^{(2)}_{\mathbf{q}_1,\mathbf{q}_2,\tilde{\mathbf{q}}_2}\, s_{\mathbf{q}_1} s_{\mathbf{q}_2} s_{\tilde{\mathbf{q}}_2}^*+ F^{(3)}_{\mathbf{q}_1,\mathbf{q}_2,\tilde{\mathbf{q}}_3)}\, s_{\mathbf{q}_1} s_{\mathbf{q}_2}^* s_{\tilde{\mathbf{q}}_3}^* \Bigr) 
\\
&-\sum_{\xi=1}^3 \alpha^{\text{m}}_\xi e^{-i\xi\omega_\text{d}t}\Bigg\lbrace
X_{\mathbf{k}}^{(\xi)}  s_{\mathbf{k}-\xi\mathbf{k}_0} +\tilde{X}_{\mathbf{k}}^{(\xi)}  s_{\xi\mathbf{k}_0-\mathbf{k}}^*
    + \sum_{\sigma =1,2} \left(G^{(1)}_{\sigma,\mathbf{k}-\xi \mathbf{k}_0, \mathbf{k}} a_{\sigma, \mathbf{k}-\xi\mathbf{k}_0}
    + G^{(2)}_{\sigma, \xi \mathbf{k}_0-\mathbf{k}, \mathbf{k}}  a_{\sigma, \xi\mathbf{k}_0-\mathbf{k}}^*\right)
\\
&\phantom{\sum_\xi \alpha_\xi^{(m)}e^{-i\xi\omega_\text{d}t}}+\int d^2\mathbf{q} \Big(
Y^{(\xi)}_{\mathbf{k}\mathbf{q}} s_{\mathbf{q}} s_{\mathbf{k}-(\xi\mathbf{k}_0 + \mathbf{q})}
+
\tilde{Y}^{(\xi)}_{\mathbf{k}\mathbf{q}} s_{\mathbf{q}} s_{(\xi\mathbf{k}_0 + \mathbf{q})-\mathbf{k}}^*
+
F^{(3)}_{\xi\mathbf{k}_0,\mathbf{q},\xi\mathbf{k}_0-(\mathbf{k}+\mathbf{q})} \, s_\mathbf{q}^* s_{\xi\mathbf{k}_0-(\mathbf{k}+\mathbf{q})}^*\Big)\Bigg\rbrace
\\
&-\sum_{\xi=1}^3 \alpha^{\text{m}*}_\xi e^{i\xi\omega_\text{d}t} \Bigg\lbrace
W_{\mathbf{k}}^{(\xi)}s_{\mathbf{k}+\xi\mathbf{k}_0} + \sum_{\sigma =1,2} G^{(3)}_{\sigma, \mathbf{k}+\xi \mathbf{k}_0, \mathbf{k}}  a_{\sigma, \mathbf{k}+\xi\mathbf{k}_0} +\int d^2\mathbf{q}\Big(
F^{(2)}_{\mathbf{q},(\xi\mathbf{k}_0+\mathbf{k})-\mathbf{q},\xi\mathbf{k}_0} s_\mathbf{q} s_{(\xi\mathbf{k}_0+\mathbf{k})-\mathbf{q}}
\\
&\phantom{-\sum_{\xi=1}^3 \alpha^{\text{m}*}_\xi e^{i\xi\omega_\text{d}t}}
+
\left( F^{(3)}_{\mathbf{q},\xi\mathbf{k}_0,\mathbf{q}-(\xi\mathbf{k}_0+\mathbf{k})} +  F^{(3)}_{\mathbf{q},\mathbf{q}-(\xi\mathbf{k}_0+\mathbf{k}),\xi\mathbf{k}_0} \right) s_{\mathbf{q}} s_{\mathbf{q}-(\xi\mathbf{k}_0+\mathbf{k})}^*
\Big)
\Bigg\rbrace
    \\
    &+ \sum_{\xi,\zeta=1}^3 \alpha^{\text{m}}_\xi \alpha^{\text{m}}_\zeta e^{-i(\xi+\zeta) \omega_\text{d} t}
    \Big( V_{\mathbf{k}}^{(\zeta\xi)}s_{\mathbf{k}-(\xi+\zeta)\mathbf{k}_0}
+
F^{(2)}_{\zeta\mathbf{k}_0,\xi\mathbf{k}_0,(\xi+\zeta)\mathbf{k}_0-\mathbf{k}}\, s_{(\xi+\zeta)\mathbf{k}_0-\mathbf{k}}^*
    \Big)
    \\
&+\sum_{\xi,\zeta=1}^3 \alpha^{\text{m}}_\xi \alpha^{\text{m}*}_\zeta e^{-i(\xi-\zeta) \omega_\text{d} t}
\Big(
\tilde{V}_{\mathbf{k}}^{(\zeta\xi)} s_{\mathbf{k}+(\zeta-\xi)\mathbf{k}_0}
+
\left( F^{(3)}_{\xi\mathbf{k}_0,\zeta\mathbf{k}_0,(\xi-\zeta)\mathbf{k}_0-\mathbf{k}} + F^{(3)}_{\xi\mathbf{k}_0,(\xi-\zeta)\mathbf{k}_0-\mathbf{k},\zeta\mathbf{k}_0} \right) s_{(\xi-\zeta)\mathbf{k}_0-\mathbf{k}}^*
\Big)\\
    &+\sum_{\xi,\zeta=1}^3 \left( F^{(3)}_{(\xi+\zeta)\mathbf{k}_0+\mathbf{k},\xi\mathbf{k}_0,\zeta\mathbf{k}_0} \right) \alpha^{\text{m}*}_\xi \alpha^{\text{m}*}_\zeta e^{i(\xi+\zeta)\omega_\text{d} t} \, s_{(\xi+\zeta)\mathbf{k}_0+\mathbf{k}}.
\end{split}
\end{align}
\end{widetext}
Here we have used the short-hand notations $\tilde{\mathbf{q}}_{1,2,3}$ defined in Appendix~\ref{app:magnon_couoplig_rates}, and defined the following auxiliary functions for compactness:
\begin{multline}
    Y^{(\xi)}_{\mathbf{k}\mathbf{q}} \equiv F^{(1)}_{\mathbf{q},\mathbf{k}-(\xi\mathbf{k}_0+\mathbf{q}),\xi\mathbf{k}_0} +\\+ F^{(1)}_{\mathbf{q},\xi\mathbf{k}_0,\mathbf{k}-(\xi\mathbf{k}_0+\mathbf{q})} + F^{(1)}_{\xi\mathbf{k}_0,\mathbf{q},\mathbf{k}-(\xi\mathbf{k}_0+\mathbf{q})},
\end{multline}
\begin{equation}
    \tilde{Y}^{(\xi)}_{\mathbf{k}\mathbf{q}}\equiv F^{(2)}_{\mathbf{q},\xi\mathbf{k}_0,(\xi\mathbf{k}_0+\mathbf{q})-\mathbf{k}} +  F^{(2)}_{\xi\mathbf{k}_0,\mathbf{q},(\xi\mathbf{k}_0+\mathbf{q})-\mathbf{k}},
\end{equation}
\begin{equation}   V_{\mathbf{k}}^{(\zeta\xi)}\equiv  F^{(1)}_{\zeta\mathbf{k}_0,\mathbf{k}-(\xi+\zeta)\mathbf{k}_0,\xi\mathbf{k}_0} + 2 F^{(1)}_{\zeta\mathbf{k}_0,\xi\mathbf{k}_0,\mathbf{k}-(\xi+\zeta)\mathbf{k}_0},
\end{equation}
\begin{equation}
\tilde{V}_{\mathbf{k}}^{(\zeta\xi)} \equiv  F^{(2)}_{\mathbf{k}+(\zeta-\xi)\mathbf{k}_0,\xi\mathbf{k}_0,\zeta\mathbf{k}_0} + F^{(2)}_{\xi\mathbf{k}_0,\mathbf{k}+(\zeta-\xi)\mathbf{k}_0,\zeta\mathbf{k}_0},
\end{equation}
\begin{equation}
    X_{\mathbf{k}}^{(\xi)}\equiv  2 T^{(2)}_{\xi\mathbf{k}_0,\mathbf{k}-\xi\mathbf{k}_0}+\sum_{\sigma =1,2} ( G^{(1)}_{\sigma,\xi \mathbf{k}_0, \mathbf{k}} \, \alpha^{\text{p}}_{\xi,\sigma}/\alpha^{\text{m}}_{\xi} ),
\end{equation}
\begin{equation}
    \tilde{X}_{\mathbf{k}}^{(\xi)} \equiv T^{(1)}_{ \xi\mathbf{k}_0-\mathbf{k}, \xi\mathbf{k}_0} + \sum_{\sigma =1,2} ( G^{(3)}_{\sigma, \xi \mathbf{k}_0, \mathbf{k}} \, \alpha^{\text{p}}_{\xi,\sigma}/\alpha^{\text{m}}_{\xi}) ,
\end{equation}
\begin{equation}
    W_{\mathbf{k}}^{(\xi)}\equiv T^{(1)}_{ \xi\mathbf{k}_0, \mathbf{k}+\xi\mathbf{k}_0 } + \sum_{\sigma =1,2} ( G^{(2)}_{\sigma,\xi \mathbf{k}_0, \mathbf{k}} \, \alpha^{\text{p}*}_{\xi,\sigma}/\alpha^{\text{m}*}_{\xi}).
\end{equation}

To simplify the above expressions we first neglect all terms involving products of two amplitudes $\{s_\mathbf{k}s_\mathbf{k'}, s_\mathbf{k}s_\mathbf{k'}^*,s_\mathbf{k}a_\mathbf{k'}, s_\mathbf{k}a_\mathbf{k'}^*, s_\mathbf{k}^*a_\mathbf{k'}, s_\mathbf{k}^*a_\mathbf{k'}^*\}$, as these terms are proportional to nonlinear coupling coefficients and thus much smaller than
certain linear terms (namely the ones proportional to magnon and phonon frequencies, to magnon-phonon linear coupling rate, or to a nonlinear coupling rate times a driving amplitude  $\alpha_\xi^{(m)} \gg s_\mathbf{q}$). This approximation is thus valid for driven systems such as the ones considered in this work. Second, we note that in the remaining equations, the terms on the right-hand side can be classified into three families: (i) diagonal terms (i.e., terms proportional to $-i\omega^\text{p}_\mathbf{k} - \gamma^\text{p}/2$ and $-i\omega^\text{m}_\mathbf{k} - \gamma^\text{m}/2$) and potentially strong resonant couplings proportional to $g_{\sigma\mathbf{k}}$ and $\tilde{g}_{\sigma\mathbf{k}}$; (ii) coupling terms proportional to other amplitudes $s_{\mathbf{k}'}$ and $a_{\mathbf{k}'}$, and (iii) coupling terms proportional to complex conjugates of other amplitudes $s^*_{\mathbf{k}'}$ and $a^*_{\mathbf{k}'}$. Terms (ii) represent energy exchange between modes and cannot result in any parametric instability. In other words, if the amplitudes are $0$ at initial time, they cannot increase by the dynamical evolution generated by these terms. Conversely, terms (iii) represent parametric processes which can result in the build-up of non zero amplitudes. Processes (i) and (iii) are thus the most relevant for the dynamics of parametric instability, while processes (ii) can be neglected as they will not induce strong modifications. By neglecting these terms we arrive to Eqs.~\eqref{eq:parametric_amplification_EOM_s} and \eqref{eq:parametric_amplification_EOM_a}.

\section{Solutions of three-mode instability equations}\label{app:3mode_stability}
In this Appendix we summarize the properties of the steady-state solutions of Eqs.~\eqref{eq:3-mode_eq4}-\eqref{eq:3-mode_eq6} introduced in Sec.~\ref{subsec:3mode_system}. For simplicity and generality, we simplify labels as $\{\mathbf{k}_0,\mathbf{k},\mathbf{k}_0-\mathbf{k}\}\to\{1,2,3\}$ and rename the coupling rates $\{ 2 T^{(2)}_{\mathbf{k},\mathbf{k}_0-\mathbf{k}},T^{(1)}_{\mathbf{k}_0-\mathbf{k},\mathbf{k}_0}
,T^{(1)}_{\mathbf{k},\mathbf{k}_0}\}\to \{g_1,g_2,g_3\}$, thus casting Eqs.~\eqref{eq:3-mode_eq4}-\eqref{eq:3-mode_eq6} as
\begin{align}
    \dot{s}_1 = &A_1 s_{1} +g_1 s_2 s_3 + \alpha, \label{AppE:e1}
    \\[5pt]
        \dot{s}_2 = &A_2 s_2 +  g_2s_1s_3^*,\label{AppE:e2}
    \\[5pt]
    \dot{s}_3 = &A_3 s_3 +  g_3 s_1s_2^*.\label{AppE:e3}
\end{align}
Here we have also renamed
\begin{equation}
    A_j\equiv -i\Delta_j-\frac{\gamma_j}{2}\equiv \vert A_j\vert e^{i\theta_j}.
\end{equation}
We assume the couplings $g_j$ are real numbers, and $g_2,g_3>0>g_1$. The driving strength $\alpha$ is an arbitrary complex number. We recall that $\Delta_j\equiv \omega_j-\Omega_j$ with $\Omega_1=\omega_d$ and $\Omega_2+\Omega_3=\omega_d$.

\subsection{Steady-state solutions}

We now focus on determining the steady-state solutions by setting the left-hand side of Eqs.~\eqref{AppE:e1}-\eqref{AppE:e3} to zero. The following trivial solution is straightforward,
\begin{equation}
    s_2=s_3=0 \hspace{0.3cm} ;\hspace{0.3cm} s_1=-\frac{\alpha}{A_1}.
\end{equation}
Let us look for non-trivial steady-state solutions for which $s_2,s_3\ne 0$. In this case we can combine Eqs.~\eqref{AppE:e2} and \eqref{AppE:e3} to find the alternative solution
\begin{equation}\label{s1mod}
    \vert s_1\vert^2 = \frac{A_2A_3^*}{g_2g_3},
\end{equation}
which does not depend on the drive $\alpha$. Since the denominator is real and positive, the existence of this solution imposes a condition on the detunings, $\text{Im}[A_2A_3^*]=0$. Together with the relation $\Omega_2+\Omega_3=\omega_d$, this condition fully determines the two frequencies
\begin{equation}
    \left[
    \begin{array}{c}
         \Omega_2  \\
         \Omega_3 
    \end{array}
    \right]=\frac{1}{\gamma_2+\gamma_3}\left[
    \begin{array}{c}
         \gamma_3\omega_2-\gamma_2\omega_3+\gamma_2\omega_d  \\
         \gamma_2\omega_3-\gamma_3\omega_2+\gamma_3\omega_d 
    \end{array}
    \right].
\end{equation}

To determine the complex phase of $s_1$ we combine Eqs.~\eqref{AppE:e1}-\eqref{AppE:e3} into
\begin{align}
\begin{split}
    A_1 s_1+\alpha &= \frac{g_2 g_3 s_1^2 (A_1^* s_1^*+\alpha^*)}{A_2 A_3 }  
    \\ &=(A_1^* s_1^*+\alpha^*)e^{i2(\phi_1-\theta_3)}
\end{split}
\end{align}
where in the last step we have used Eq.~(\ref{s1mod}) and written $s_1=\vert s_1\vert e^{i\phi_1}$. Rearranging this expression we find a quadratic equation for the variable $e^{i\phi_1}$, whose solution is
\begin{multline}\label{phase1}
    e^{i\phi_1}=e^{i(\arg{\alpha}+\theta_3)}\Bigg(i\frac{\vert s_1 A_1\vert\sin(\theta_1+\theta_3)}
    {\vert\alpha\vert}
    \\
    -\sqrt{1-\frac{\vert s_1 A_1\vert^2\sin^2(\theta_1+\theta_3)}
    {\vert\alpha^2\vert}}\Bigg),
\end{multline}
where we define $A_1\equiv\vert A_1\vert e^{i\theta_1}$.
The second solution of the quadratic equation is discarded as it can be shown to lead to an unphysical solution where $\vert s_2\vert^2<0$.
With the above considerations, the expressions for the mode amplitudes of modes $2$ and $3$ are easily obtained as
\begin{equation}\label{s2mod}
    \vert s_2\vert^2=\frac{A_3}{g_1g_3}\left(\frac{\alpha}{s_1}+A_1\right),
\end{equation}
\begin{equation}
    \vert s_3\vert^2=\frac{A_2}{g_1g_2}\left(\frac{\alpha}{s_1}+A_1\right),
\end{equation}
which correspond to Eqs.~\eqref{s1constant}-\eqref{s2constant} in the main text. The phase condition Eq.~(\ref{phasecondition}) also directly follows.

As a final task we determine under which parameter regime does the above non-trivial solution exist. This regime follows from imposing that Eq.~(\ref{s2mod}) is positive. Introducing explicitly the phase expression Eq.~(\ref{phase1}) and manipulating the expression, this condition results in the inequality
\begin{equation}
    \cos(\theta_1+\theta_3)< \sqrt{\frac{g_2g_3 \vert \alpha\vert^2 }{\vert A_2 A_3\vert \vert A_1 \vert^2}-\sin^2(\theta_1+\theta_3)}
\end{equation}
which must be fulfilled. Note that $\cos\theta_j<0$ as $\text{Re}[A_j]<0$, and $\cos(\theta_1+\theta_3)>0$. For resonant drive $\theta_1=\pi$ and we obtain the condition
\begin{equation}
     \vert\alpha\vert^2  >\frac{\vert A_2 A_3\vert\vert A_1\vert^2}{g_2g_3} =  \alpha_c^2 ,
\end{equation}
where $\alpha_c$ is the critical threshold obtained in Sec.~\ref{sec:parametrc_instability}. 
In other words, the non-trivial steady-state solution only exists above the critical threshold.

\bibliographystyle{apsrev4-2}

\bibliography{biblio}

\end{document}